\documentclass{aa}
\usepackage{graphicx}
\usepackage{bm}
\usepackage{natbib}
\usepackage{appendix}
\usepackage{float}
\usepackage{subcaption}
\usepackage[utf8]{inputenc}
\usepackage{multirow}
\usepackage{footmisc}
\usepackage{xcolor}
\usepackage[normalem]{ulem}
\usepackage[breaklinks=true]{hyperref}
\hypersetup{colorlinks=true,linkcolor=blue,citecolor=blue,filecolor=blue,urlcolor=blue}
\usepackage{txfonts}
\usepackage{placeins}
\DeclareUnicodeCharacter{2212}{-}
\DeclareUnicodeCharacter{2212}{-}
\DeclareUnicodeCharacter{0308}{-}
\DeclareUnicodeCharacter{0302}{-}

\makeatletter
\renewcommand*\aa@pageof{, page \thepage{} of \pageref*{LastPage}}
\makeatother
\phantomsection
\newcommand{\paperone}{\citealt{sbaffoni+25} (hereafter Paper~I)}
\newcommand{\paperI}{\citetalias{sbaffoni+25}}

\begin{document} 
\defcitealias{sbaffoni+25}{Paper~I}

   \title{Galaxy And Mass Assembly (GAMA): Deconstructing the galaxy stellar mass function by star formation and environment}

   \author{A. Sbaffoni
          \inst{1,2}
          \and
          J. Liske \inst{1,2}
          \and
          A. S. G. Robotham \inst{3}
          \and
          L. J. M. Davies \inst{3}
          \and
          S. P. Driver \inst{3}
          \and
          E. N. Taylor \inst{4}
          }

   \institute{Hamburger Sternwarte, Universität Hamburg, Gojenbergsweg 112, 21029 Hamburg, Germany
         \and
             Cluster of Excellence Quantum Universe, Universität Hamburg, Luruper Chaussee 149, 22761 Hamburg, Germany
        \and
            International Centre for Radio Astronomy Research (ICRAR), University of Western Australia, Crawley, WA 6009, Australia
        \and
            Centre for Astrophysics and Supercomputing, Swinburne University of Technology, Hawthorn, VIC 3122, Australia
            }

    \titlerunning{title running here}
    \authorrunning{Sbaffoni A. et al.}

  \abstract
   {Using the equatorial Galaxy and Mass Assembly (GAMA) dataset, we investigate how the low-redshift galaxy stellar mass function (GSMF) varies across different galaxy populations and as a function of halo mass. We find that: (i) The GSMF of passive and star-forming galaxies are well described by a double and a single Schechter function, respectively, although the inclusion of a second component for the star-forming population yields a more accurate description. 
   Furthermore, star-forming galaxies dominate the low-mass end of the total GSMF, whereas passive galaxies mainly shape the intermediate-to-high-mass regime. (ii) The GSMF of central galaxies dominates the high-mass end, whereas satellites and ungrouped galaxies shape the intermediate-to-low-mass regime. Additionally, we find a relative increase in the abundance of low-mass galaxies moving from dense group environments to isolated systems. (iii) More massive halos host more massive galaxies, have a higher fraction of passive systems, and show a steeper decline in the number of intermediate-mass galaxies. Finally, our results reveal larger differences between passive and star-forming GSMFs than predicted by a phenomenological quenching model, but generally confirm the environmental quenching trends for centrals and satellites reported in other works.}

   \keywords{galaxies: evolution -- galaxies: fundamental parameters -- galaxies: distances and redshifts -- galaxies: luminosity function, mass function -- galaxies: stellar content -- cosmology: large-scale structure of Universe}

    \maketitle

\section{Introduction}
The rate at which new stars form in galaxies is a crucial aspect of understanding the initial formation and subsequent evolution of galaxies. Galaxies grow primarily through two mechanisms, namely star formation \citep[SF;][]{kennicutt+98a} and hierarchical merging \citep{baugh+06}, which are not independent. The first process is closely related to a galaxy’s internal properties, such as its atomic and molecular gas content \citep{kennicutt+98a,keres+05}, stellar \citep{brinchmann+04,daddi+07,elbaz+07,noeske+07} and dust mass \citep{dacunha+10}, morphology \citep{kauffmann+03,guglielmo+15}, activity of active galactic nuclei \citep[AGN;][]{netzer+09,thacker+14} and metallicity \citep{ellison+08,mannucci+10,lara-lopez+13}. In contrast, galaxy mergers are driven by external factors, particularly the local environment in 6D phase space, which determines how galaxies interact in pairs, groups, or clusters \citep{mcintosh+08,ellison+10,deravel+11}. In addition, galaxy mergers and other close interactions with nearby galaxies can enhance or inhibit SF \citep{davies+15}. Measuring the star formation rate (SFR) across different galaxy types and environments is therefore essential to understanding how galaxies accumulate mass through SF and how their surroundings shape this process.

The advent of large-scale galaxy surveys, such as the Two-degree Field Galaxy Redshift Survey (2dFGRS; \citealt{colless+01}) and the Sloan Digital Sky Survey (SDSS; \citealt{abazajian+09}) has revealed that galaxies in the local Universe can be broadly classified into two populations: at fixed stellar mass, the early-type elliptical galaxies can be interpreted as old, red, quiescent (or passive) systems, which have more spheroidal morphologies and little or no active SF, whereas the late-type spiral galaxies as blue star-forming systems, which have typically disc-like morphologies and are actively converting gas into new stars \citep[i.e. high SFR;][]{shen+03,blanton+03,baldry+04,balogh+04a,bell+04a,brinchmann+04,ellis+05,driver+06,papovich+12,taylor+15}. Additionally, early-type galaxies are on average more massive and more commonly found in groups or clusters than late-type galaxies \citep{dressler+80,kauffmann+04,kauffmann+03,blanton+05,baldry+06,vanderwel+08}.

Since colour is more easily measured than morphology in large imaging surveys, it is now preferable to classify galaxies as belonging to the red or blue sequence rather than as early- or late-type systems. In this context, the colour-magnitude diagram (CMD) is a fundamental diagnostic tool. On the one hand, the colour provides key insights into galaxies' stellar populations, acting as a proxy for the dust-corrected, luminosity-weighted mean stellar age. This, in turn, reflects the average specific SFR (sSFR) over long ($\sim$ Gyr) timescales. In other words, a galaxy’s colour is directly tied to its SF activity, dust content, and chemical enrichment history, making it easier to interpret within theoretical models. On the other hand, a galaxy’s absolute magnitude represents its total integrated starlight and often serves as a proxy for its stellar mass. As a result, the CMD effectively traces the evolution of SF as a function of stellar mass. However, a key challenge arises from the significant overlap in the (optical) colour distributions of the red and blue populations \citep{baldry+04,taylor+15}. This introduces both a practical issue in separating the two populations and a deeper conceptual ambiguity in defining red and blue galaxies. 

A clearer bimodality emerges when considering the SFR or sSFR versus stellar mass \citep{balogh+04a,moustakas+13,davies+16,davies+19a}, separating galaxies into passive and star-forming ones, and suggesting that these populations may represent distinct and possibly sequential evolutionary stages. In particular, investigations of this bimodality across different redshifts show that the passive population has nearly doubled in terms of stellar mass, stellar mass density, and number density over the past $\sim 7$ Gyr, although the two populations seem to be roughly equivalent in terms of total stellar mass at z $\sim 1$ \citep{bell+04a,arnouts+07,foltz+18}. However, the existence of two distinct colour distributions implies that the transition from blue to red must occur over a broad range of stellar masses and on relatively short timescales \citep{balogh+04b}. Besides stellar mass, environmental factors play a crucial role in shaping the transition from actively star-forming to passive quiescent systems. \paperone{} provided a detailed overview of the various definitions of environments used in the literature. These include the geometric classification of galaxies (into voids, sheets, filaments, and clusters, groups, or knots), the distinction between grouped and ungrouped galaxies (also referred to as field galaxies), the classification based on halo mass, and the local density estimates averaged over a given scale. In particular, many studies demonstrate that galaxies in groups or clusters, at a given stellar mass, are more likely to be red \citep{kauffmann+04,baldry+06,vanderburg+18,reeves+21} and quiescent \citep{dressler+80,blanton+05,woo+13} than galaxies in the field. Additionally, increased local environmental density is often correlated with lower SFR \citep{schaefer+17,schaefer+19}, a lower fraction of star-forming galaxies \citep{barsanti+18}, and changes in their stellar kinematics \citep{vandesande+21}.

Cosmological models incorporate quenching mechanisms to reproduce the observed galaxy colour bimodality \citep{bell+03,baldry+04,balogh+04b,taylor+15}, and in particular the evolving mass functions of red and blue galaxies \citep{bell+04b,arnouts+07,drory+09,peng+10,ilbert+10,brammer+11}. In these models, quenching primarily affects more massive galaxies and/or galaxies in groups or clusters, by either depleting their existing gas reservoirs or preventing the inflow of new material. %In the local Universe, environmental effects and internal quenching mechanisms can be fully decoupled \citep{baldry+06,peng+10,kovac+14}.
In the local Universe, environmental effects and internal quenching mechanisms have been argued to act independently of each other \citep{peng+10,kovac+14}. Specifically, \cite{peng+10} distinguish between mass quenching, which depends on SFR and predominantly affects massive galaxies regardless of their environment, and environmental quenching, which acts independently of stellar mass and preferentially affects galaxies in groups or clusters \citep{davies+16,kawinwanichakij+17}. Therefore, the observed quenching results from a combination of these two mechanisms.

The exact mechanisms responsible for quenching are still under debate. Mass quenching is commonly attributed to feedback mechanisms, including supernovae and galactic winds \citep{oppenheimer+10}, differences between hot and cold accretion modes due to the presence or absence of persistent shocks in infalling gas \citep{keres+05,dekel+06,cattaneo+08,vandenbosch+08}, and AGN activity, especially in high-mass systems \citep{benson+03,croton+06,menci+06,bower+06,bower+08,tremonti+07,somerville+08,davies+25a}. Although different cosmological models implement these mechanisms in various ways, they generally agree that mass quenching correlates strongly with halo mass, primarily through the suppression of gas cooling in massive haloes.

At the same time, other processes have been proposed as drivers of environmental quenching. These include cold gas stripping through tidal interactions and ram pressure \citep{gunn+72,moore+99,brough+13,brown+17,poggianti+17,barsanti+18}, or hot gas removal through strangulation or starvation \citep{larson+80,moore+99,nichols+11,peng+15}, as well as galaxy harassment and dry mergers \citep{moore+96,bialas+15}. These processes lead to the depletion of star-forming gas, particularly in galaxy groups and clusters \citep{barsanti+18,trussler+20,sotilloramos+21}. These environmental mechanisms primarily affect satellite galaxies, whereas central galaxies remain largely unaffected. Centrals, as being the most massive galaxies in their group and located at the centre of their halo, retain their gas reservoirs because of minimal tidal interactions, harassment, or stripping. As a result, centrals and satellites undergo different quenching mechanisms, leading to different passive fractions, even when controlling for other factors \citep{vandenbosch+08,weinmann+09,peng+12,wetzel+12,knobel+13,robotham+14,grootes+17,davies+25b}. This framework represents the accepted model of environmental quenching, where satellite galaxies experience additional suppression of SF in groups and clusters \citep{wetzel+13,treyer+18,davies+25b}, particularly when they are significantly less massive than their central galaxy.

An important approach to studying environmental quenching involves linking the evolutionary histories of galaxies to the environments defined by their host dark matter (DM) halos. Group catalogues are powerful tools in this context, as they provide a crucial bridge between astrophysical observations and DM halos predicted by the $\Lambda$CDM cosmological paradigm \citep{press+74,zheng+24}. 
Since galaxy groups represent the observable counterparts of DM haloes, they provide direct insight into the physical processes shaping these structures over cosmic time. For example, these catalogues allow for the study of DM dynamics \citep{plionis+06,robotham+08}, and reveal how galaxies are distributed within halos \citep{cooray+02,yang+03,cooray+06,robotham+06,robotham+10b}. In particular, using the GAMA galaxy group catalogue \citep[G$^{3}$C;][]{robotham+11}, we can investigate galaxy evolution as a function of halo mass, especially in statistically significant low-mass regimes. This represents a significant improvement over earlier spectroscopic surveys such as SDSS and 2dFGRS, which, being mostly single-pass, suffer from spectroscopic incompleteness in groups and clusters. In contrast, GAMA, with at least six passes per sky unit, achieves high completeness across all angular scales \citep{robotham+10a,liske+15}.%\citep{robotham+10a,driver+11}.

The galaxy stellar mass function \citep[GSMF;][]{bell+03,baldry+08,baldry+12,wright+17,driver+22}, which describes the number density of galaxies as a function of stellar mass, is one of the most fundamental measurements in extragalactic astronomy, containing valuable information on the assembly history of stellar mass and the evolution of SFR through cosmic time. Quenching mechanisms influence both the shape and the evolution of the GSMF, making it a powerful tool for understanding the physical processes responsible for the observed CMD bimodality \citep{bell+03,baldry+04,faber+07,martin+07,schawinski+14}. Abundance matching between the theoretical halo mass function (HMF) and the observed GSMF reveals that the fraction of baryonic mass converted into stars increases with mass, reaching a peak before declining again \citep{marinoni+02,shankar+06,baldry+08,conroy+09,guo+10,moster+10}. Therefore, cosmological models of galaxy formation must account for the preferred mass for SF efficiency, the relatively shallow low-mass slope compared to the HMF, and the exponential cutoff at high masses \citep{oppenheimer+10}. In other words, assuming a direct correlation between stellar and halo mass results in an overestimation in the number of both low- and high-mass galaxies. This discrepancy between the GSMF and the HMF is thought to be regulated by the feedback mechanisms discussed above. 

The mass distributions of star-forming and passive galaxies, as well as those of centrals and satellites, show significant differences. \cite{peng+10} proposed a phenomenological model in which the combination of mass and environment quenching naturally produces a quasi-static single Schechter function for star-forming galaxies, governed entirely by the mass quenching rate, and a double Schechter function for passive galaxies, with its components corresponding to the two distinct quenching mechanisms. In a follow-up study, \cite{peng+12} showed that environmental quenching affects only satellite galaxies, whereas central galaxies are subject to mass-driven quenching. As a result, the GSMF of passive centrals is well described by a single Schechter function, while that of passive satellites requires a double Schechter form because of the combined effects of mass and satellite quenching. In addition, halo mass is a fundamental driver of galactic properties, including their stellar mass. \cite{vazquez+20} showed that the mass functions of centrals and satellites vary systematically with halo mass, and that the red galaxy fraction increases with halo mass, highlighting the ongoing suppression of SF in galaxy groups. 

This study aims to clarify the respective contributions of mass and environmental quenching in shaping the GSMF. To this end, we present a detailed characterisation of the mass functions for star-forming and passive galaxies, centrals and satellites, and their dependence on halo mass, using GAMA, which offers robust measurements of galaxy properties and group environments at low redshift.

This paper is organised as follows. In Sect.~\ref{2}, we first describe the precise GAMA data products that have been used for the present analysis. In Sects.~\ref{3} and \ref{4}, we then explain the complexities of our target selection and the derivation of different galaxy subsamples, respectively. In Sect.~\ref{5}, we describe our method of deriving the GSMF. In Sect.~\ref{6}, we present our results separately for each galaxy subsample. In Sect.~\ref{7}, we compare our results to previous studies that also investigated the dependence of the GSMF on the galaxy type and environment. Finally, in Sect.~\ref{8}, we draw our conclusions. Throughout this paper, we assume a `737' cosmology, with $(H_{0}, \Omega_{\rm M}, \Omega_{\rm \Lambda}) = (70, 0.3, 0.7)$, corresponding to the same cosmological model used in most GAMA studies on the GSMF. For all physical quantities that depend on $H_{0}$ we include this dependency using $h_{70} = H_{0}/(70$~km~s$^{-1}$~Mpc$^{-1})$.

\section{Data}\label{2}
Our data are part of both the GAMA\footnote{\url{https://www.gama-survey.org/}}~II \citep{liske+15} and the GAMA III \citep{bellstedt+20a,driver+22} surveys. GAMA is a spectroscopic and multi-wavelength imaging survey of $\sim$$238\,000$ galaxies down to $r<19.8$~mag over $\sim$$286$~deg$^{2}$ of sky, split into 5 survey regions, including 3 equatorial fields of $12 \times 5$~deg$^{2} = 60$~deg$^{2}$ each, and out to a redshift of $\sim$$0.6$. Thanks to its depth, imaging resolution, area coverage and high spectroscopic completeness, GAMA provides an exceptionally comprehensive overview of the low redshift galaxy population across a wide range of physical scales. In contrast to the previous phases, GAMA III (also known as GAMA-KiDS-VIKING or GKV) did not add any spectroscopic data to the survey. Instead, the input catalogue of the equatorial survey regions was (retroactively) replaced by a new, GAMA-derived photometric catalogue based on KiDS and VIKING imaging data. 

As in \paperI, we only consider the three equatorial GAMA survey regions G09, G12 and G15, for which the overall redshift completeness is $\sim$$98.5\%$ down to the magnitude limit of $r = 19.8$~mag. We again highlight that this exceptionally high average redshift completeness is maintained even in the densest environments such as pairs, groups and clusters of galaxies (\paperI, Figure 1).

In the rest of this section, we describe the various GAMA data products that we use in our work. These include stellar masses (\ref{2.1}), star formation rates (\ref{2.2}) and the group catalogue (\ref{2.3}).

\subsection{Stellar masses}\label{2.1}
The GAMA collaboration has updated both their preferred multi-band photometry, now derived using the the source finding and image analysis software \textsc{ProFound} (\citeauthor{robotham+18} \citeyear{robotham+18} and published by \citeauthor{bellstedt+20a} \citeyear{bellstedt+20a}), and their preferred method of deriving stellar masses, now using the code \textsc{ProSpect} \citep{robotham+20}. \textsc{ProSpect} is a generative galaxy spectral energy distribution (SED) package, which is used to fit the far-UV to far-IR ($0.15 - 500~\mu$m; \citeauthor{driver+16} \citeyear{driver+16}) photometry of GAMA III galaxies.

Using \textsc{ProSpect} instead of the method presented in \cite{taylor+11} on the previously preferred \textsc{LAMBDAR} photometry yields systematically higher stellar masses, with a median offset of $0.06$ dex and scatter of $0.13$ dex. This is attributed to the greater flexibility of \textsc{ProSpect} in modelling star formation history (SFH), allowing for older stellar populations and consequently higher $M/L$ values, which in turn require more stellar mass to reproduce the observed flux (\citeauthor{robotham+20} \citeyear{robotham+20}, Fig. 33). When combining our new \textsc{ProFound} photometry with our new stellar mass estimation method using \textsc{ProSpect}, the offset relative to \cite{taylor+11} increases to $0.10$ dex, with a scatter of $0.11$ dex, indicating improved consistency with the data and enhanced sensitivity to older stellar populations (\citeauthor{robotham+20} \citeyear{robotham+20}, Fig. 34).

For the stellar mass measurements and uncertainties we hence make use of the table \textit{ProSpectv03} \citep{bellstedt+20b}, which provides stellar masses, star formation rates, metallicities, and dust parameters derived with the \textsc{ProSpect} SED-fitting code for all GAMA III galaxies with a redshift in the equatorial survey regions. The values in this catalogue were derived using $H_{0} = 67.8 $~km~s$^{-1}$~Mpc$^{-1}$, $\Omega_{\rm M} = 0.308$ and $\Omega_{\rm \Lambda} = 0.692$ (consistent with a Planck 15 cosmology: Planck Collaboration XIII \citeyear{planck+15}).

\subsection{Star formation rates}\label{2.2}
The H$\alpha$ emission represents a direct tracer of a galaxy’s present-day SFR, reflecting SF activity over short timescales (less than $10-20$~Myr) and being thus minimally affected by the galaxy’s past SFH \citep{kennicutt+98a}. Therefore, for the SFRs we make use of the H$\alpha$-derived values as described in \cite{davies+16}.

Following the approach of \cite{hopkins+03}, the stellar absorption-corrected H$\alpha$ luminosity can be written as:
\begin{equation}\label{1}
\begin{split}
    \frac{L_{\rm H\alpha}}{1 \text{W}} = \frac{EW_{\rm H\alpha} + EW_{\rm c}}{1 \AA} \times \frac{10^{-0.4(M_{\rm r} - 34.1)}}{1 \text{W}/\text{Hz}} \\ \times \frac{3 \times 10^{18} \text{Hz}}{6564.1 \AA (1+z)^{\rm 2}} \frac{1 \AA}{1 \text{Hz}} \times \, \biggr(\frac{F_{\rm H\alpha}/F_{\rm H\beta}}{2.86}\biggl)^{\rm 2.36}.
\end{split}
\end{equation}
In the equation above, EW$_{\rm H\alpha}$ represents the H$\alpha$ equivalent width, EW$_{\rm c}$ the equivalent width correction for stellar absorption, $M_{r}$ the galaxy rest-frame absolute $r$-band AB magnitude and $F_{\rm H\alpha}/F_{\rm H\beta}$ the Balmer decrement. Here, we use a single value for the stellar absorption corrections, i.e. EW$_{\rm c} = 2.5~\AA$ \citep{gunawardhana+11,gunawardhana+13,hopkins+13,davies+16}. We note that the third factor converts the continuum luminosity from per unit frequency (as determined by the second factor) to per unit wavelength.

In this way, the SFR$_{\rm H\alpha}$ are determined from \cite{kennicutt+98b} as
\begin{equation}
    \frac{\rm{SFR}_{\rm H\alpha}}{1 M_\odot \rm yr^{-1}}
    = \frac{L_{\rm H\alpha}}{1.27 \times 10^{34} \text{W}} \times 1.53,
\end{equation}
where the last factor converts from a Salpeter to a Chabrier initial mass function (IMF) \citep{driver+13}. SFR$_{\rm H\alpha}$ uncertainties are computed by propagating the observational errors of EW, flux and $r$-band absolute magnitude in equation \ref{1}.

We note that \textsc{ProSpect} (see Sect.~\ref{2.1}) provides not only stellar masses but also SFRs. However, we choose to use the H$\alpha$-based SFRs rather than the SED-based SFRs from \textsc{ProSpect}, because the bimodality between star-forming and passive galaxies is more clearly defined in the H$\alpha$ SFR–stellar mass plane than in the SED SFR–stellar mass plane. Consequently, we only use the SED-based SFRs in cases where the H$\alpha$-based SFR provides only an inconclusive upper limit (see Sect.~\ref{4.1}).

\subsection{Group catalogue}\label{2.3}
As in \paperI, we make use of the same tables \textit{G3CGalv08} and \textit{G3CFoFGroupv09} provided by the GAMA galaxy group catalogue (G$^{3}$C) \citep{robotham+11}. We again emphasise that the G$^{3}$C lists two different estimates for the dynamical group halo masses, namely \texttt{MassA} and \texttt{MassAfunc}, as well as for the luminosity-based ones.

\paperI{} shows that the dynamical group halo masses yield more conclusive results. They also find that the results are largely insensitive to the particular dynamical halo mass estimator used (Figs. 13 and 15). Thus, in this work we only make use of \texttt{MassAfunc} in studying the dependence of the GSMF on group halo mass, as it explicitly accounts for group multiplicity and redshift.

\begin{figure}
\centering
\includegraphics[width=0.5\textwidth]{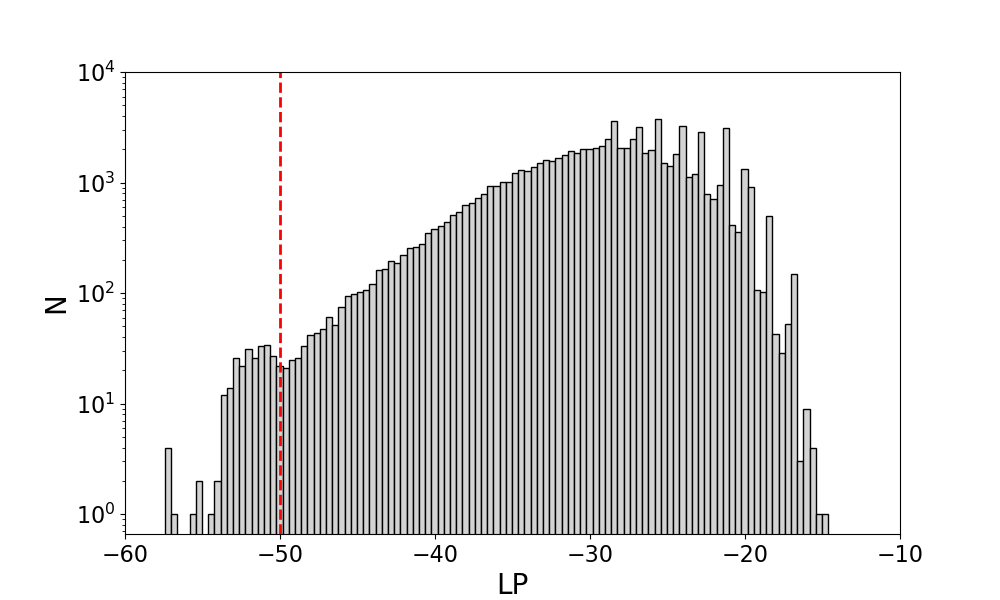}
\caption{Distribution of the LP parameter. The dashed red line marks our lower limit (LP $= -50$) that has been applied to exclude galaxies with poor SED fits, caused by unreliable photometric data.}
\label{fig:distr}
\end{figure}

\section{Sample selection}\label{3}
In this section we describe the selection defining our parent sample of galaxies. The final selection, which results from the application of the redshift-dependent stellar mass limit, is described in Section~\ref{5.2}.

As in \paperI, the starting point is represented by the table \textit{TilingCatv46}, which defines the class of each survey target (SURVEY\_CLASS) as well as various redshift quality parameters ($nQ$ and $nQ2\_{\rm FLAG}$). Following the same approach, our initial sample selection is defined as follows:

\begin{enumerate}
    \item[(i)] Survey regions G09, G12 and G15;
    \item[(ii)] $r<19.8$ mag;
    \item[(iii)] SURVEY\_CLASS $\geq 4$ to select only main survey targets, excluding additional filler targets from the sample;
    \item[(iv)] $nQ \geq 3$ or ($nQ = 2$ and $nQ2\_{\rm FLAG} \geq 1$) to select targets with reliable redshifts;
    \item[(v)] $z \leq 0.213$\footnote{As in our previous work, this redshift cut is based on different measurements for different galaxies: the CMB frame redshift $z_{\rm CMB}$ coming from the table \textit{DistanceFramev14} for ungrouped galaxies; the median redshift $z_{\rm FOF}$ of the galaxy's group coming from \textit{G3CFoFGroupv09} for a grouped galaxy.}.
\end{enumerate}

The same considerations made in \paperI{} regarding the exclusion of sources due to their $nQ$ and $nQ2\_{\rm FLAG}$ values, as well as the determination of the selection function that precisely bounds the parent sample data, still apply here. See Sect. 3 of \paperI{} for a detailed discussion.

Our parent galaxy sample thus contains a total of $89\,451$ galaxies across G09, G12 and G15, of which $39\,903$ galaxies belong to $11\,820$ groups. In the rest of this section we now define three additional restrictions that have to be applied to our sample.

\begin{table*}
\caption{Summary of our parent sample selection process: for each step in the selection, we list the selection criterion, the GAMA table used in this step, as well as the numbers of remaining galaxies and groups after the selection was applied.}
\begin{center}
\begin{tabular}{c|c|c|c}
\hline
Selection step & GAMA table & No of galaxies & No of groups (galaxies in groups) \\
\hline
\hline 
$r<19.8$ mag & \textit{TilingCatv46} & $184\,081$ & $23\,654$ ($75\,029$) \\
SURVEY$_{\rm CLASS} \geq 4$ & \textit{TilingCatv46} & & \\
$nQ \geq3$ or ($nQ = 2$ and $nQ2_{\rm FLAG} \geq 1$) & \textit{TilingCatv46} & $182\,993$ & $23\,652$ ($74\,721$) \\
\hline
$z_{\rm CMB} \leq 0.213$ & \textit{DistancesFramesv14} & & \\
$z_{\rm FOF} \leq 0.213$ & \textit{G3CFoFGroupv09} & $89\,451$ & $11\,820$ ($39\,903$) \\
\hline
\hline
Stellar mass availability & \textit{ProSpectv03} & $83\,673$ & $11\,645$ ($37\,318$) \\
LP $\geq -50$ & \textit{ProSpectv03} & $83\,431$ & $11\,642$ ($37\,202$)\\
\hline
$z_{\rm CMB} \geq0.02$ & \textit{DistancesFramesv14} & $82\,981$ & $11\,579$ ($37\,066$) \\
\hline
H$\alpha$ star formation rate availability & \textit{Davies et al. 2016} & $82\,936$ & $11\,579$ ($37\,040$) \\
\hline
\end{tabular}
\end{center}
\label{tab:ss}
\end{table*}

\subsection{Stellar mass availability}\label{3.1.1}
A mismatch of $5778$ galaxies between our spectroscopic and the available stellar mass sample described in Section~\ref{2.1}, due to different selection processes, reduces our sample to $83\,673$ objects. Furthermore, the \textsc{ProSpect} catalogue provides the logarithm of the posterior (LP) at the best-fit values as a goodness of fit statistic. In Fig.~\ref{fig:distr} we show the distribution of the LP parameter for our galaxy sample. For LP $< -50$ we find a small secondary bump in the distribution. A visual inspection of these sources reveals problematic SED fits, often driven by unreliable photometric measurements. This confirms that the model struggles to reproduce the observed data for these objects. Hence, we impose here a lower limit of LP $\geq -50$ (dashed red line) in order to discard those galaxies where the data most invalidate the model. After the LP cut, our galaxy sample consists of $83\,431$ total galaxies, of which $37\,202$ galaxies belong to $11\,642$ groups.

\subsection{Lower redshift limit}\label{3.1.2}
To be consistent with \paperI, in this step of the selection we impose the same lower redshift limit of $z = 0.02$ on our sample, in addition to the upper limit of $z = 0.213$ already applied above. Our galaxy sample now consists of $82\,981$ total galaxies and $37\,066$ galaxies identified within $11\,579$ groups.

\subsection{H$\alpha$ star formation rate availability}\label{3.1.3}
Another small mismatch of $45$ galaxies is found between our spectroscopic sample and the available SFR sample described in Section~\ref{2.2}. Our final parent sample thus consists of a total of $82\,936$ galaxies and $37\,040$ galaxies belonging to the same groups.

We summarise the complete selection process of the parent sample in Table~\ref{tab:ss}. The final step of the selection process, i.e.\ the application of the redshift-dependent stellar mass limit, is described in Section~\ref{5.2}.

\section{Defining new subsamples}\label{4}
In this section we now determine two additional selections that we use in our analysis. In particular, we describe how we distinguish between star-forming and passive galaxies (\ref{4.1}) as well as between centrals and satellites (\ref{4.2}).

\subsection{Selecting star-forming/passive galaxies}\label{4.1}
In the literature, several approaches provide a separation between blue, star-forming, and red, passive galaxies, based on colour, SFR, morphology, or other galactic properties \citep{davies+19a}. A common method, for instance, relies on specific colour cuts \citep{bell+03,baldry+04,peng+10}. 

However, at lower stellar masses, dust attenuation and complex SFHs make this distinction more problematic. To address this issue, \cite{taylor+15} develop a statistical model that describes the observed galaxy distribution as the sum of two overlapping components, referred to as R-type and B-type, each defined by its own stellar mass function, colour-mass relation, and intrinsic scatter. 
They show that the previously observed differences between red and blue galaxy populations arise from inconsistent and arbitrary definitions, as hard cuts in CMDs lead to contamination and incompleteness, particularly at low masses where the two populations significantly overlap. While this method is flexible and highly effective, it is challenging to apply, as it consists of a complex 40-parameter probabilistic model tuned for nearby $z<0.12$ galaxies. 

Interestingly, \cite{davies+19b} 
use a distinction based on an offset from the star-forming sequence (SFS). More precisely, they first consider a preliminary division at SFR$/M = 10^{-10.5}$~yr$^{-1}$; then, they perform two different least-squares regression fits to galaxies above and below this threshold (defining the star-forming and passive fits, respectively); finally, they identify the minimum density points along cross-sections perpendicular to the two fits, and the interpolation of those minima is taken as the final dividing line. The choice not to consider a separation at $1$ dex below the SFS, as performed in other previous works, is due to the fact that within GAMA the SFS and the passive cloud show two different slopes. Therefore, this approach just provides a more robust division between the two populations.

\begin{figure}
\vspace{-0.3cm}
\centering
\includegraphics[width=0.5\textwidth]{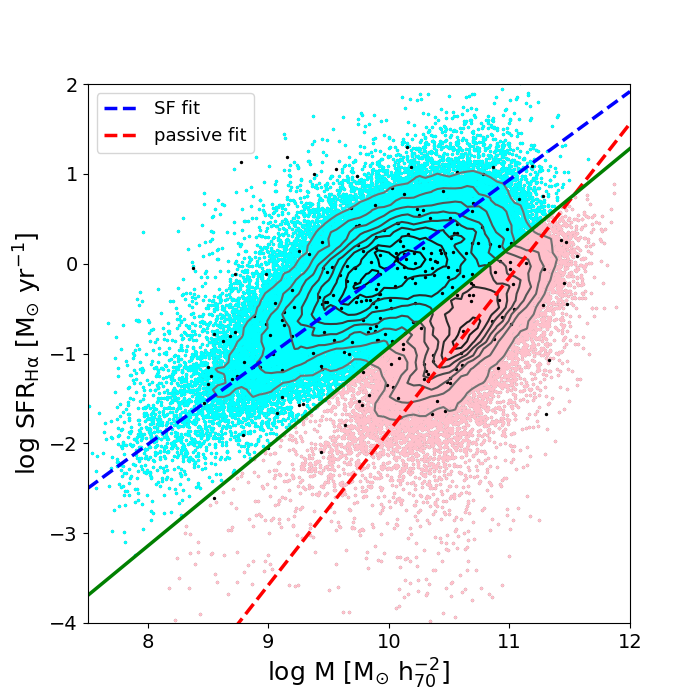}
\caption{Selection of star-forming and passive galaxies using the $\log M-\log$ SFR plane. The dashed blue and red lines display the star-forming and passive population fits, respectively, whereas the solid green line displays our final dividing line. Star-forming galaxies are shown in cyan and passive galaxies in pink. Black dots refer to the $242$ sources with LP $<-50$ (Sect. \ref{3.1.1}).}
\label{fig:separ}
\end{figure}

Here, we implement the separation between SF and passive samples following a similar approach to that of \cite{davies+19b}. Since we do not want to be limited to their fixed, somewhat arbitrary initial division at SFR$/M = 10^{-10.5}$~yr$^{-1}$, we here implement an iterative procedure. At each iteration, the two regression fits and the dividing line are redetermined, using the current dividing line to define the star-forming and passive samples for the next step. The process continues until the dividing line converges, with a tolerance of $10^{-2}$. This slightly modified approach gives a final dividing line between the star-forming and passive populations which is independent of the choice of the initial division. In Fig.~\ref{fig:separ} we show the distribution of our parent sample in the $\log M-\log$ SFR plane, split by star-forming (cyan dots) and passive (pink dots) galaxies. The star-forming and passive fits are marked as dashed blue and red lines, respectively, whereas the green solid line represents our final division. The density contours in Fig.~\ref{fig:separ} show that the dividing line accurately follows the minimum ridge. Our procedure yields a final separation that is very similar to the one obtained by \cite{davies+19b}.

We note that the black dots identify the $242$ sources with LP $<-50$ (Sect.~\ref{3.1.1}): this is to show that, since they are distributed approximately randomly in the $\log M-\log$ SFR plane, their exclusion from our sample is expected not to introduce any bias. We also note that for $4165$ galaxies our H$\alpha$-based SFR measurements provide only an upper limit. Among these, $128$ have an upper limit that lies above our dividing line, and therefore their classification is uncertain. For these galaxies, we use the SED-based SFR measurements from \textsc{ProSpect} instead. This results in $51$ of these $128$ galaxies being classified as star-forming. The remaining $77$ galaxies are classified as passive.

From our total parent sample of $82\,936$ objects, we thus identify $55\,914$ star-forming and $27\,022$ passive galaxies. Of these, $20\,769$ and $16\,271$ are grouped galaxies, respectively.

\subsection{Grouped central and satellite galaxies}\label{4.2}
The G$^{3}$C provides two different approaches for the determination of the central galaxy in a group. In particular, the catalogue identifies the source  with the highest $r$-band luminosity (i.e. the brightest group galaxy, or BGG hereafter) and the source closest to the iterative centre of the group. This centre is defined through an iterative procedure where, at each iteration, the $r$-band centre of light is calculated and the most distant galaxy rejected until only two galaxies remain. Then, the brightest $r$-band galaxy is selected as central. See Sect. 4.2 of  \cite{robotham+11} for detailed explanations. Despite these two approaches producing similar results, \cite{robotham+11} show that the iterative method always offers the most accurate agreement with the exact group centre. Therefore, as done by \cite{vazquez+20}, in this work we adopt this strategy to identify the central galaxy in each group.

From our final sample of $37\,040$ grouped galaxies, the iterative procedure identifies $11\,038$ centrals and $26\,002$ satellites. We here highlight that $\sim$$97\%$ of those centrals are also BGGs. Therefore, we do not expect our results to differ when selecting centrals according to the BGG approach.

\section{Method}\label{5}
In this section we describe the methods we used in our work: the Modified Maximum Likelihood estimator for the construction of the GSMFs (\ref{5.1}) and the stellar mass completeness limit for the derivation of the selection function (\ref{5.2}).

\subsection{Analytical parametrisation of the GSMFs}\label{5.1}
In this work, we adopt the well-established parametrisation that represents the GSMFs with either a single or a double Schechter function. The latter is given by:
\begin{equation}\label{doubleschechter}
    \phi(M)dM = e^{-\frac{M}{M^{\star}}} \Biggl( \phi_{1}^{\star} \biggl( \frac{M}{M^{\star}} \biggr)^{\alpha_{1}} + \phi_{2}^{\star} \biggl( \frac{M}{M^{\star}} \biggr)^{\alpha_{2}} \Biggr) \frac{dM}{M^{\star}} ,
\end{equation}
where $\phi_{1}^{\star}$, $\phi_{2}^{\star}$ and $\alpha_{1}$, $\alpha_{2}$ describe the normalisation and slope parameters, respectively, for the two components. The single Schechter function is simply obtained by neglecting the second term in parentheses.

Following the approach of \paperI, we fit the Schechter function using the modified maximum likelihood (MML) estimation, which was comprehensively documented by \cite{obreschkow+18}. See Sect. 5.1 of \paperI{} for a detailed explanation of the MML approach. 

Following the approach of \cite{weigel+16}, we do not impose any a priori assumption about whether the individual GSMFs are better described by a single or a double Schechter function. However, in all but one case, the decision between the two functional forms is straightforward and does not require a sophisticated model selection procedure. The only exception is the star-forming galaxy sample, for which we employ the statistical model selection techniques described in Sect.~\ref{7.1} to determine the preferred functional form.

\subsection{Stellar mass completeness limit}\label{5.2}
As in \paperI, we follow the same approach presented by \cite{pozzetti+10} for deriving the stellar mass limit $M_{\rm lim}(z)$ above which our sample is complete at a given redshift. See Sect. 5.2 of \paperI{} for a thorough explanation of this approach.

\begin{figure*}
\captionsetup[subfigure]{labelformat=empty}
\centering
\vspace{-0.4cm}

\begin{subfigure}[b]{0.47\textwidth}
    \centering
    \includegraphics[width=\textwidth]{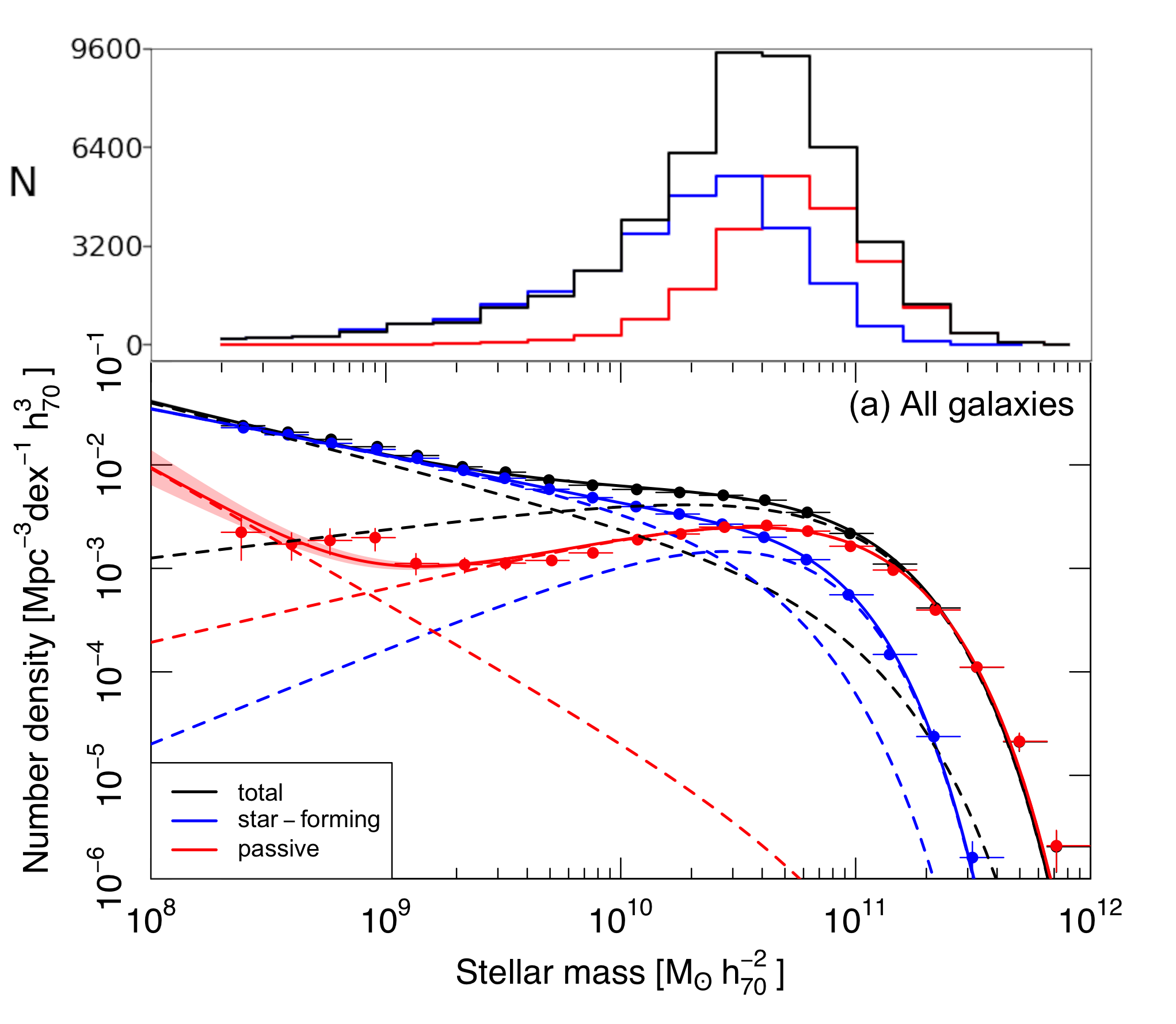}
    \label{fig:6.1.1TOT}
\end{subfigure}

\vspace{-0.2cm}

\begin{subfigure}[b]{0.46\textwidth}
    \centering
    \includegraphics[width=\textwidth]{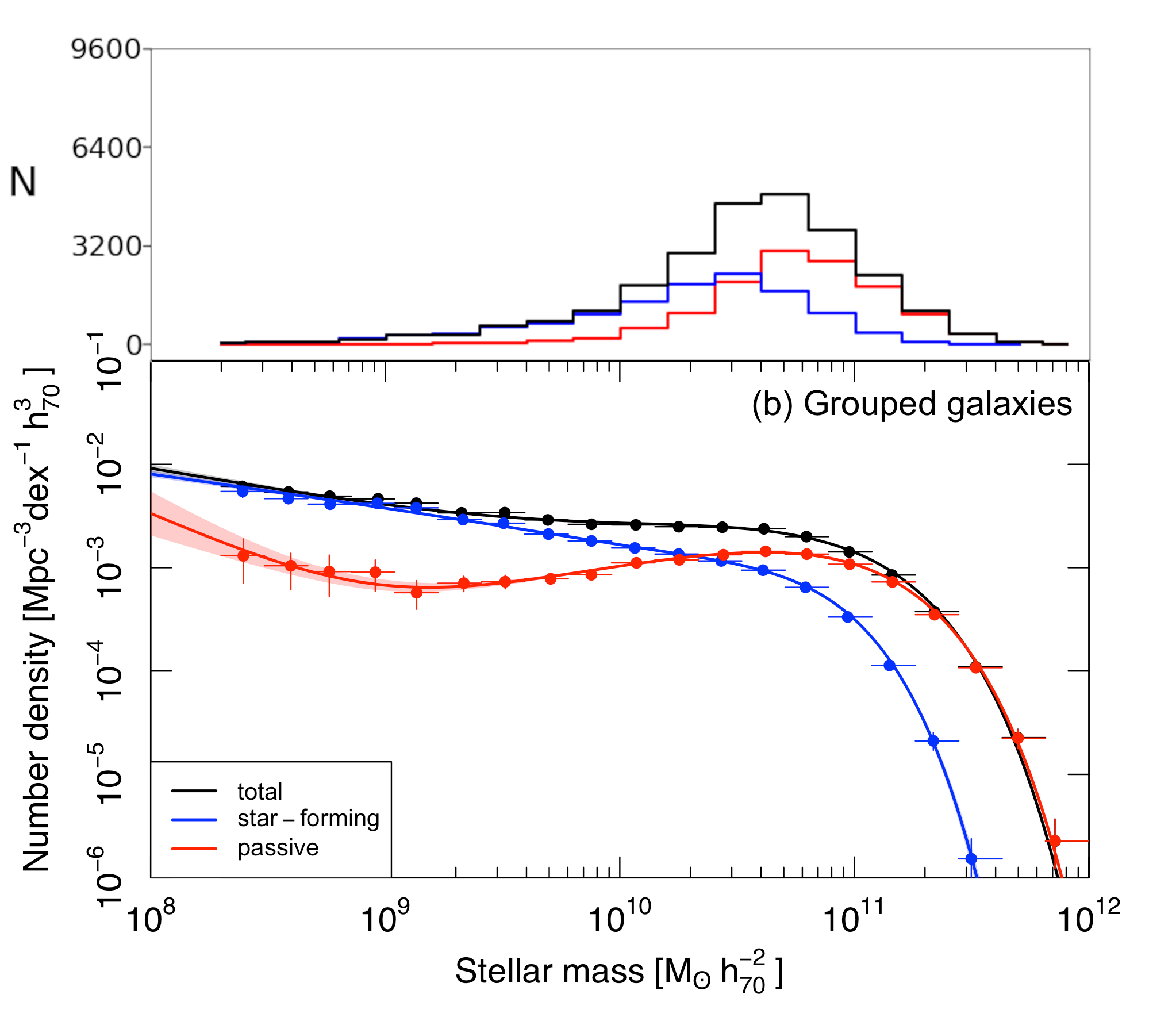}
    \label{fig:6.1.2TOT}
\end{subfigure}
\hfill
\begin{subfigure}[b]{0.46\textwidth}
    \centering
    \includegraphics[width=\textwidth]{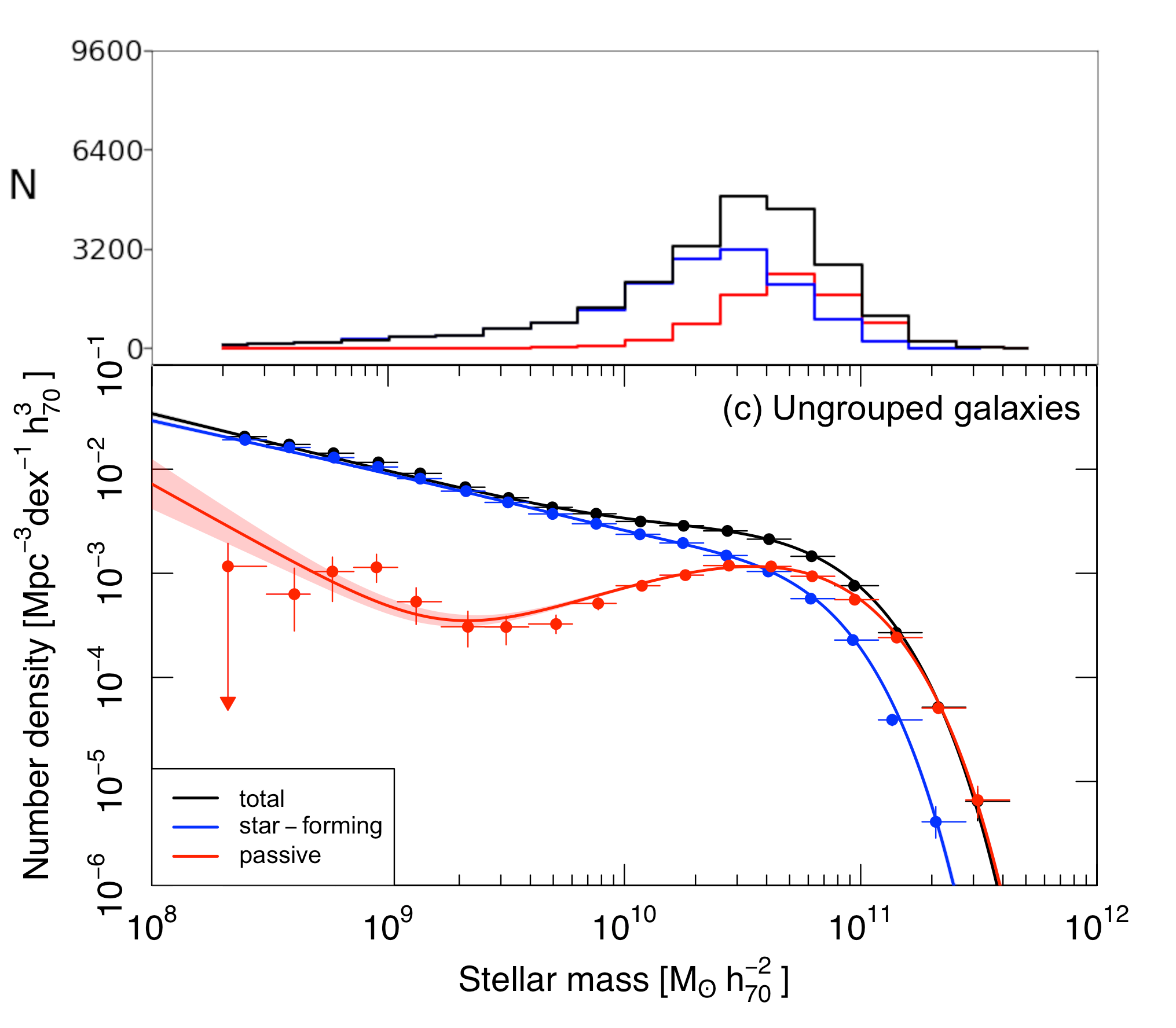}
    \label{fig:6.1.3TOT}
\end{subfigure}

\caption{GSMFs for our total, star-forming and passive galaxy populations, as indicated in the legend. Panel (a) shows the full sample, whereas panels (b) and (c) show grouped and ungrouped galaxies, respectively. In each sub-figure, the lower panel shows the GSMFs, and the upper panel displays the raw number of galaxies as a function of stellar mass in each sample, as indicated. Dashed lines in panel (a) show the two individual Schechter components of each GSMF.}
\label{fig:6.1}
\end{figure*}

We once again emphasise that, after determining $M_{\rm lim}(z)$, the (small) remaining incompleteness of the selected subsample is not random, but rather depends on luminosity and stellar mass-to-light ratio ($M/L$), with faint, high-$M/L$ galaxies being the most affected. For the purposes of this work, it would be inappropriate to adopt the selection function of the full sample as representative for the subsamples discussed in Section \ref{4}, since each of these exhibits its own $L_r$–$M/L_r$ distribution, resulting in different levels of incompleteness. Unlike \paperI, we thus re-derive $M_{\rm lim}(z)$ individually for each subsample, that is, we adopt distinct selection functions for star-forming and passive galaxies (Section \ref{4.1}) as well as for centrals and satellites (Section \ref{4.2}). In practice, however, we find that using a single selection function (that of the full sample) produces essentially the same results. As in \paperI, we also impose a lower stellar mass limit of $\log [M / (M_\odot h_{70}^{-1})] > 8.3$ to guarantee accurate GSMF measurements across all subsamples. Accordingly, we compute the GSMF of each subsample by providing the comoving distances, stellar masses and their uncertainties, selection function, and the chosen fitting function (i.e.\ double Schechter) using the MML framework.

\begin{figure}[!h]
\centering
\vspace{-0.45cm}
\includegraphics[width=0.49\textwidth]{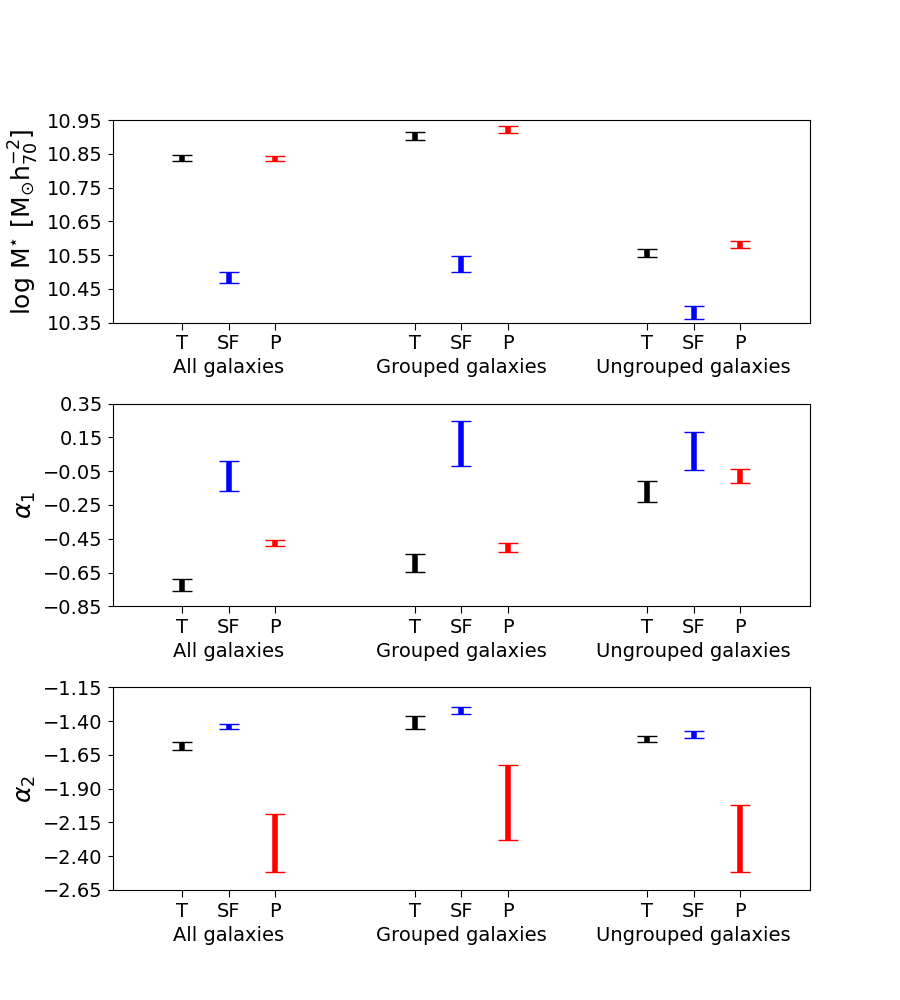}
\caption{Best-fit double Schechter function parameters of the GSMFs shown in Fig.~\ref{fig:6.1}, using the same colour-coding. The $x$-axes distinguish between all, grouped, and ungrouped galaxies; within each sample, the three ticks correspond to the total (T), star-forming (SF), and passive (P) subsamples. The corresponding corner plots illustrating the correlations among the fitted parameters are provided in Figs.~\ref{fig:appendix1}--\ref{fig:appendix3}.}
\label{fig:6.1.3}
\end{figure}

\section{Results}\label{6}
In this section we present our results on the variation of the GSMF as a function of the subsamples defined in Sect.~\ref{4}: star-forming and passive galaxies (Sect.~\ref{6.1}), and centrals and satellites (Sect.~\ref{6.2}). Furthermore, we study the dependence of the GSMF on group halo mass $M_{\rm{halo}}$ (Sect.~\ref{6.3}).

\subsection{How the GSMF differs in star-forming and passive galaxies}\label{6.1}
We now investigate the GSMF in star-forming and passive galaxies. As discussed in Sect. \ref{4.1}, we implement a modified version of the method by \cite{davies+19b} to separate the two populations in the $\log M-\log$ SFR plane, using an iterative procedure that updates the linear regression fits and dividing line at each step until convergence. This approach removes any dependence on an arbitrary initial sSFR threshold.

In Fig.~\ref{fig:6.1} we show the GSMFs for both our star-forming and passive galaxy populations, first using our full parent sample and then considering only galaxies inside and outside of groups (panels (a), (b) and (c), respectively). In panel (a), we also display the two Schechter components of each mass function as dashed lines. For the star-forming population, we note that the transition between the two components occurs at relatively high stellar masses compared to both the passive and global populations. In other words, in the mass range where the two Schechter components of the star-forming subsample intersect, both the passive and the total GSMFs are entirely dominated by the intermediate-mass component. As a result, the shape of the star-forming GSMF from low to intermediate masses is effectively governed by the low-mass component, while the intermediate-mass component primarily accounts for the exponential cut-off at the high-mass end. This behaviour makes the star-forming GSMF resemble a single Schechter function, which explains why previous works have typically found that a single component is a good description for such samples. Still, we find that the inclusion of a second component yields a slightly more accurate description, based on commonly used model selection criteria (see Sect.~\ref{7.1} for a more detailed discussion of this issue). 

In contrast, for the passive galaxy population, we confirm the need for a double Schechter function, as we clearly observe a low-mass upturn consistent with several previous findings \citep{peng+10,baldry+12,weigel+16}, although some other studies do not report this upturn \citep[e.g.][]{taylor+15,moffett+16}. This may be due in part to the method used to distinguish between star-forming and passive galaxies. The question thus arises whether the observed upturn may be due to an increased classification uncertainty towards the low-mass end. To investigate this, we visually inspected a random subsample of $20$ galaxies with stellar masses below $10^{9}~M_{\odot}$, examining both their $grz$ colour images and spectra. We find that $\sim$$20\%$ of the galaxies classified as passive in this mass range show clear signs of ongoing SF. Although this result demonstrates that some level of contamination of the passive subsample by star-forming galaxies is present, it is not sufficient to account for the pronounced low-mass upturn that we observe in the passive GSMF (see panel (a) of Fig.~\ref{fig:6.1}). 

The best-fit double Schechter parameters are tabulated in Table \ref{tab:ds1} and shown in Fig.~\ref{fig:6.1.3}. Following the approach of \paperI, in these and similar tables and figures throughout this paper we only present our results regarding $M^{\star}$, $\alpha_{1}$ and $\alpha_{2}$, since we are more interested in any change of the shape of the mass function and less in its normalisation.
As double Schechter fits can exhibit strong parameter degeneracies, we further examine the joint likelihood distributions of $M^{\star}$, $\alpha_{1}$ and $\alpha_{2}$. The $1$-$\sigma$, $2$-$\sigma$ and $3$-$\sigma$ likelihood contours for the double Schechter parameters of all GSMFs presented in this paper are shown in Appendix~\ref{app:cornerplots}. Any claimed differences in the fitted parameters are based on these corner plots rather than on the best-fit values and their error bars alone.

We also test the robustness of our GSMF fits against variations in the dividing line used to separate star-forming and passive galaxies in Sect.~\ref{4.1}, considering the 1$\sigma$ confidence region around the fiducial slope and intercept. The resulting uncertainties in the double Schechter parameters are small ($\sim 0.02$ dex for $\log M^\star$, 0.03 and 0.01 for $\alpha_1$ and $\alpha_2$, respectively), indicating that our conclusions are not significantly affected by these errors (see Appendix~\ref{app:robustness} for details).

When considering our full parent sample as well as grouped and ungrouped galaxies, we find that the total GSMF is dominated by passive and star-forming galaxies at the high- and low-mass end, respectively (cf. Fig. \ref{fig:6.1}). Notably, the characteristic mass $M^{\star}$ of the passive population almost coincides with that of the total sample in each case; in contrast, the low-mass slopes $\alpha_{2}$ of the star-forming population are just slightly shallower compared to those of the total sample (cf. Fig. \ref{fig:6.1.3}). We attribute these small discrepancies in $M^{\star}$ and $\alpha_{2}$ to the use of different selection functions for the passive and star-forming subsamples, compared to the total populations. On the other hand, the star-forming population exhibits systematically lower $M^{\star}$ values compared to the total sample, while the passive population shows steeper (i.e. lower) $\alpha_{2}$ values. Interestingly, the intermediate-mass slope $\alpha_1$ of the total population is, in each case, steeper than those of the passive and star-forming components, and more closely resembles that of the passive population. These results confirm that star-forming and passive galaxies play a more prominent role in shaping the total GSMF at the low- and intermediate-to-high-mass regimes, respectively.

\begin{figure*}
\captionsetup[subfigure]{labelformat=empty}
\centering
\vspace{-0.3cm}
\begin{subfigure}[b]{0.475\textwidth}
    \centering
    \includegraphics[width=\textwidth]{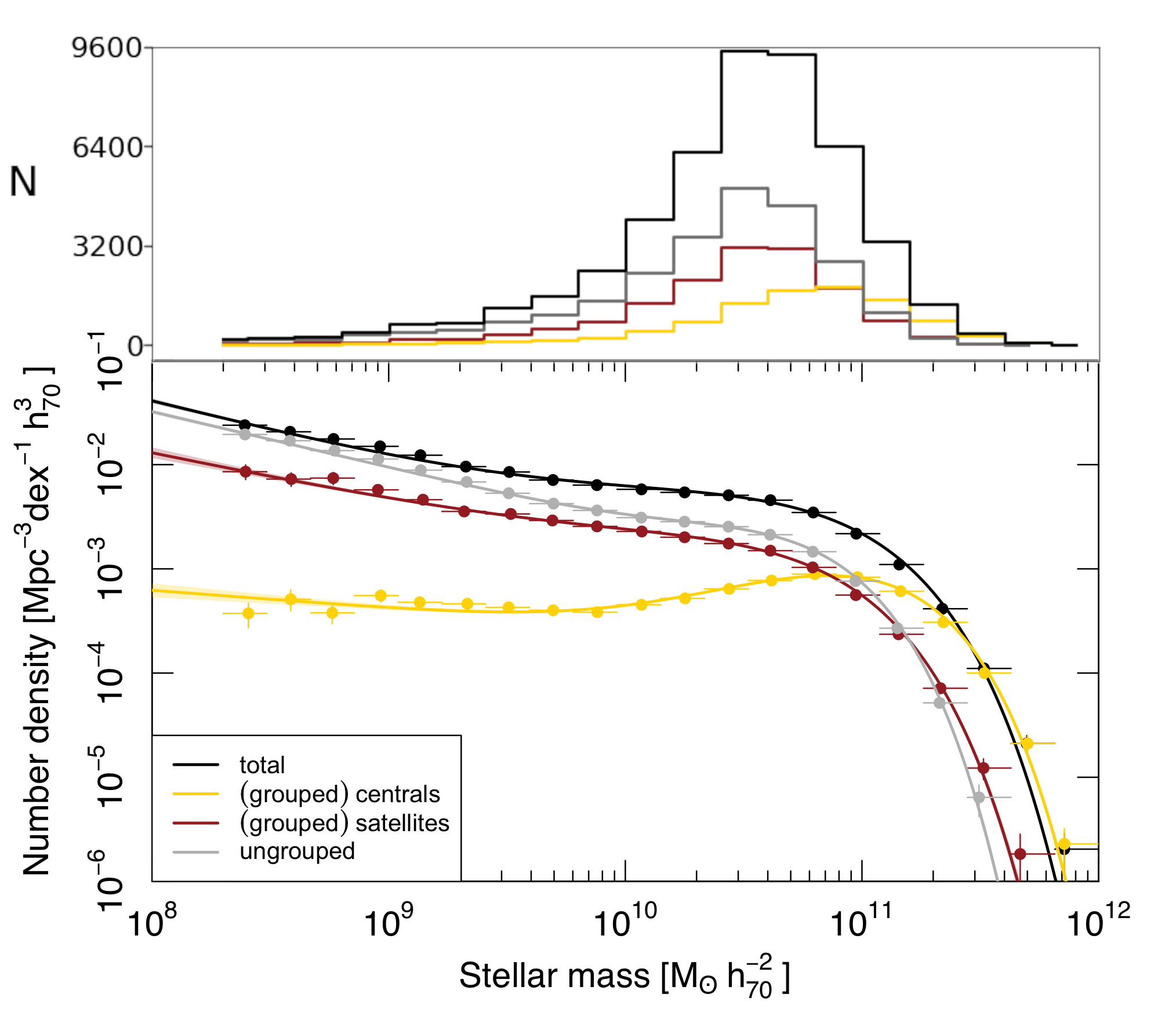}
    \label{fig:6.2.1TOT}
\end{subfigure}
\hfill
\begin{subfigure}[b]{0.475\textwidth}
    \centering
    \includegraphics[width=\textwidth]{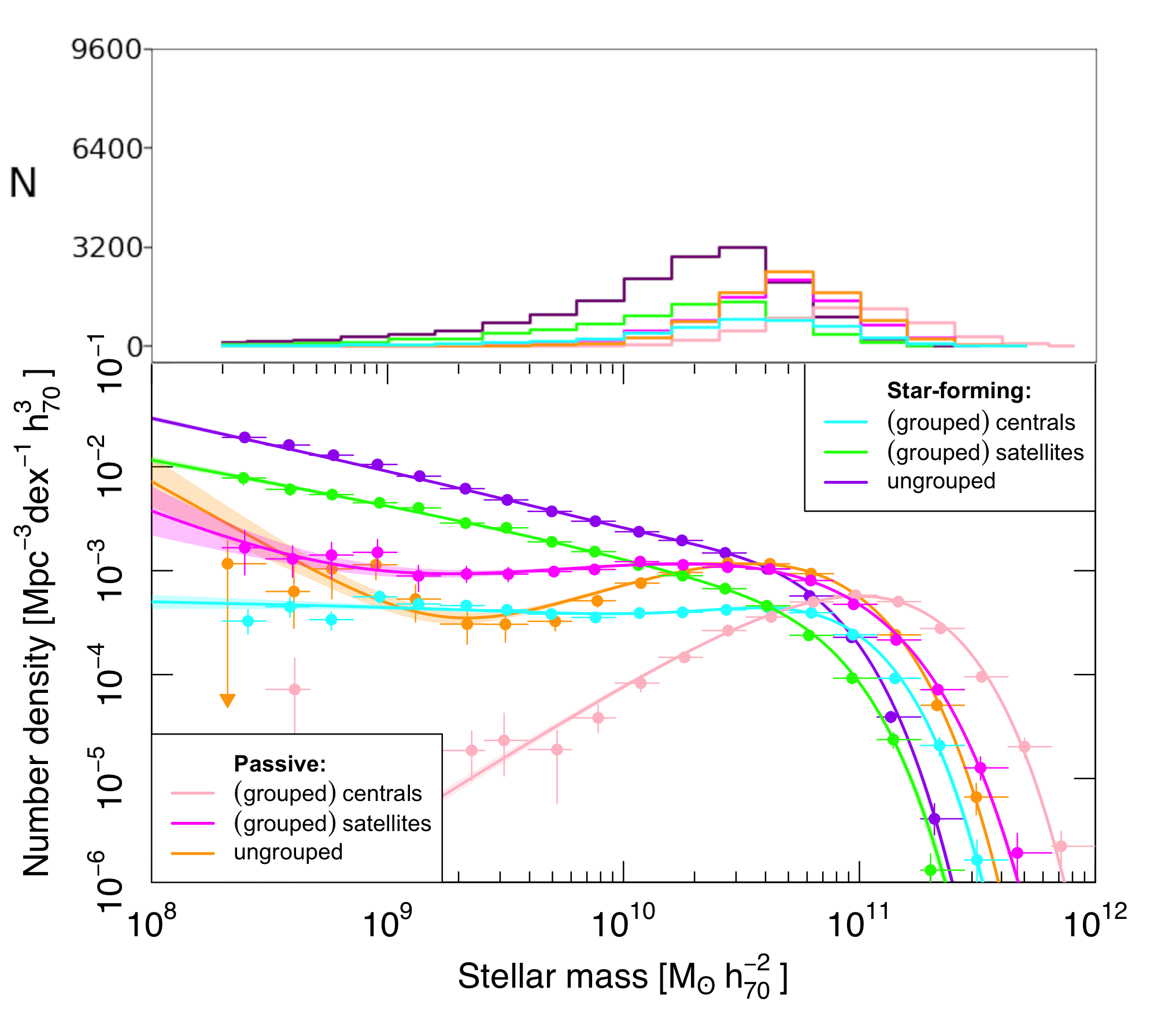}
    \label{fig:6.2.2TOT}
\end{subfigure}

\caption{GSMFs for our total, central, satellite, and ungrouped galaxy populations, as indicated in the legend. The left-hand panel shows the GSMFs for each entire population, while the right-hand panel shows their subdivision into star-forming and passive galaxies. In each sub-figure, the lower panel shows the GSMFs, and the upper panel displays the raw number of galaxies as a function of stellar mass in each sample, as indicated.}
\label{fig:6.2}
\end{figure*}

\begin{figure}[!h]
\centering
\vspace{-0.65cm}
\includegraphics[width=0.49\textwidth]{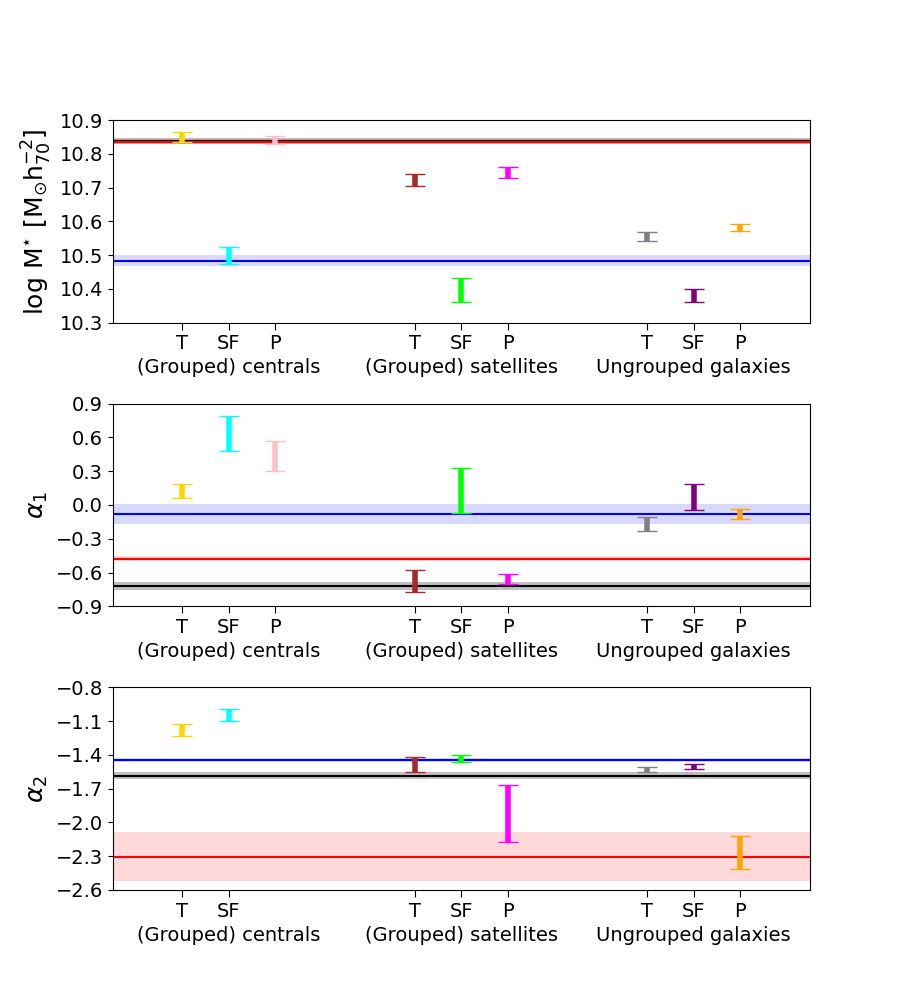}
\caption{Best-fit double Schechter function parameters of the GSMFs shown in Fig.~\ref{fig:6.2}, using the same colour-coding. The $x$-axes distinguish between central, satellite, and ungrouped galaxies; within each sample, the three ticks correspond to the total (T), star-forming (SF), and passive (P) subsamples. The black, blue and red horizontal bands show the results for the full parent, all-star-forming and all-passive samples, respectively, taken from Fig.~\ref{fig:6.1.3}. These reference lines are meant to be compared only with the T, SF, and P values in each subsample, respectively. The corresponding corner plots illustrating the correlations among the fitted parameters are provided in Figs.~\ref{fig:appendix4} and \ref{fig:appendix5}.}
\label{fig:6.2.3}
\end{figure}

By comparing the grouped and ungrouped subsamples\footnote{In contrast to \paperI, we do not apply any volume corrections here. Therefore, these are the actual mass functions, not the conditional ones.}, we find that both star-forming and passive galaxies outside of groups show systematically lower $M^{\star}$ values than their counterparts residing in groups, as already expected from the results in \paperI. This confirms that galaxies in groups are more likely to grow into more massive systems. In contrast, we observe that $\alpha_1$ for passive galaxies is shallower outside of groups, indicating a relatively higher abundance of intermediate-mass passive objects. Furthermore, the low-mass slope $\alpha_2$ is slightly steeper for both star-forming and passive galaxies outside of groups, suggesting a relatively lower abundance of low-mass galaxies.

Finally, we note that uncertainties on $\alpha_1$ are relatively large for star-forming galaxies. This is because the two Schechter components intersect at relatively high masses, with the intermediate-mass component contributing significantly only at the high-mass end. As a result, its slope is poorly constrained. For passive galaxies, instead, the large errors on $\alpha_2$ are due to the limited number of low-mass objects available. These results still hold in Sect.~\ref{6.3}, where we further investigate the dependence of the GSMF on $M_{\rm halo}$ for both our star-forming and passive subsamples.

\subsection{How the GSMF differs in central and satellites galaxies}\label{6.2}
We now investigate the GSMF in central and satellite galaxies. As discussed in Sect. \ref{4.2}, centrals and satellites are identified using the iterative method from the G$^{3}$C, which selects as central the brightest galaxy among the two closest to the group's centre of light. Following \cite{vazquez+20}, ungrouped galaxies are excluded from the central galaxy population.

In Fig.~\ref{fig:6.2} we show the GSMFs for our central, satellite, and ungrouped galaxy populations, first considering the entire samples and then their subdivision into star-forming and passive galaxies (left- and right-hand panels, respectively). Their best-fit double Schechter parameters are tabulated in Table \ref{tab:ds1} and shown in Fig.~\ref{fig:6.2.3}. 

\begin{table*}
\centering
\caption{Best-fit double Schechter function parameters of the GSMFs of all, grouped, ungrouped, central, and satellite galaxies for different subsamples, as indicated.}
\begin{center}
\noindent\begin{tabular}{c||c|c|c}      
\hline
All galaxy sample & $\log M^{\star}$ & $\alpha_{1}$ & $\alpha_{2}$\\
& $(M_{\odot} \, h_{70}^{-2})$ &  &\\ 
\hline
\hline 
total &  10.84 $\pm$ 0.01 & $-$0.72 $\pm$ 0.03 & $-$1.58 $\pm$ 0.03\\
star-forming & 10.48 $\pm$ 0.02 & $-$0.08 $\pm$ 0.09 & $-$1.44 $\pm$ 0.02\\ 
passive &  10.83 $\pm$ 0.01 & $-$0.48 $\pm$ 0.02 & $-$2.30 $\pm$ 0.22\\ 
\hline
Grouped galaxy sample & $\log M^{\star}$ & $\alpha_{1}$ & $\alpha_{2}$\\
& $(M_{\odot} \, h_{70}^{-2})$ &  &\\ 
\hline
\hline 
total &  10.90 $\pm$ 0.01 & $-$0.59 $\pm$ 0.05 & $-$1.41 $\pm$ 0.05\\
star-forming & 10.52 $\pm$ 0.02 & 0.11 $\pm$ 0.13 & $-$1.32 $\pm$ 0.03\\ 
passive & 10.92 $\pm$ 0.01 & $-$0.50 $\pm$ 0.03 & $-$2.00 $\pm$ 0.28\\ 
\hline
Ungrouped galaxy sample & $\log M^{\star}$ & $\alpha_{1}$ & $\alpha_{2}$\\
& $(M_{\odot} \, h_{70}^{-2})$ &  &\\ 
\hline
\hline 
total &  10.56 $\pm$ 0.01 & $-$0.17 $\pm$ 0.06 & $-$1.53 $\pm$ 0.02\\
star-forming & 10.38 $\pm$ 0.02 & 0.07 $\pm$ 0.11 & $-$1.50 $\pm$ 0.02\\ 
passive & 10.58 $\pm$ 0.01 & $-$0.08 $\pm$ 0.04 & $-$2.27 $\pm$ 0.15\\ 
\hline
Central galaxy sample & $\log M^{\star}$ & $\alpha_{1}$ & $\alpha_{2}$\\
& $(M_{\odot} \, h_{70}^{-2})$ &  &\\ 
\hline
\hline 
total &  10.85 $\pm$ 0.01 & 0.12 $\pm$ 0.06 & $-$1.18 $\pm$ 0.05\\
star-forming & 10.50 $\pm$ 0.03 & 0.64 $\pm$ 0.16 & $-$1.05 $\pm$ 0.05\\ 
passive & 10.84 $\pm$ 0.01 & 0.44 $\pm$ 0.13 & ...\\
\hline
Satellite galaxy sample & $\log M^{\star}$ & $\alpha_{1}$ & $\alpha_{2}$\\
& $(M_{\odot} \, h_{70}^{-2})$ &  &\\ 
\hline
\hline 
total &  10.72 $\pm$ 0.02 & $-$0.67 $\pm$ 0.10 & $-$1.48 $\pm$ 0.07\\
star-forming & 10.40 $\pm$ 0.03 & 0.13 $\pm$ 0.20 & $-$1.43 $\pm$ 0.03\\ 
passive & 10.74 $\pm$ 0.01 & $-$0.66 $\pm$ 0.05 & $-$1.92 $\pm$ 0.25\\ 
\hline
\end{tabular}
\end{center}
\label{tab:ds1}
\end{table*}

When considering the entire samples, we find that the total GSMF is dominated by centrals and ungrouped galaxies at the high- and low-mass ends, respectively (cf. Fig. \ref{fig:6.2}, left-hand panel). Notably, the characteristic mass $M^{\star}$ of central galaxies closely matches that of the total sample, with satellites and ungrouped galaxies showing progressively lower values. Small differences in the fitted parameters across different samples may arise from the use of different selection functions, but these do not affect the main trends discussed below.  
In contrast, the low-mass slope $\alpha_{2}$ of ungrouped galaxies is very similar to that of the total sample, while satellites and centrals exhibit increasingly shallower slopes. This results in a decreasing trend for both $M^{\star}$ and $\alpha_{2}$ from centrals to ungrouped galaxies (cf. Fig. \ref{fig:6.2.3}). Interestingly, the intermediate-mass slope $\alpha_1$ of satellites is in striking agreement with that of the total population, whereas both centrals and ungrouped galaxies show significantly higher (i.e. shallower) values. This may suggest that satellites play the dominant role in shaping the total GSMF in this stellar mass regime.

When splitting into star-forming and passive galaxies, we observe consistent trends across all populations. In each case, passive galaxies exhibit higher $M^{\star}$ values and steeper $\alpha_2$ slopes compared to their star-forming counterparts (cf. Fig. \ref{fig:6.2}, right-hand panel). The only exception is the $\alpha_2$ values for central galaxies, since the passive subpopulation is best fit by a single Schechter function, for which only $\alpha_1$ is defined. Specifically, we observe a decreasing trend for $M^{\star}$ from centrals to satellites to ungrouped galaxies, within both the star-forming and passive subpopulations. Notably, the $M^{\star}$ value of the central star-forming population is comparable to that of the all-star-forming sample, and the same holds for the passive central population with respect to the all-passive sample (cf. Table \ref{tab:ds1}).  

This finding demonstrates that central star-forming and passive galaxies dominate the high-mass end of the corresponding total GSMFs. A similar trend is observed for $\alpha_{2}$ within the star-forming subpopulation, with increasingly steeper slopes from centrals to satellites to ungrouped galaxies. For the passive subpopulation, although $\alpha_{2}$ is not available for centrals, the slope still steepens from satellites to ungrouped galaxies. In particular, the $\alpha_{2}$ value of the all-star-forming galaxy sample lies between those of the satellite and ungrouped populations, showing good agreement with both. In contrast, the all-passive galaxy sample exhibits a value of $\alpha_{2}$ that is in striking agreement with that of the ungrouped passive population. These findings suggest that both satellite and ungrouped galaxies shape the low-mass end of the star-forming GSMF, whereas the passive GSMF in this regime is primarily shaped by ungrouped galaxies. For $\alpha_1$, the trend remains consistent with the previous results only within the star-forming subpopulation, with increasingly steeper values from centrals to satellites to ungrouped galaxies. In particular, the $\alpha_1$ value of the all-star-forming galaxy sample is comparable to those of the satellites and ungrouped star-forming populations. For passive galaxies, the value of $\alpha_1$ is closest to that of passive satellites, with centrals and ungrouped galaxies showing significantly higher values (i.e. shallower) values. 
These findings suggest that both satellite and ungrouped galaxies shape the intermediate-mass regime of the star-forming GSMF, whereas the passive GSMF in this regime is predominantly shaped by satellite galaxies alone.

By comparing the grouped (both central and satellite) and ungrouped subsamples, we confirm the trends found in Sect.~\ref{6.1} regarding $M^{\star}$ and $\alpha_2$. Isolated galaxies exhibit lower $M^{\star}$ and steeper $\alpha_2$ values compared to galaxies in groups.

\begin{figure*}
\captionsetup[subfigure]{labelformat=empty}
\centering

\begin{subfigure}[b]{0.5\textwidth}
    \centering
    \includegraphics[width=\textwidth]{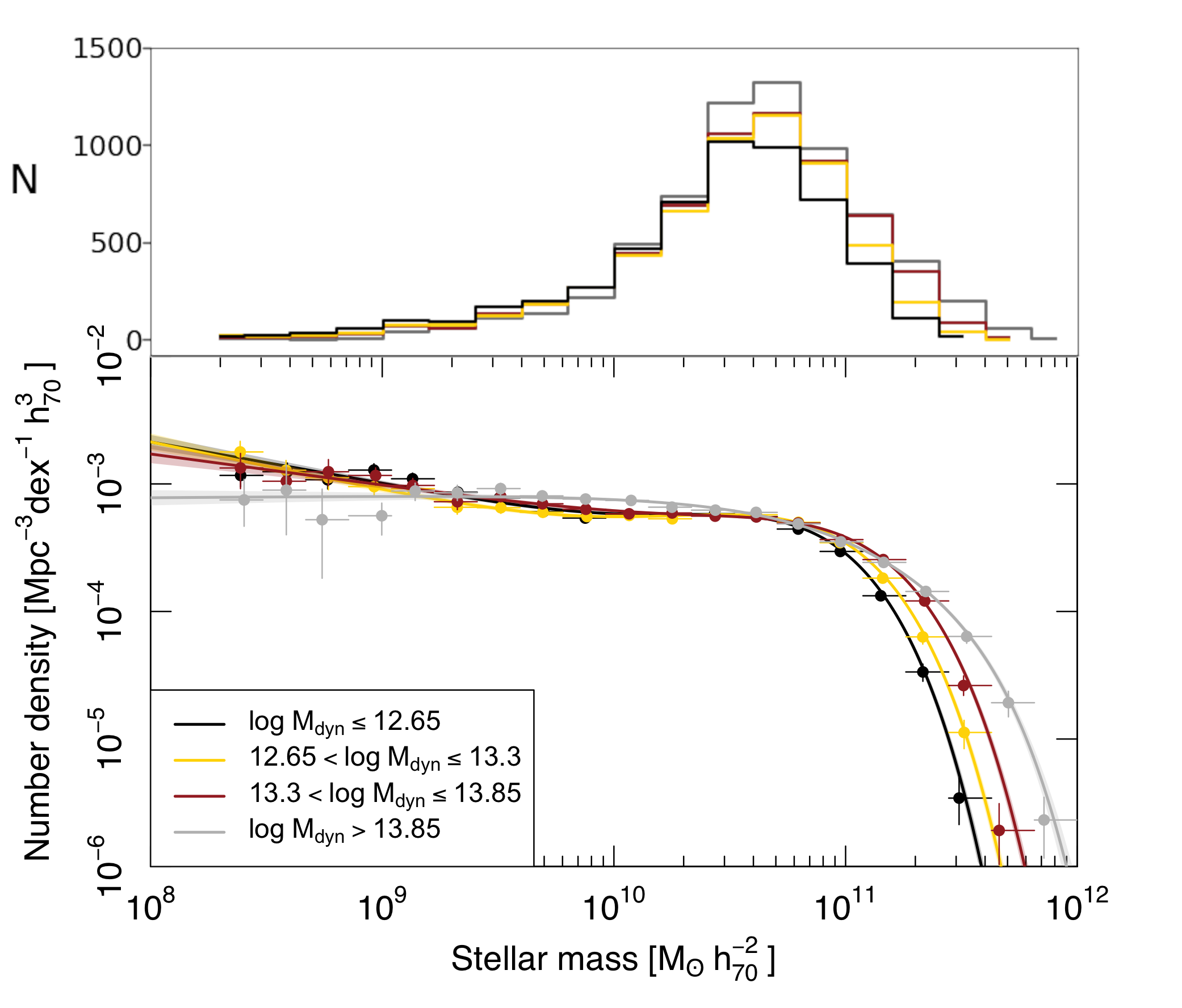}
    \caption{(a) All grouped galaxies}
    \label{fig:6.3.1a}
\end{subfigure}

\vspace{0.2cm}

\begin{subfigure}[b]{0.49\textwidth}
    \centering
    \includegraphics[width=\textwidth]{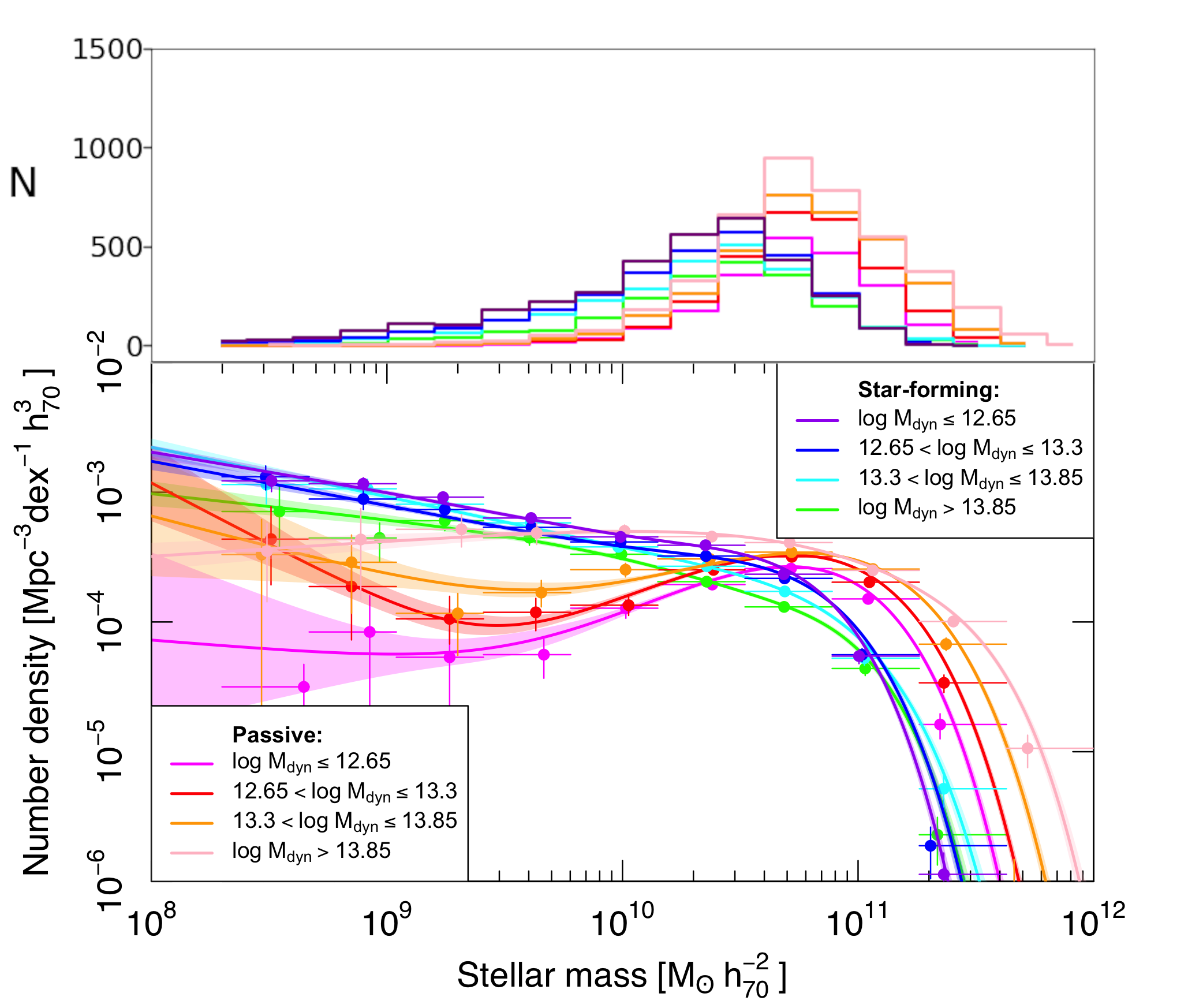}
    \caption{(b) Grouped passive vs star-forming galaxies}
    \label{fig:6.3.1b}
\end{subfigure}
\hfill
\begin{subfigure}[b]{0.49\textwidth}
    \centering
    \includegraphics[width=\textwidth]{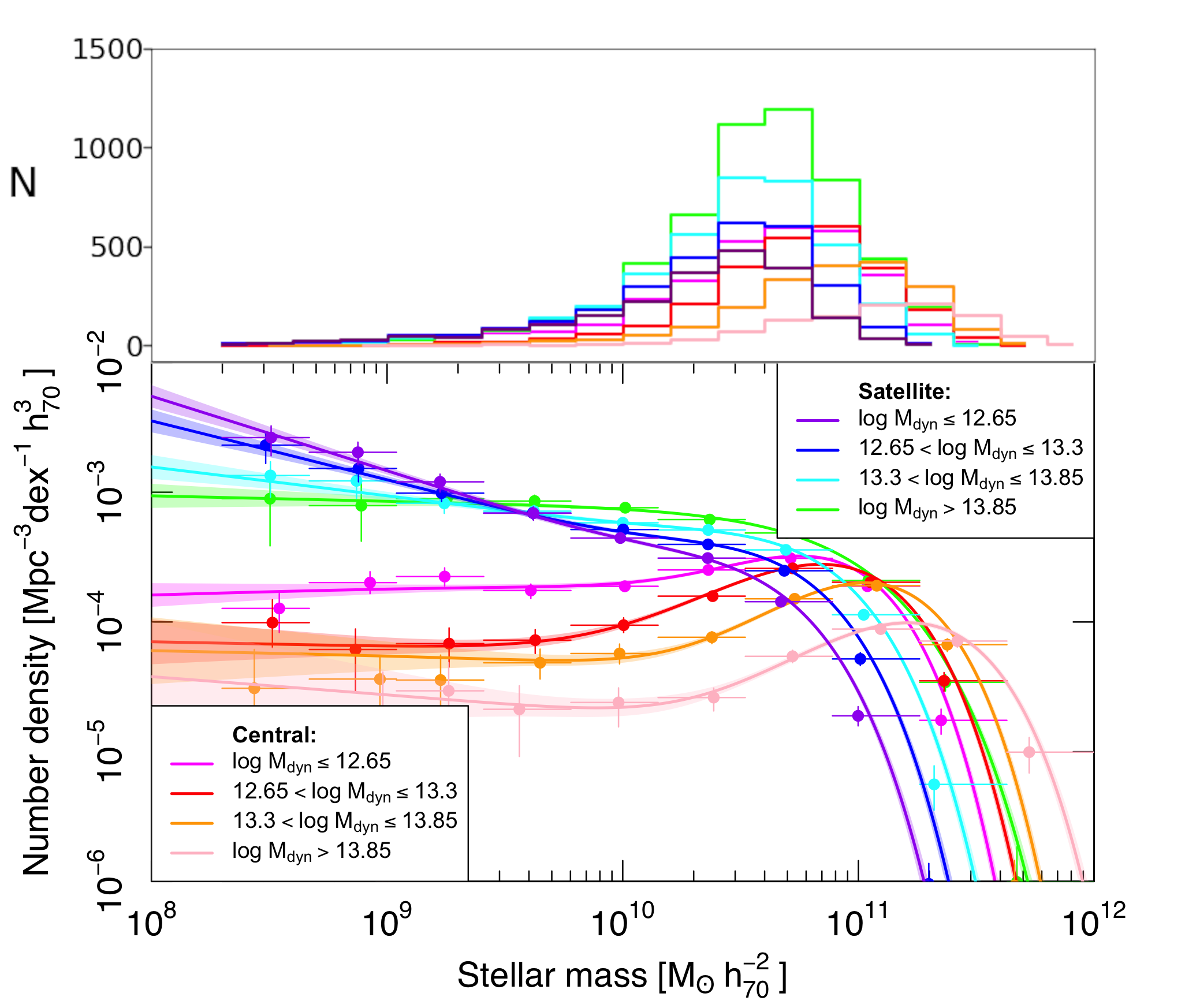}
    \caption{(c) Central vs satellite galaxies}
    \label{fig:6.3.1c}
\end{subfigure}

\caption{GSMFs of our grouped galaxy subsample colour-coded by $M_{\rm{halo}}$, as indicated in the legend. Panel (a) shows the GSMFs as a function of $M_{\rm dyn}$, while panels (b) and (c) further distinguish between passive and star-forming galaxies, and between central and satellite galaxies, respectively. In each sub-figure, the lower panel shows the GSMFs, and the upper panel displays the raw number of galaxies as a function of stellar mass in each sample, as indicated.}
\label{fig:6.3}
\end{figure*}

\subsection{GSMF dependence on group halo mass}\label{6.3}
To study the dependence of the GSMF on group halo mass $M_{\rm halo}$, we are forced to discard  $1055$/$11\,579$ ($9.1\%$) of our groups for which the G$^3$C does not report any $M_{\rm dyn}$ values because the measured velocity dispersion of these groups is smaller than its error. These are overwhelmingly groups with multiplicity $N_{\rm FOF} = 2$. Our total sample now consists of $10\,524$ groups containing $34\,931$ galaxies.

\begin{figure*}
\centering
\includegraphics[width=1.02\textwidth]{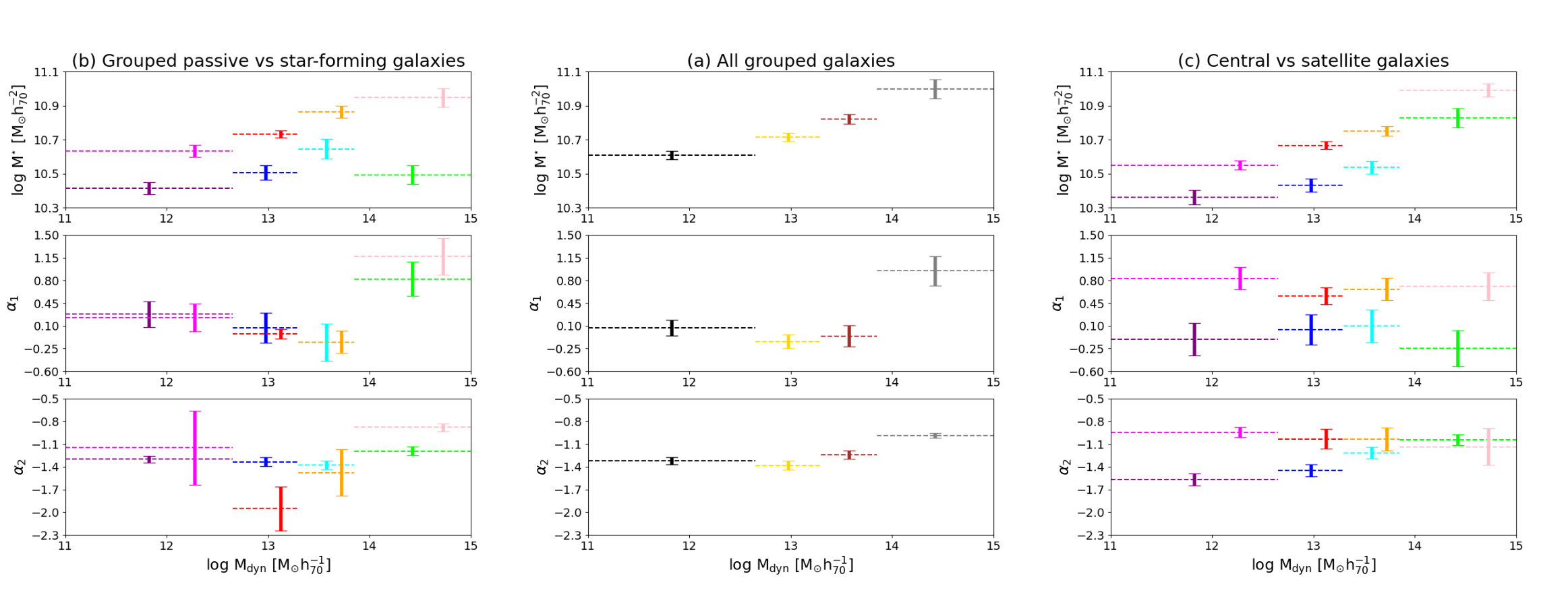}
\caption{Best-fit double Schechter function parameters of the GSMFs shown in Fig.~\ref{fig:6.3}, using the same colour-coding by $M_{\rm{halo}}$. Panel (a) is displayed in the centre, while panels (b) and (c) are on the left and right, respectively. For clarity, the vertical error bars corresponding to the passive galaxies in the left-hand panel and the centrals in the right-hand panel have been slightly offset to the right. The corresponding corner plots illustrating the correlations among the fitted parameters are provided in Figs.~\ref{fig:appendix6}--\ref{fig:appendix8}.}
\label{fig:6.3.2}
\end{figure*}

We now bin galaxies according to the mass of the group that they belong to into four different bins in $\log [M_{\rm halo} / (M_{\odot} \, h^{-1}_{70})]$: $\leq 12.65$, $12.65$--$13.3$, $13.3$--$13.85$, and $> 13.85$. This choice is made to maintain approximately the same number of galaxies per bin. Our resulting GSMFs, colour-coded by $M_{\rm halo}$, are shown in Fig.~\ref{fig:6.3}. Panel (a) shows the GSMFs as a function of $M_{\rm dyn}$, while panels (b) and (c) further distinguish between passive and star-forming galaxies, and between central and satellite galaxies, respectively. Their best-fit double Schechter parameters are tabulated in Table \ref{tab:ds2} and shown in Fig.~\ref{fig:6.3.2}.

When we consider our entire grouped galaxy sample, we note that the characteristic mass $M^{\star}$ increases systematically with $M_{\rm dyn}$, indicating that more massive halos tend to host more massive galaxies. Both the intermediate-mass slope $\alpha_1$ and the low-mass slope $\alpha_2$ exhibit a similar trend, namely a mild steepening when moving from low- to intermediate-mass halos, followed by a progressive shallowing as $M_{\rm dyn}$ continues to increase. We note that the exceptionally high value of $\alpha_1$ in the most massive halo bin reflects that the GSMF can be more accurately described by a single Schechter function, as the intermediate-mass component has negligible influence on the overall shape. Unsurprisingly, our findings are essentially identical with the main results of \paperI{}, where we investigated the dependence of the GSMF on $M_{\rm halo}$ using four different halo mass estimators, despite the use of a different photometry and revised stellar mass estimates in the present analysis.

When we further distinguish between passive and star-forming galaxies, we observe distinct behaviours in the evolution of $M^{\star}$. Passive galaxies follow a consistent trend similar to the overall sample, with $M^{\star}$ increasing steadily as $M_{\rm dyn}$ grows. Star-forming galaxies show a similar trend, except for the highest halo mass bin where $M^{\star}$ actually decreases. This decline may reflect environmental effects suppressing SF in the most massive halos. However, passive galaxies show consistently higher $M^{\star}$ values than star-forming galaxies, confirming their tendency to be more massive at fixed halo mass. Regarding $\alpha_2$, both populations exhibit a consistent trend similar to the overall sample, namely a steepening from low to intermediate halo masses, followed by a shallowing as $M_{\rm dyn}$ continues to increase. However, this trend is more pronounced in passive galaxies, which show a stronger steepening and subsequent flattening. In contrast, star-forming galaxies display a milder but more extended steepening, with the shallowing occurring only at the highest halo mass bin. Similarly, $\alpha_1$ shows a comparable evolution for both star-forming and passive galaxies. In both cases, there is a steepening with increasing $M_{\rm dyn}$, followed by a shallowing at the most massive halos. As previously noted for the entire grouped galaxy sample, the exceptionally high values of $\alpha_1$ in the highest halo mass bin likely indicate that the intermediate-mass component contributes negligibly to the shape of the GSMF. Consequently, those mass functions are better represented by a single Schechter function, with the intermediate-mass slope effectively losing its physical meaning.

When we separate galaxies into centrals and satellites, $M^{\star}$ still increases steadily with $M_{\rm dyn}$, regardless of the population, indicating that more massive halos host more massive galaxies. However, central galaxies show consistently higher values than satellites, confirming their tendency to be more massive at fixed halo mass. Regarding $\alpha_2$, central galaxies exhibit a mild but progressive steepening as $M_{\rm dyn}$ increases. This indicates that the relative abundance of low-mass centrals gradually grows in more massive halos. In contrast, satellite galaxies show the opposite behaviour, namely $\alpha_2$ becomes progressively shallower with increasing $M_{\rm dyn}$, indicating a relative decrease in the number of low-mass satellites in more massive halos. Concerning $\alpha_1$, central galaxies follow a trend similar to that observed in the overall sample, with a steepening from low to intermediate halo masses, followed by a progressive shallowing as $M_{\rm dyn}$ continues to increase. In contrast, satellites exhibit the opposite trend, with $\alpha_1$ becoming shallower as $M_{\rm dyn}$ increases and then steepening in the highest halo mass bin. These distinct behaviours for central and satellite galaxies presumably reflect different quenching mechanisms shaping these galaxy populations.

\begin{table*}
\centering
\caption{Best-fit double Schechter function parameters of the GSMFs of grouped galaxies for different subsamples, as indicated.}
\begin{center}
\noindent\begin{tabular}{c||c|c|c}      
\hline
All grouped galaxy sample & $\log M^{\star}$ & $\alpha_{1}$ & $\alpha_{2}$\\
& $(M_{\odot} \, h_{70}^{-2})$ &  &\\ 
\hline
\hline 
$\log (\frac{M_{\rm dyn}}{M_{\odot} h^{-1}_{70}}$) $\leq 12.65$ &  10.61 $\pm$ 0.02 & 0.07 $\pm$ 0.12 & $-$1.32 $\pm$ 0.05\\
$12.65 < \log (\frac{M_{\rm dyn}}{M_{\odot} h^{-1}_{70}}$) $\leq 13.3$ &  10.71 $\pm$ 0.02 & $-$0.14 $\pm$ 0.11 & $-$1.38 $\pm$ 0.06\\ 
$13.3 < \log (\frac{M_{\rm dyn}}{M_{\odot} h^{-1}_{70}}$) $\leq 13.85$ &  10.82 $\pm$ 0.03 & $-$0.06 $\pm$ 0.16 & $-$1.24 $\pm$ 0.06\\ 
$\log (\frac{M_{\rm dyn}}{M_{\odot} h^{-1}_{70}}) > 13.85$ &   11.00 $\pm$ 0.06 & 0.95 $\pm$ 0.23 & $-$0.99 $\pm$ 0.03\\
\hline
Grouped passive galaxy sample & $\log M^{\star}$ & $\alpha_{1}$ & $\alpha_{2}$\\
& $(M_{\odot} \, h_{70}^{-2})$ &  &\\ 
\hline
\hline 
$\log (\frac{M_{\rm dyn}}{M_{\odot} h^{-1}_{70}}$) $\leq 12.65$ &  10.63 $\pm$ 0.03 & 0.23 $\pm$ 0.21 & $-$1.15 $\pm$ 0.49\\
$12.65 < \log (\frac{M_{\rm dyn}}{M_{\odot} h^{-1}_{70}}$) $\leq 13.3$ &  10.73 $\pm$ 0.02 & $-$0.02 $\pm$ 0.07 & $-$1.95 $\pm$ 0.29\\ 
$13.3 < \log (\frac{M_{\rm dyn}}{M_{\odot} h^{-1}_{70}}$) $\leq 13.85$ &  10.86 $\pm$ 0.03 & $-$0.15 $\pm$ 0.17 & $-$1.48 $\pm$ 0.31\\ 
$\log (\frac{M_{\rm dyn}}{M_{\odot} h^{-1}_{70}}) > 13.85$ &   10.95 $\pm$ 0.06 & 1.17 $\pm$ 0.29 & $-$0.88 $\pm$ 0.05\\
\hline
Grouped star-forming galaxy sample & $\log M^{\star}$ & $\alpha_{1}$ & $\alpha_{2}$\\
& $(M_{\odot} \, h_{70}^{-2})$ &  &\\ 
\hline
\hline 
$\log (\frac{M_{\rm dyn}}{M_{\odot} h^{-1}_{70}}$) $\leq 12.65$ &  10.41 $\pm$ 0.04 & 0.28 $\pm$ 0.20 & $-$1.30 $\pm$ 0.04\\
$12.65 < \log (\frac{M_{\rm dyn}}{M_{\odot} h^{-1}_{70}}$) $\leq 13.3$ &  10.51 $\pm$ 0.04 & 0.07 $\pm$ 0.23 & $-$1.34 $\pm$ 0.06\\ 
$13.3 < \log (\frac{M_{\rm dyn}}{M_{\odot} h^{-1}_{70}}$) $\leq 13.85$ &  10.64 $\pm$ 0.06 & $-$0.15 $\pm$ 0.29 & $-$1.38 $\pm$ 0.05\\ 
$\log (\frac{M_{\rm dyn}}{M_{\odot} h^{-1}_{70}}) > 13.85$ &   10.49 $\pm$ 0.05 & 0.82 $\pm$ 0.27 & $-$1.19 $\pm$ 0.06\\
\hline
Central galaxy sample & $\log M^{\star}$ & $\alpha_{1}$ & $\alpha_{2}$\\
& $(M_{\odot} \, h_{70}^{-2})$ &  &\\ 
\hline
\hline 
$\log (\frac{M_{\rm dyn}}{M_{\odot} h^{-1}_{70}}$) $\leq 12.65$ &  10.55 $\pm$ 0.03 & 0.83 $\pm$ 0.17 & $-$0.95 $\pm$ 0.06\\
$12.65 < \log (\frac{M_{\rm dyn}}{M_{\odot} h^{-1}_{70}}$) $\leq 13.3$ &  10.67 $\pm$ 0.02 & 0.56 $\pm$ 0.13 & $-$1.03 $\pm$ 0.13\\ 
$13.3 < \log (\frac{M_{\rm dyn}}{M_{\odot} h^{-1}_{70}}$) $\leq 13.85$ &  10.75 $\pm$ 0.03 & 0.67 $\pm$ 0.17 & $-$1.04 $\pm$ 0.15\\ 
$\log (\frac{M_{\rm dyn}}{M_{\odot} h^{-1}_{70}}) > 13.85$ &   10.99 $\pm$ 0.04 & 0.71 $\pm$ 0.21 & $-$1.14 $\pm$ 0.24\\
\hline
Satellite galaxy sample & $\log M^{\star}$ & $\alpha_{1}$ & $\alpha_{2}$\\
& $(M_{\odot} \, h_{70}^{-2})$ &  &\\ 
\hline
\hline 
$\log (\frac{M_{\rm dyn}}{M_{\odot} h^{-1}_{70}}$) $\leq 12.65$ &  10.36 $\pm$ 0.04 & $-$0.11 $\pm$ 0.25 & $-$1.57 $\pm$ 0.08\\
$12.65 < \log (\frac{M_{\rm dyn}}{M_{\odot} h^{-1}_{70}}$) $\leq 13.3$ &  10.43 $\pm$ 0.04 & 0.04 $\pm$ 0.23 & $-$1.45 $\pm$ 0.08\\ 
$13.3 < \log (\frac{M_{\rm dyn}}{M_{\odot} h^{-1}_{70}}$) $\leq 13.85$ &  10.54 $\pm$ 0.04 & 0.10 $\pm$ 0.25 & $-$1.22 $\pm$ 0.08\\ 
$\log (\frac{M_{\rm dyn}}{M_{\odot} h^{-1}_{70}}) > 13.85$ &   10.83 $\pm$ 0.06 & $-$0.25 $\pm$ 0.28 & $-$1.04 $\pm$ 0.07\\
\hline
\end{tabular}
\end{center}
\label{tab:ds2}
\end{table*}

\section{Discussion}\label{7}
In this section we discuss and compare our results on the variation of the GSMF in star-forming and passive galaxies (Sect.~\ref{7.1}), centrals and satellites (Sect.~\ref{7.2}), and as a function of group halo mass (Sect.~\ref{7.3}), in the context of other similar studies. 

\subsection{How the GSMF differs in star-forming and passive galaxies}\label{7.1}
\cite{peng+10} present an empirical model describing galaxy evolution based on observations from the Sloan Digital Sky Survey (SDSS) and zCOSMOS, emphasising the roles of mass and environment. The study demonstrates that the effects of mass and environment on galaxy quenching are fully separable up to redshift $z\sim1$. This suggests distinct mechanisms are at play, where mass quenching is related to a galaxy's SFR, and environment quenching is linked to the growth of large-scale cosmic structures. 
The combination of these two quenching processes, plus some additional quenching due to merging, naturally produces (1) a quasi-static single Schechter mass function for star-forming galaxies with an exponential cut-off at a value $M^{\star}$ that is set uniquely by the constant of proportionality between the SF and mass quenching rates, and (2) a double Schechter function for passive galaxies. Remarkably, the characteristic mass of the star forming population remains constant up to redshift $z\sim2$, indicating a universal efficiency of mass quenching over cosmic time. For the passive population, the dominant component (at high masses) is produced by mass quenching and shares the same $M^{\star}$ as star-forming galaxies but has a low-mass slope $\alpha_{1,\rm P}$ differing by approximately $\alpha_{1,\rm P} - \alpha_{\rm SF} \approx 1$. The other component is produced by environmental effects and has the same $M^{\star}$ and $\alpha_{2,\rm P}$ as star-forming galaxies, but lower amplitude depending on the environment, with high-density environments showing a stronger component. 

Several key predictions of the empirical model proposed by \citet{peng+10} are not supported by our results. The only qualitative agreement that we observe is in the functional form of the GSMFs. Specifically, we confirm that both the total and the passive populations are best described by a double Schechter function fit, whereas the star-forming population is well approximated by a single Schechter function. Although we show that a double Schechter fit provides a marginally better representation, the transition between the two components occurs at very high stellar masses. As a consequence, the star-forming GSMF is dominated by its low-mass component across most of its mass range, and its overall shape closely resembles a single Schechter function, consistent with \citet{peng+10}. Nonetheless, our analysis shows that adding a second component produces a slight improvement in the fit, possibly reflecting early signs of quenching at the intermediate-mass regime of the star-forming population. However, except for this structural similarity, our results diverge significantly from the predictions of \citet{peng+10}.

\begin{figure}[!h]
\centering
\includegraphics[width=0.5\textwidth]{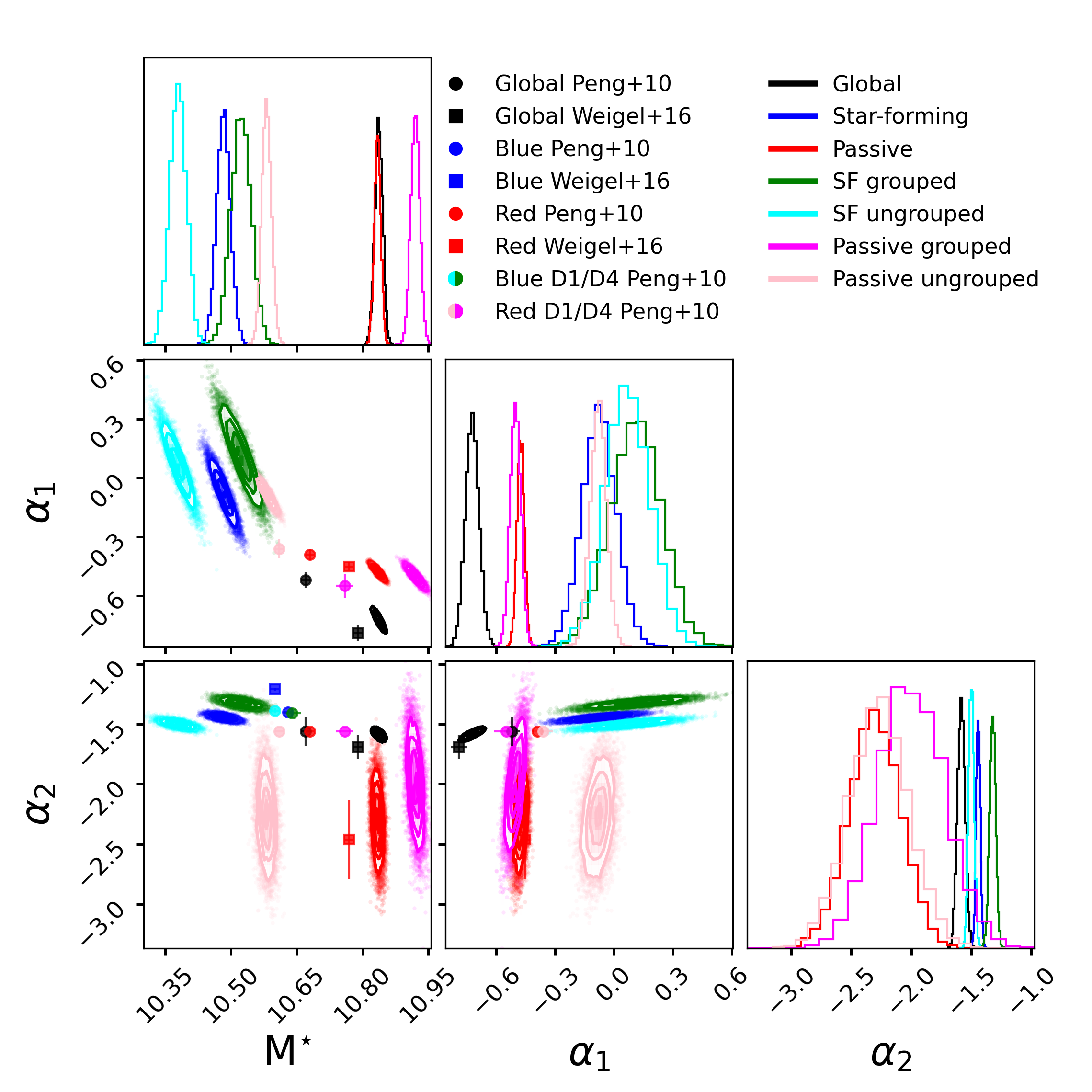}
\caption{Comparison between our best-fit double Schechter function parameters and literature values for different subsamples, as indicated in the legend. The likelihood contours show the $1$-$\sigma$, $2$-$\sigma$ and $3$-$\sigma$ confidence regions for our fits. For each distribution, $10^{4}$ random samples were generated. Symbols indicate the best-fit values from previous studies: circles for \citet{peng+10} and squares for \citet{weigel+16}.}
\label{fig:contours1}
\end{figure}

Comparisons between the best-fit double Schechter function parameters from \citet{peng+10} and our results are shown in Table~\ref{tab:ds4} and Fig.~\ref{fig:contours1}. To investigate the role of environment in shaping the GSMF, we also compare our measurements for grouped and ungrouped galaxies with the high- and low-density bins (D4 and D1, respectively) presented by \citet{peng+10}. While our environmental classification is based on group membership within the GAMA survey, \citet{peng+10} use local overdensity quartiles derived from the zCOSMOS sample, where density is estimated via the $5^{\rm th}$ nearest neighbour method. 

\begin{table*}
\centering
\caption{Best-fit double Schechter function parameters of the GSMFs of blue/star-forming and red/passive galaxies for different works, as indicated.}
\begin{center}
\noindent\begin{tabular}{c||c|c|c}      
\hline
Sample & $\log M^{\star}$ & $\alpha_{1}$ & $\alpha_{2}$\\
& $(M_{\odot} \, h_{70}^{-2})$ &  &\\ 
\hline
\hline 
Global -- \cite{peng+10} &  10.67 $\pm$ 0.01 & $-$0.52 $\pm$ 0.04 & $-$1.56 $\pm$ 0.12\\
Global -- \cite{weigel+16} &  10.79 $\pm$ 0.01 & $-$0.79 $\pm$ 0.04 & $-$1.69 $\pm$ 0.10\\
Global -- this work &  10.84 $\pm$ 0.01 & $-$0.72 $\pm$ 0.03 & $-$1.58 $\pm$ 0.03\\
\hline 
Blue -- \cite{peng+10} &  10.63 $\pm$ 0.01 & ... & $-$1.40 $\pm$ 0.01\\
Blue -- \cite{weigel+16} &  10.60 $\pm$ 0.01 & ... & $-$1.21 $\pm$ 0.01\\
Star-forming -- this work & 10.48 $\pm$ 0.02 & $-$0.08 $\pm$ 0.09 & $-$1.44 $\pm$ 0.02\\ 
\hline 
Red -- \cite{peng+10} &  10.68 $\pm$ 0.01 & $-$0.39 $\pm$ 0.03 & ($-$1.56)\\ 
Red -- \cite{weigel+16} &  10.77 $\pm$ 0.01 & $-$0.45 $\pm$ 0.02 & $-$2.46 $\pm$ 0.33\\ 
Passive -- this work &  10.83 $\pm$ 0.01 & $-$0.48 $\pm$ 0.02 & $-$2.30 $\pm$ 0.22\\
\hline
\hline 
Blue D1 -- \cite{peng+10} & 10.60 $\pm$ 0.01 & ... & $-$1.39 $\pm$ 0.02\\
Star-forming ungrouped -- this work & 10.38 $\pm$ 0.02 & 0.07 $\pm$ 0.11 & $-$1.50 $\pm$ 0.02\\ 
\hline 
Blue D4 -- \cite{peng+10} & 10.64 $\pm$ 0.02 & ... & $-$1.41 $\pm$ 0.04\\
Star-forming grouped -- this work & 10.52 $\pm$ 0.02 & 0.11 $\pm$ 0.13 & $-$1.32 $\pm$ 0.03\\ 
\hline 
Red D1 -- \cite{peng+10} &  10.61 $\pm$ 0.01 & $-$0.36 $\pm$ 0.05 & ($-$1.56)\\ 
Passive ungrouped -- this work & 10.58 $\pm$ 0.01 & $-$0.08 $\pm$ 0.04 & $-$2.27 $\pm$ 0.15\\ 
\hline
Red D4 -- \cite{peng+10} &  10.76 $\pm$ 0.02 & $-$0.55 $\pm$ 0.06 & ($-$1.56)\\
Passive grouped -- this work & 10.92 $\pm$ 0.01 & $-$0.50 $\pm$ 0.03 & $-$2.00 $\pm$ 0.28\\ 
\hline
\end{tabular}
\tablefoot{Ellipses in the $\alpha_1$ column denote that a single Schechter function was used. Values of $\alpha_2$ shown in parentheses without uncertainties were not fitted but just fixed to the global value adopted in the corresponding study.}
\end{center}
\label{tab:ds4}
\end{table*}

First, contrary to their model’s prediction that the characteristic stellar mass $M^{\star}$ should be the same for star-forming and passive galaxies, regardless of environment, we find much larger discrepancies between the two populations. Whereas \citet{peng+10} report a difference of $\Delta M^{\star} \approx 0.05$ dex between their full star-forming and passive samples, and a maximum offset of $\Delta M^{\star} \approx 0.16$ dex when comparing their star-forming D1 and passive D4 bins, our results show significantly larger variations, with $\Delta M^{\star} \approx 0.35$ dex for our full samples and up to $\Delta M^{\star} \approx 0.54$ dex between our grouped and ungrouped subsamples. Similarly, the low-mass slope $\alpha_2$, which in \citet{peng+10} is nearly identical for star-forming and passive galaxies, both for the full samples and across different environments, differs substantially in our results. The empirical framework proposed by \citet{peng+10} assumes that mass and environment quenching act independently, and that the characteristic mass $M^{\star}$ remains constant because the mass-quenching rate scales directly with the SFR. The strong dependences of both $M^{\star}$ and $\alpha_2$ on environment revealed here suggest that these assumptions break down: environmental conditions seem to modulate the efficiency of mass quenching, indicating that the two processes are physically coupled rather than independent. Similarly, the differences in $M^{\star}$ between star-forming and passive populations also imply that mass quenching cannot be simply proportional to SFR, and that additional factors, such as structural properties or local density, likely influence the quenching process.

Regardless of our environmental classification (i.e. full sample, grouped, or ungrouped galaxies), we find that the $\alpha_2$ value for passive galaxies is significantly steeper than those of star-forming and total populations, which instead display relatively similar values. In the \citet{peng+10} model, environment quenching is expected to be a mass-independent process that only changes the normalisation of the passive GSMF. The fact that we observe systematically steeper slopes for passive galaxies in denser environments once again indicates that environmental quenching may depend on stellar mass, modifying the shape of the mass function itself.

Interestingly, one prediction from \cite{peng+10} that still holds in our results is the relation $\alpha_{1,\rm P} - \alpha_{\rm SF} \approx 1$, which we confirm for both our full sample ($= 0.96$) and our grouped galaxy population ($= 0.82$). However, this trend does not extend to ungrouped galaxies, for which $\alpha_{1,\rm P} - \alpha_{\rm SF} = 1.42$, suggesting that the connection between star-forming and passive slopes may be more sensitive to the environment than previously thought.

Together, the persistence of $\alpha_{1,\rm P} - \alpha_{\rm SF} \approx 1$ and the agreement in the overall shapes of the GSMFs indicate that the empirical framework proposed by \citet{peng+10} still provides a good phenomenological description of galaxy populations. However, the deviations in $M^{\star}$ and the environmental trends discussed above suggest a more intricate interplay between mass and environment quenching than assumed in their model.

We caution here that our classification of galaxies into star-forming and passive populations is based on direct measurements of SF activity via H$\alpha$ emission, while \cite{peng+10} use broad-band photometric colours. These choices may partly account for the discrepancies that we observe in our results.

Finally, we have tested whether the differences in the $M^{\star}$ values between star-forming and passive galaxies could be driven by our choice of the fitting function (i.e. using a double instead of a single Schechter parametrisation). When fitting the star-forming population with a single Schechter function, as done by \citet{peng+10}, the inferred $M^{\star}$ shifts to higher values, partially reducing the discrepancies with the passive population. Specifically, $M^{\star}$ increases from $10^{10.48}$ to $10^{10.74}~M_{\odot}$ for the full sample, from $10^{10.52}$ to $10^{10.80}~M_{\odot}$ for the grouped subsample, and from $10^{10.38}$ to $10^{10.68}~M_{\odot}$ for the ungrouped subsample. Although a significant mismatch with the $M^{\star}$ value of the passive population persists in all cases, a large part of the star-forming/passive $M^{\star}$ discrepancy rests on our choice of representing the star-forming population with a double Schechter function. Hence, the question arises as to how robust this choice is to variations of our star-forming/passive classification method.

In Sect.~\ref{6.1} we already investigated the sensitivity of the fitted double Schechter function parameters to variations of our dividing line used to separate star-forming and passive galaxies. We now also test the robustness of the model selection in favour of the double Schechter function. For each of the 100 pairs extracted from the 1$\sigma$ confidence region of our fiducial dividing line in the $\log M-\log$ SFR plane, we perform the GSMF fit using both a single and a double Schechter function, and compute different model selection statistics such as the log-likelihood ratio test (LRT), the Bayesian information criterion (BIC; \citealt{schwarz+78}), the Akaike information criterion (AIC; \citealt{akaike+74}), and the Bayesian evidence\footnote{The LRT compares the goodness-of-fit of two nested models, where one (e.g. the single Schechter) is a special case of the other (e.g. the double Schechter) and is defined as: $\mathcal{R} = -2 \ln \mathcal{L}_\text{single} + 2 \ln \mathcal{L}_\text{double}$, where $\ln \mathcal{L}_\text{single}$ and $\ln \mathcal{L}_\text{double}$ represent the maximum log-likelihoods obtained from fitting the single and double Schechter functions, respectively. This statistic quantifies the improvement in fit when moving from the simpler to the more complex model. Hence, starting from the maximum log-likelihood $\ln \mathcal{L}$, the BIC is defined as \(\mathrm{BIC} = -2 \ln \mathcal{L} + k \ln n\), while the AIC as \(\mathrm{AIC} = 2k - 2 \ln \mathcal{L}\), where \(k\) is the number of model parameters and \(n\) the number of data points. The differences in BIC and AIC values between single and double Schechter models, RBIC and RAIC respectively, quantify the relative preference for one model over the other. Finally, the Bayesian evidence provides a fully Bayesian approach by integrating the likelihood over the prior distribution of the model parameters.}.

\begin{figure}[!h]
\centering
\includegraphics[width=0.49\textwidth]{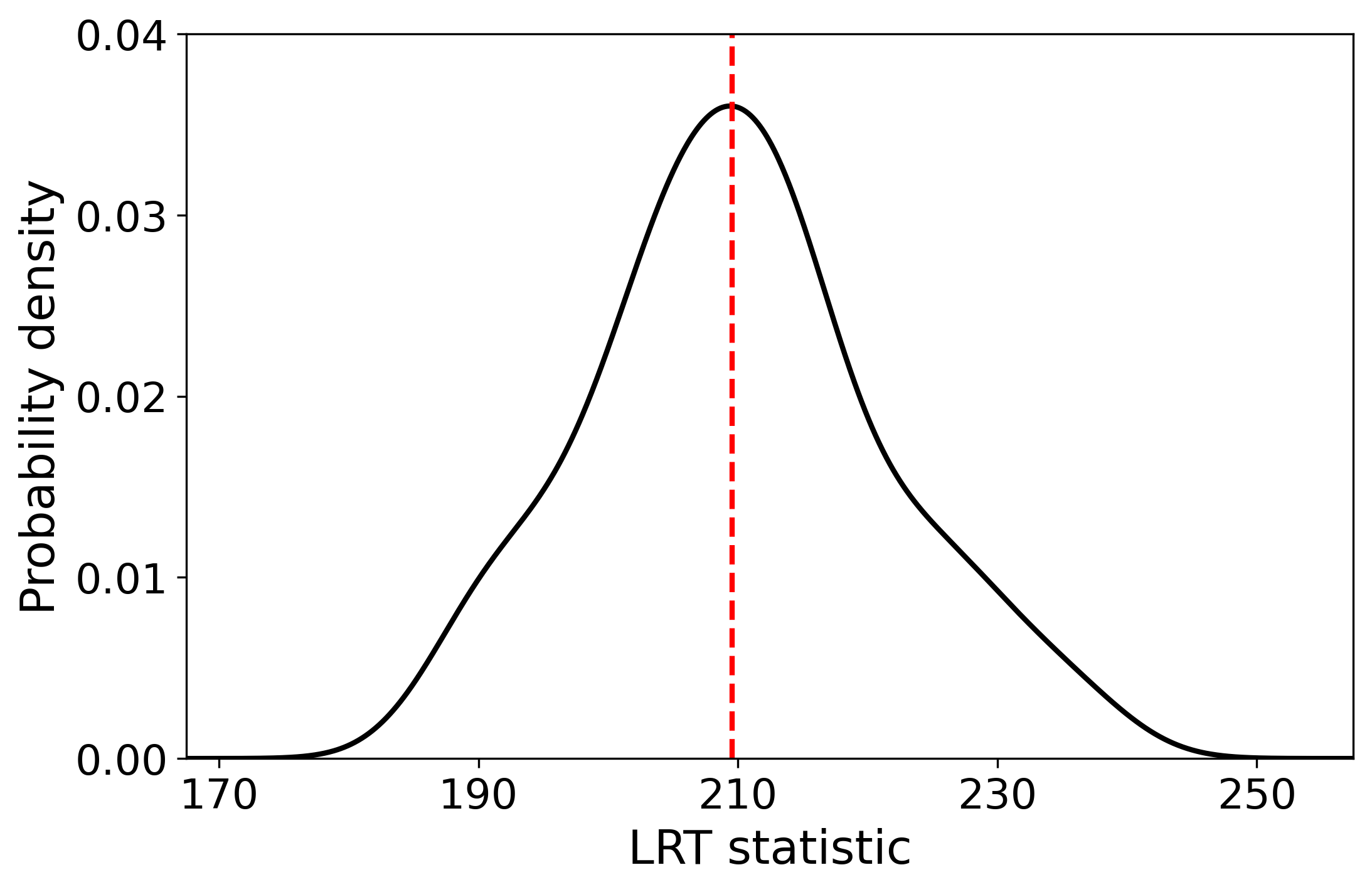}
\caption{Distribution of the LRT statistic obtained from 100 star-forming subsamples. The red dashed line marks our observed LRT value, which lies near the peak of the simulated distribution. This compatibility suggests that the data are consistent with the double Schechter model representing the true distribution, whereas the single Schechter function fails to provide an accurate description.}
\label{fig:lrt_distr}
\end{figure}

The LRT values, as well as the differences in BIC and AIC between the single and double Schechter models, are consistently very high. According to the scale proposed by \cite{jeffreys+39}, such values provide decisive evidence in support of the more complex model (i.e. double Schechter). Likewise, the strongly negative differences in Bayesian evidence further confirm that the double Schechter representation is robust to small changes in the classification process. In particular, we explore the distribution of the LRT statistic derived from the 100 different star-forming subsamples, and shown in Fig.~\ref{fig:lrt_distr}. Our observed LRT value, shown as a red dashed line, lies near the peak of the simulated distribution, which exhibits no significant tails towards lower or higher values. This finding indicates that, regardless of how the dividing line is perturbed within the 1$\sigma$ region, the double Schechter model reliably provides a significantly better fit to the data. Therefore, our preference for a double Schechter function is well justified and not an artefact of how we select star-forming and passive galaxies. We thus conclude that the discrepancies that we observe with the results of \cite{peng+10} cannot be attributed to the functional form alone, but rather arise from real deviations in the assumptions of their model. 

\cite{taylor+15} investigate the GSMFs for their blue/star-forming (B-type) and red/passive (R-type) populations, based on a sample of $M_{\star} >10^{8.7}~M_{\odot}$ and $z<0.12$ galaxies from the GAMA survey.
When modelling the GSMFs, they use a double Schechter function for both subsamples. For the B-type population, the secondary component detects a small deficit of galaxies observed in the range $10^{10}$ -- $10^{10.3}~M_{\odot}$, which corresponds to the apparent upturn observed in the overall GSMF. This results in a slightly more complex shape that cannot be fully described by a single Schechter function. For the R-type population, in contrast, the secondary component becomes negligible above $10^{9.5}~M_{\odot}$, failing to reproduce the apparent low-mass upturn in the red GSMF observed by \cite{peng+10}. This discrepancy is mainly due to broader selection criteria for the red population, which includes a significant number of star-forming galaxies with young stellar populations. Such misclassifications are particularly common at low masses, where red galaxies are intrinsically rare, and a simple colour cut would classify the reddest tail of the blue population as red galaxies.
It is important to note, however, that \cite{taylor+15} do detect a similar low-mass upturn in their R-type GSMF. This upturn is more pronounced when the B/R classification is based on the intrinsic, rather than the restframe, CMD. However, they do not consider this feature to be robust. 

We also confirm that the double Schechter function provides a better representation of both blue/star-forming and red/passive galaxies. However, we do not find strong evidence that the apparent upturn in the overall GSMF is primarily driven by the blue population. In contrast to \cite{taylor+15}, we clearly observe a low-mass upturn in the red population, which is more evident for ungrouped than grouped galaxies, but still present when considering the full passive sample (cf. Fig.~\ref{fig:6.1}). Although we confirm that the low-mass Schechter component of the red GSMF becomes negligible above $M_{\star} \approx 10^{9.5}~M_{\odot}$, the upturn we observe may also be due to the fact that our sample extends down to $M_{\star} \approx 10^{8.3}~M_{\odot}$, somewhat deeper than the limit of $10^{8.7}~M_{\odot}$ adopted by \cite{taylor+15}.  
While the R-type GSMF remains relatively flat below $10^{10.5}~M_{\odot}$ and is systematically lower at low masses compared to the red GSMFs available in the literature, we find excellent agreement at the very low-mass end in terms of number density between our all-passive GSMF and the R-type GSMF of \cite{taylor+15}. This consistency is particularly evident when the B/R classification is based on the intrinsic, rather than the restframe, CMD. The smooth decline of the R-type fraction toward lower stellar masses observed by \cite{taylor+15} indicates that, 
while more massive galaxies are more likely to host older stellar populations, quenching cannot be attributed to stellar mass alone. 

\cite{weigel+16} present a comprehensive study of the GSMF at $0.02 \leq z \leq 0.06$, using data from the SDSS DR7. Their analysis investigates how the GSMF depends on various galaxy properties and environmental parameters. To this end, the main sample is divided into more than 130 subsamples based on morphology, colour, sSFR, central/satellite classification, halo mass, and local density.
For each subsample, they independently evaluate both single and double Schechter fits and use a likelihood ratio test to identify the model that best describes the data. The classification of blue and red galaxies does not follow the method of \cite{taylor+15}, but rather relies on two linear cuts in the CMD, which identify blue, red, and intermediate green galaxies lying between the two cuts. Blue galaxies are generally well described by a single Schechter function with a relatively steep low-mass slope, whereas red galaxies show an upturn at the low-mass end, which requires a double Schechter fit. \cite{weigel+16} 
find their blue and red GSMFs to be systematically lower than those presented by \cite{peng+10}, likely because they explicitly identify a green valley population, which reduces the number densities of both blue and red galaxies. Nevertheless, their GSMFs are more consistent with those of \cite{taylor+15} and with our own, despite their use of a single Schechter function.

Direct comparisons of the best-fit double Schechter parameters (Table~\ref{tab:ds4} and Fig.~\ref{fig:contours1}) show that the total and red GSMFs presented by \cite{weigel+16} are remarkably consistent with our mass functions, and significantly closer than those reported by \cite{peng+10}. In particular, the values of $M^{\star}$ for both the total and red populations differ by $\approx$$0.05$ dex, whereas the slopes $\alpha_1$ and $\alpha_2$ are consistent within the uncertainties. This agreement is particularly notable given the differences in methodology and classification criteria. For the blue population, however, the discrepancy is more pronounced. \cite{weigel+16} find a higher $M^{\star}$ and a significantly shallower $\alpha_2$. This difference is attributed to our use of a double Schechter function, which provides a better fit to the mass function, particularly around the intermediate-mass regime, and likely results in a lower $M^{\star}$ and a steeper $\alpha_2$.

\subsection{How the GSMF differs in central and satellite galaxies}\label{7.2}
In a follow-up study, \cite{peng+12} investigate the environmental effects on galaxy evolution, focusing on the quenching of satellite galaxies within SDSS groups.
They find that the environmental quenching effects identified in \cite{peng+10} can be entirely attributed to the satellites, while central galaxies (including isolated singletons) are affected only by internal, mass-dependent processes. For centrals, the red fraction depends solely on stellar mass and shows no dependence on environmental factors such as local overdensity or halo mass. 
In contrast, the red fraction of satellites increases with both stellar mass and local overdensity, but interestingly not with global group properties like richness or the parent DM halo mass. Moreover, the fraction of blue central galaxies that are quenched when they become satellites (the so-called "satellite quenching efficiency") is found to be largely independent of stellar mass but strongly dependent on local overdensity (which reflects a galaxy’s position within the group) rather than the overall DM halo mass. The phenomenological model adopted in \cite{peng+10} also predicts the GSMFs of star forming and passive centrals and satellites. In particular, the star forming population, whether it consists of centrals or satellites, follow a single Schechter function with the same $M^{\star}$, which is independent of halo mass (above $10^{12} M_{\odot}$). This confirms the universality of the mass-quenching process, which operates through the same physical mechanism for both types of galaxies, unaffected by their environment. For the passive population, centrals are well described by a single Schechter, whereas satellites require a double Schechter function, with one component attributed to mass quenching and the other to satellite quenching. 

If we adopt a similar classification, i.e. defining as centrals both central galaxies in groups and isolated singletons, we find that our results are broadly consistent with this framework. In terms of functional form, we show that the GSMF of the passive central population is well described by a single Schechter function, in line with the results of \cite{peng+12}, who associate this shape with only mass quenching. Similarly, the GSMFs of the star-forming population, whether centrals or satellites, can also be reasonably fitted with a single Schechter function. Although a double Schechter may offer a slightly more accurate fit, the $\alpha_1$ values being very close to zero suggest that the intermediate-mass component does not play a significant role. Therefore, the only subsample that clearly requires a double Schechter function fit is the passive satellite population, as also shown by \cite{peng+12}, where the excess at low masses is attributed to environment (satellite) quenching.

\begin{table*}
\centering
\caption{Best-fit double Schechter function parameters of the GSMFs of central and satellite galaxies for different works, as indicated.}
\begin{center}
\noindent\begin{tabular}{c||c|c|c}      
\hline
Sample & $\log M^{\star}$ & $\alpha_{1}$ & $\alpha_{2}$\\
& $(M_{\odot} \, h_{70}^{-2})$ &  &\\ 
\hline
\hline
Centrals -- \cite{weigel+16} &  10.80 $\pm$ 0.01 & -0.51 $\pm$ 0.09 & $-$1.16 $\pm$ 0.05\\
Centrals -- this work &  10.85 $\pm$ 0.01 & 0.12 $\pm$ 0.06 & $-$1.18 $\pm$ 0.05\\
Satellites -- \cite{weigel+16} &  10.71 $\pm$ 0.02 & $-$0.84 $\pm$ 0.08 & $-$1.83 $\pm$ 0.23\\
Satellites -- this work &  10.72 $\pm$ 0.02 & $-$0.67 $\pm$ 0.10 & $-$1.48 $\pm$ 0.07\\
\hline 
\hline 
Blue centrals -- \cite{peng+12} &  10.61 $\pm$ 0.01 & ... & $-$1.32 $\pm$ 0.02\\ 
Star-forming centrals and ungrouped -- this work & 10.50 $\pm$ 0.02 & $-$0.10 $\pm$ 0.10 & $-$1.44 $\pm$ 0.02\\ 
\hline 
Red centrals -- \cite{peng+12} &  10.70 $\pm$ 0.02 & $-$0.33 $\pm$ 0.04 & ...\\ 
Passive centrals and ungrouped -- this work &  10.84 $\pm$ 0.01 & $-$0.26 $\pm$ 0.08 & ...\\
\hline
Blue satellite -- \cite{peng+12} & 10.59 $\pm$ 0.02 & ... & $-$1.56 $\pm$ 0.03\\
Star-forming satellite -- this work & 10.40 $\pm$ 0.03 & 0.13 $\pm$ 0.20 & $-$1.43 $\pm$ 0.03\\ 
\hline 
Red satellite -- \cite{peng+12} & 10.61 $\pm$ 0.02 & $-$0.49 $\pm$ 0.08 & ($-$1.5)\\
Passive satellite -- this work & 10.74 $\pm$ 0.01 & $-$0.66 $\pm$ 0.05 & $-$1.92 $\pm$ 0.25\\ 
\hline
\end{tabular}
\end{center}
\label{tab:ds5}
\end{table*}

\begin{figure}[!h]
\centering
\includegraphics[width=0.5\textwidth]{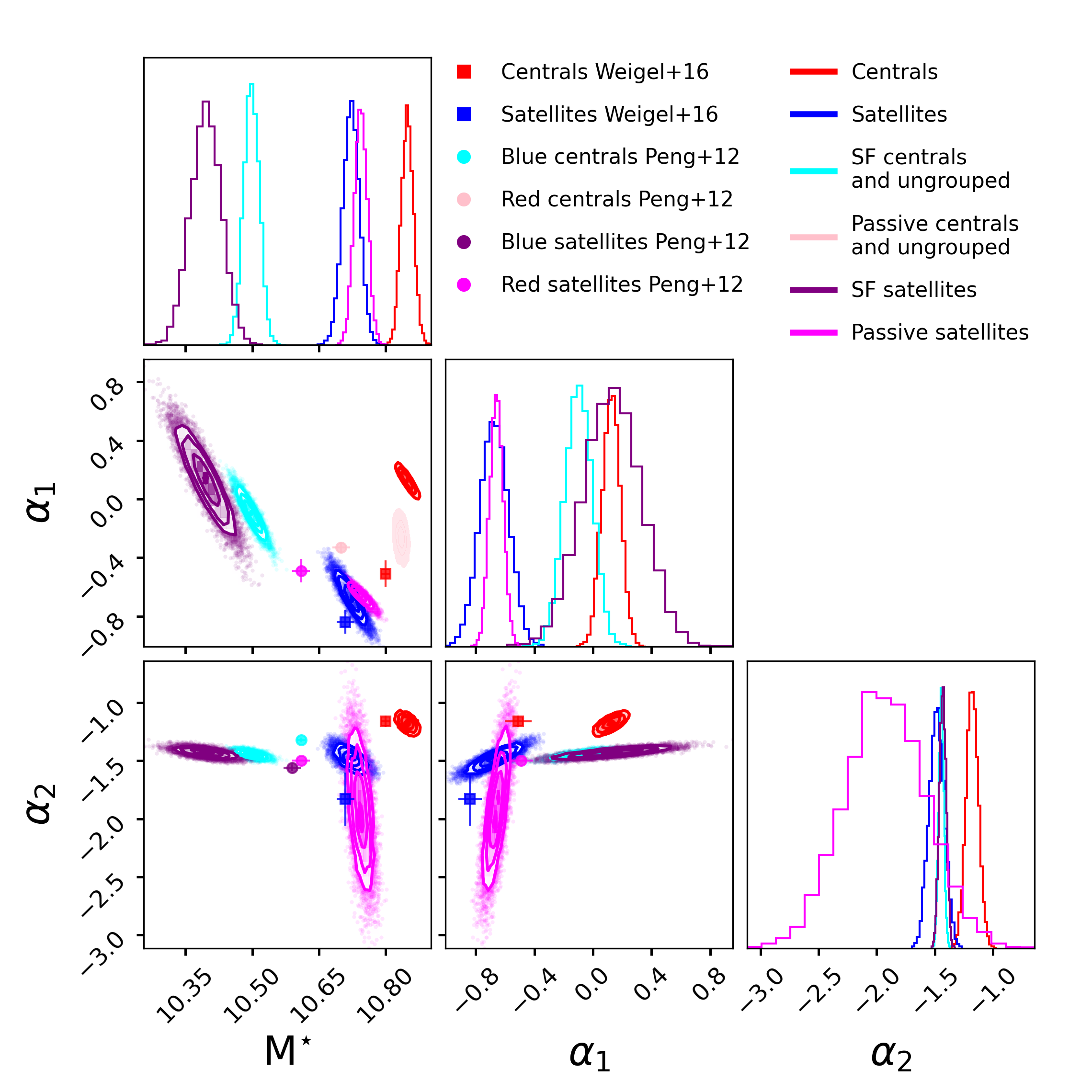}
\caption{Same as Fig.~\ref{fig:contours1}, but for different subsamples as indicated in the legend.}
\label{fig:contours2}
\end{figure}

In Table~\ref{tab:ds5} and Fig.~\ref{fig:contours2}, we also provide comparisons of the best-fit double Schechter function parameters from \cite{peng+12} with our results. Although the values differ, possibly due to differences in photometry, stellar mass estimation methods, and GSMF measurement techniques, the observed trends are in good qualitative agreement. The characteristic stellar mass $M^{\star}$ systematically increases from satellite to central galaxies, and from blue to red populations. In both studies, red centrals exhibit the highest values, followed by red satellites, blue centrals, and blue satellites. We find a difference in $M^{\star}$ of $0.10$ dex between centrals and satellites, for both red and blue populations, compared to $0.09$ dex and $0.02$ dex, respectively, as reported by \cite{peng+12}. For the low-mass slopes $\alpha_2$ of the blue populations, we find similar values for blue centrals and satellites. In contrast, \cite{peng+12} show a shallower and a steeper $\alpha_2$ for blue centrals and satellites, respectively, indicating a stronger environmental dependence in the low-mass regime. This discrepancy might be attributed to the different fitting procedures: while we adopt a double Schechter function, \cite{peng+12} use a single Schechter, which potentially influences the derived low-mass slope values. Interestingly, our intermediate-mass slope $\alpha_1$ for red central galaxies is fully consistent with the value shown by \cite{peng+12}, within the uncertainties, confirming that a single Schechter function adequately describes this subsample. Finally, for the red satellite galaxies, we find that our values of $\alpha_1$ and $\alpha_2$ are significantly steeper than those reported by \cite{peng+12}. This indicates a more pronounced decline in the GSMF at both intermediate and low masses in our red satellite population. Nevertheless, we confirm that a double Schechter function remains the most appropriate fit, consistent with \cite{peng+12}.

In their study, \cite{weigel+16} also find that the total GSMFs of both central and satellite galaxies are better described by a double Schechter function. For centrals, this is due to an excess of high-mass galaxies, which requires an additional high-mass component. Satellites, on the other hand, show a pronounced upturn at the low-mass end, which also requires an additional steeper component. This result is in contrast with \cite{yang+09}, who suggest that the satellite GSMF can be sufficiently fit by a single Schechter function.

Direct comparisons of the best-fit double Schechter parameters for central and satellite galaxies (Table~\ref{tab:ds5} and Fig.~\ref{fig:contours2}) show overall good agreement between our results and those of \cite{weigel+16}. The characteristic mass $M^{\star}$ differs by less than $0.05$ dex in each subsample. Similarly, $\alpha_1$ for satellites and $\alpha_2$ for centrals are consistent within the uncertainties. However, we find a notably higher (i.e. less negative) value of $\alpha_1$ for central galaxies, indicating a flatter slope at intermediate masses.  For satellites, instead, our $\alpha_2$ is significantly shallower than in \cite{weigel+16}, suggesting a less pronounced upturn at the low-mass end. These differences may be attributed to the different sample selection and group identification method. Nonetheless, the overall agreement in $M^{\star}$ and the general shape of the GSMFs confirm the consistency of the main trends across both studies.

When splitting both centrals and satellites into blue and red galaxies, \cite{weigel+16} model all GSMFs using a single Schechter function. This approach is consistent with our results and those of \cite{peng+12} for centrals and blue satellites, but not red satellites, for which a double Schechter is needed to capture the low-mass upturn. These differences may be attributed to their use of two colour cuts to explicitly define three separate populations (blue, green, and red) rather than just a binary classification. As a result, a significant fraction of objects falls into the green valley, reducing the number densities of both the blue and red samples. This also makes their results not directly comparable to our work or that of \cite{peng+12}, where no green population is explicitly defined. We note that \cite{weigel+16} adopt a double Schechter function for both green centrals and green satellites. This suggests that the steep low-mass component that we observe in the red satellite subpopulation may instead be captured by their green subsample, making direct comparisons more challenging.

A complementary analysis by \cite{davies+19b}, based on the G$^{3}$C, investigates the passive fractions of central and satellite galaxies as a function of both stellar and halo mass. By taking the appropriate ratios of the GSMFs presented in Sect.~\ref{6.2}, we are able to reproduce all of their results. Specifically, we confirm that the passive fractions of both centrals and satellites increase with halo mass when considering all stellar masses, and also increase with stellar mass when considering all halo masses. Moreover, satellites exhibit higher passive fractions than centrals at fixed stellar mass for $\log[M/(M_{\odot} \, h^{-2}_{70})] < 11$, consistent with the environmental quenching scenario for low-mass satellites. Even when controlling for both stellar and halo mass, we find that satellite galaxies have systematically higher passive fractions than centrals, particularly at low stellar masses and in high-mass halos. These results are expected, as we adopt a similar method for separating star-forming and passive galaxies (see Sect.~\ref{4.1}). This consistency confirms the robustness of our classification and supports the reliability of our derived GSMFs and their environmental trends.

\subsection{GSMF dependence on group halo mass}\label{7.3}
\cite{weigel+16} also investigate how the GSMF varies as a function of halo mass $M_{\rm halo}$. Although their analysis is based on only three bins, the identified trends are fully consistent with ours. In particular, they report a systematic increase of the characteristic mass $M^{\star}$ with $M_{\rm halo}$ and, for the low-mass slope $\alpha_2$, a steepening from low to intermediate halo masses, followed by a shallower slope in the most massive halos. Moreover, they find that a single Schechter function better describes the GSMF in the highest halo mass bin. This confirms the behaviour we also observe, where the intermediate-mass component becomes negligible and the double Schechter fit effectively reduces to a single component. These trends for $M^{\star}$ and $\alpha_2$ with $M_{\rm halo}$ are preserved when \cite{weigel+16} split their sample into red and blue galaxies, although their classification explicitly identifies a green valley population in between. Specifically, they find a systematic increase in $M^{\star}$ as $M_{\rm halo}$ increases for both red and blue galaxies, with red galaxies still showing higher $M^{\star}$ values than their blue counterparts at fixed halo mass. Similarly, $\alpha_2$ shows a steepening from low to intermediate halo masses, followed by a shallower value in the most massive halos. These trends are in excellent agreement with our findings when we divide galaxies into passive/red and star-forming/blue populations. Despite differences in data sets, sample selection, classification methods, and halo mass bins, our results on $M^{\star}$ and $\alpha_2$  seem to be generally comparable with those of \cite{weigel+16} across corresponding halo mass regimes.

In another study, \cite{vazquez+20} investigate how the GSMFs of central and satellite galaxies depend on the host DM halo mass, using data from the GAMA survey. Central GSMFs are well described by log-normal functions, especially at high masses, for both blue and red populations. However, at low halo masses, they observe a slight excess of low-mass centrals, indicating that the log-normal function underestimates their abundance in this regime. In contrast, satellite GSMFs are generally well described by single Schechter functions. Nonetheless, at high halo masses, a slight excess of high-mass satellites is observed. Central galaxies show a clear increase in their characteristic stellar mass $M^{\star}$ with halo mass. At fixed halo mass, red centrals are systematically $\approx$$0.2$ dex more massive than blue centrals. Similarly, satellite galaxies exhibit an increase in their characteristic stellar mass as halo mass grows. This reflects the mass–richness correlation in galaxy groups, where more massive haloes host both more numerous and more massive satellites. While the general trends are consistent with previous studies, \cite{vazquez+20} find a stronger dependence of the characteristic stellar mass of both centrals and satellites on halo mass compared to \cite{yang+09}, likely due to differences in group definitions and halo mass estimates. In addition, they do not confirm the systematic steepening of the satellite GSMF low-mass slope with halo mass reported by \cite{yang+09}, in either blue or red populations. Instead, they find the steepest slopes in the lowest-mass haloes, suggesting that faint, low-mass satellite galaxies are relatively more abundant in low-mass environments. Finally, they show that the fraction of red galaxies increases with halo mass. Further evidence for environmental quenching comes from the observed increment in the red-to-blue galaxy ratio within groups since $z \sim 0.3$, compared to the field. This suggests that group environments continuously quench SF in infalling galaxies. 

Although \cite{vazquez+20} model central and satellite GSMFs using different functional forms than those adopted in this work, a meaningful comparison with our results is still possible. First, we confirm the increase of $M^{\star}$ with $M_{\rm halo}$ for both central and satellite galaxies. Specifically, our $M^{\star}$ values for satellites in all halo mass bins agree well with those reported by \cite{vazquez+20}, despite differences in binning and data sets. For centrals, their $M^{\star}$ values tend to be slightly higher than ours, especially in the most massive halos. Nevertheless, both studies consistently find that, at fixed halo mass, central galaxies are more massive than satellites. Regarding the low-mass end slope, both analyses report a general shallowing of the GSMF with increasing halo mass. In \cite{vazquez+20}, this trend is not strictly monotonic, as the slope slightly steepens in their highest halo mass bin, but the overall behaviour is still comparable. On the other hand, our $\alpha$ values are systematically steeper across all bins, which may be attributed to our use of a double Schechter function, as well as to differences in binning and data sets.

\section{Conclusions}\label{8}
We employed the equatorial GAMA~II dataset at $z \leq 0.213$ to construct multiple GSMFs. For our analysis, we made use of various GAMA data products: stellar masses (Sect.~\ref{2.1}; \citealp{robotham+20}), H$\alpha$-derived SFRs (Sect.~\ref{2.2}; \citealp{davies+16}) and the G$^{3}$C (Sect.~\ref{2.3}) for the identification of groups \citep{robotham+11}. Our magnitude-limited parent galaxy sample (the selection of which is summarised in Table~\ref{tab:ss}) consists of $82\,936$ galaxies and $11\,579$ groups. To account for completeness, we re-derived the redshift-dependent stellar mass limit for each subsample considered. This differs from \paperI, where a single selection function was applied. However, adopting a global selection function yields nearly identical results.

We investigated how the low-redshift GSMF varies across different galaxy populations (star-forming vs. passive in Sect.~\ref{6.1}, and centrals vs. satellites in Sect.~\ref{6.2}) and as a function of halo mass (Sect.~\ref{6.3}). We noted first that the GSMFs of star-forming and passive galaxies show distinct shapes: passive galaxies require a double Schechter function to capture the low-mass bump, while the star-forming population appears well approximated by a single Schechter function due to its high-mass cut-off being dominated by the intermediate-mass component. Nonetheless, we find that a double Schechter provides a slightly more accurate description for the star-forming sample. Our results confirm that the total GSMF is dominated by star-forming galaxies at the low-mass end and by passive galaxies in the intermediate-to-high-mass regime (cf. Fig.~\ref{fig:6.1}). We also tested the robustness of our star-forming/passive GSMFs against small variations of the star-forming/passive classification by varying the dividing line in the $\log M-\log$ SFR plane within its 1$\sigma$ confidence region. The resulting uncertainties in the double Schechter function parameters were similar or smaller to the random errors, confirming that our results are robust against small perturbations in the classification. Comparing galaxies inside and outside of groups, we found that galaxies in groups have higher $M^{\star}$ and shallower $\alpha_2$ for both our star-forming and passive populations, indicating, respectively, that more massive systems preferentially form in groups and that groups host a relatively lower abundance of low-mass galaxies.

We then examined the GSMFs of central and satellite galaxies. We found that central and ungrouped galaxies dominate the high- and low-mass end of the total GSMF, respectively (cf. Fig.~\ref{fig:6.2}). $M^{\star}$ shows a decreasing trend from centrals to satellites to ungrouped galaxies, consistently across the full, star-forming, and passive samples. A similar trend is observed for $\alpha_2$, with steeper values for isolated systems, although for passive centrals the fit requires only a single Schechter component (cf. Fig.~\ref{fig:6.2.3}). Interestingly, $\alpha_1$ of satellites closely matches that of the total population, suggesting that satellites play a key role in shaping the intermediate-mass regime of the total GSMF. When separating by SF activity, we find that central star-forming and passive galaxies dominate the high-mass ends of their respective GSMFs. In contrast, the low-mass end of the star-forming GSMF is shaped by both satellite and ungrouped galaxies, while that of the passive GSMF is primarily shaped by ungrouped galaxies. These trends confirm and extend the environmental dependencies observed in Sect.~\ref{6.1}: galaxies in groups show higher $M^{\star}$ and shallower $\alpha_2$ values compared to their ungrouped counterparts.

Finally, we studied the dependence of the GSMF on group halo mass $M_{\rm halo}$. We found that $M^{\star}$ increases with $M_{\rm halo}$ across all populations, confirming that more massive halos host more massive galaxies (cf. Fig.~\ref{fig:6.3.2}). For $\alpha_2$, we observed opposite trends in centrals and satellites: it steepens with $M_{\rm halo}$ in centrals, but becomes progressively shallower in satellites, reflecting a relative decline in the number of low-mass satellites in more massive halos. Star-forming and passive galaxies exhibit consistent trends in all double Schechter parameters, though passive galaxies are systematically more massive. Overall, these results confirm and refine those of \paperI{}, and are consistent with the trends reported by other studies \citep{weigel+16,vazquez+20}, highlighting the combined influence of halo mass and galaxy type on the shape of the GSMF.

The only qualitative agreement with the empirical model of \citet{peng+10} lies in the overall functional form: star-forming galaxies are well described by a single Schechter function (with marginal improvement from a second component), while passive and total populations require a double Schechter function. Beyond this, several key aspects of their model are not supported by our findings. First, $M^{\star}$ is not constant across different galaxy types or environments. Second, $\alpha_2$ is significantly steeper for passive galaxies, in contrast to their assumption of similar slopes for star-forming and passive galaxies. The relation $\alpha_{1,\rm P} - \alpha_{\rm SF} \approx 1$ holds in our grouped, but not ungrouped environments, suggesting a stronger environmental dependence than originally proposed. Part of the mismatch arises from differences in classification methods, as we rely on H$\alpha$-derived SFR measurements rather than photometric colours. While adopting a single Schechter fit for star-forming galaxies reduces some discrepancies, it does not entirely remove them. Additionally, model selection criteria (e.g., LRT; Fig.~\ref{fig:lrt_distr}) strongly favour a double Schechter function for the star-forming population, confirming that the observed discrepancies with \citet{peng+10} cannot be attributed to the functional form alone, but rather arise from real deviations from the assumptions of their model. 

On the other hand, our results agree well with those of \citet{weigel+16}, particularly for passive galaxies, both in terms of $M^{\star}$ and $\alpha_2$. In contrast, the lack of a low-mass upturn in the red GSMF reported by \citet{taylor+15} is not confirmed in our analysis, which reveals a clear low-mass component in passive galaxies down to $M_\star \sim 10^{8.3}~M_\odot$.

Our central and satellite populations exhibit distinct mass functions, in agreement with the environmental quenching scenario proposed by \citet{peng+12}. Passive centrals are well described by a single Schechter function, indicating that their quenching is primarily mass-driven and largely independent of environment. In contrast, passive satellites require a double Schechter function, reflecting an additional low-mass subpopulation quenched by environmental processes. The star-forming populations of both centrals and satellites are generally consistent with a single Schechter form, supporting the universality of mass quenching across different environments. Our findings qualitatively agree with previous studies \citep{peng+12,weigel+16}, despite some quantitative differences likely due to methodology and sample selection. Moreover, we reproduce the enhanced passive fractions in satellites at fixed stellar and halo mass observed by \citet{davies+19b}, supporting the importance of environment in suppressing SF. Overall, these findings demonstrate that galaxy environment plays a crucial role in shaping the low-mass end of the GSMF, particularly through satellite quenching, while mass quenching dominates for centrals.

\begin{acknowledgements}
GAMA is a joint European-Australasian project based around a spectroscopic campaign using the Anglo-Australian Telescope. The GAMA input catalogue is based on data taken from the SDSS and the UKIRT Infrared Deep Sky Survey. Complementary imaging of the GAMA regions was obtained by a number of independent survey programmes including GALEX MIS, VST KiDS, VISTA VIKING, WISE, Herschel-ATLAS, GMRT and ASKAP providing UV to radio coverage. GAMA is funded by the STFC (UK), the ARC (Australia), the AAO, and the participating institutions. The GAMA website is \url{https://www.gama-survey.org/}. AS is funded by, and JL acknowledges support by the Deutsche Forschungsgemeinschaft (DFG, German Research Foundation) under Germany's Excellence Strategy -- EXC 2121 ``Quantum Universe'' -- 390833306.
\end{acknowledgements}

\bibliographystyle{aa}
\typeout{}
\bibliography{biblio}

@ARTICLE{liske+15,
       author = {{Liske}, J. and {Baldry}, I.~K. and {Driver}, S.~P. and {Tuffs}, R.~J. and {Alpaslan}, M. and {Andrae}, E. and {Brough}, S. and {Cluver}, M.~E. and {Grootes}, M.~W. and {Gunawardhana}, M.~L.~P. and {Kelvin}, L.~S. and {Loveday}, J. and {Robotham}, A.~S.~G. and {Taylor}, E.~N. and {Bamford}, S.~P. and {Bland-Hawthorn}, J. and {Brown}, M.~J.~I. and {Drinkwater}, M.~J. and {Hopkins}, A.~M. and {Meyer}, M.~J. and {Norberg}, P. and {Peacock}, J.~A. and {Agius}, N.~K. and {Andrews}, S.~K. and {Bauer}, A.~E. and {Ching}, J.~H.~Y. and {Colless}, M. and {Conselice}, C.~J. and {Croom}, S.~M. and {Davies}, L.~J.~M. and {De Propris}, R. and {Dunne}, L. and {Eardley}, E.~M. and {Ellis}, S. and {Foster}, C. and {Frenk}, C.~S. and {H{\"a}u{\ss}ler}, B. and {Holwerda}, B.~W. and {Howlett}, C. and {Ibarra}, H. and {Jarvis}, M.~J. and {Jones}, D.~H. and {Kafle}, P.~R. and {Lacey}, C.~G. and {Lange}, R. and {Lara-L{\'o}pez}, M.~A. and {L{\'o}pez-S{\'a}nchez}, {\'A}. R. and {Maddox}, S. and {Madore}, B.~F. and {McNaught-Roberts}, T. and {Moffett}, A.~J. and {Nichol}, R.~C. and {Owers}, M.~S. and {Palamara}, D. and {Penny}, S.~J. and {Phillipps}, S. and {Pimbblet}, K.~A. and {Popescu}, C.~C. and {Prescott}, M. and {Proctor}, R. and {Sadler}, E.~M. and {Sansom}, A.~E. and {Seibert}, M. and {Sharp}, R. and {Sutherland}, W. and {V{\'a}zquez-Mata}, J.~A. and {van Kampen}, E. and {Wilkins}, S.~M. and {Williams}, R. and {Wright}, A.~H.},
        title = "{Galaxy And Mass Assembly (GAMA): end of survey report and data release 2}",
      journal = {\mnras},
         year = 2015,
        month = sep,
       volume = {452},
       number = {2},
        pages = {2087-2126},
          doi = {10.1093/mnras/stv1436},
archivePrefix = {arXiv},
       eprint = {1506.08222},
 primaryClass = {astro-ph.GA},
       adsurl = {https://ui.adsabs.harvard.edu/abs/2015MNRAS.452.2087L}
}

@ARTICLE{bellstedt+20a,
       author = {{Bellstedt}, Sabine and {Driver}, Simon P. and {Robotham}, Aaron S.~G. and {Davies}, Luke J.~M. and {Bogue}, Kamran R.~J. and {Cook}, Robin H.~W. and {Hashemizadeh}, Abdolhosein and {Koushan}, Soheil and {Taylor}, Edward N. and {Thorne}, Jessica E. and {Turner}, Ryan J. and {Wright}, Angus H.},
        title = "{Galaxy And Mass Assembly (GAMA): assimilation of KiDS into the GAMA database}",
      journal = {\mnras},
         year = 2020,
        month = aug,
       volume = {496},
       number = {3},
        pages = {3235-3256},
          doi = {10.1093/mnras/staa1466},
archivePrefix = {arXiv},
       eprint = {2005.11215},
 primaryClass = {astro-ph.GA},
       adsurl = {https://ui.adsabs.harvard.edu/abs/2020MNRAS.496.3235B}
}

@ARTICLE{bellstedt+20b,
       author = {{Bellstedt}, Sabine and {Robotham}, Aaron S.~G. and {Driver}, Simon P. and {Thorne}, Jessica E. and {Davies}, Luke J.~M. and {Lagos}, Claudia del P. and {Stevens}, Adam R.~H. and {Taylor}, Edward N. and {Baldry}, Ivan K. and {Moffett}, Amanda J. and {Hopkins}, Andrew M. and {Phillipps}, Steven},
        title = "{Galaxy And Mass Assembly (GAMA): a forensic SED reconstruction of the cosmic star formation history and metallicity evolution by galaxy type}",
      journal = {\mnras},
         year = 2020,
        month = nov,
       volume = {498},
       number = {4},
        pages = {5581-5603},
          doi = {10.1093/mnras/staa2620},
archivePrefix = {arXiv},
       eprint = {2005.11917},
 primaryClass = {astro-ph.GA},
       adsurl = {https://ui.adsabs.harvard.edu/abs/2020MNRAS.498.5581B}
}

@ARTICLE{driver+22,
       author = {{Driver}, Simon P. and {Bellstedt}, Sabine and {Robotham}, Aaron S.~G. and {Baldry}, Ivan K. and {Davies}, Luke J. and {Liske}, Jochen and {Obreschkow}, Danail and {Taylor}, Edward N. and {Wright}, Angus H. and {Alpaslan}, Mehmet and {Bamford}, Steven P. and {Bauer}, Amanda E. and {Bland-Hawthorn}, Joss and {Bilicki}, Maciej and {Bravo}, Mat{\'\i}as and {Brough}, Sarah and {Casura}, Sarah and {Cluver}, Michelle E. and {Colless}, Matthew and {Conselice}, Christopher J. and {Croom}, Scott M. and {de Jong}, Jelte and {D'Eugenio}, Franceso and {De Propris}, Roberto and {Dogruel}, Burak and {Drinkwater}, Michael J. and {Dvornik}, Andrej and {Farrow}, Daniel J. and {Frenk}, Carlos S. and {Giblin}, Benjamin and {Graham}, Alister W. and {Grootes}, Meiert W. and {Gunawardhana}, Madusha L.~P. and {Hashemizadeh}, Abdolhosein and {H{\"a}u{\ss}ler}, Boris and {Heymans}, Catherine and {Hildebrandt}, Hendrik and {Holwerda}, Benne W. and {Hopkins}, Andrew M. and {Jarrett}, Tom H. and {Heath Jones}, D. and {Kelvin}, Lee S. and {Koushan}, Soheil and {Kuijken}, Konrad and {Lara-L{\'o}pez}, Maritza A. and {Lange}, Rebecca and {L{\'o}pez-S{\'a}nchez}, {\'A}ngel R. and {Loveday}, Jon and {Mahajan}, Smriti and {Meyer}, Martin and {Moffett}, Amanda J. and {Napolitano}, Nicola R. and {Norberg}, Peder and {Owers}, Matt S. and {Radovich}, Mario and {Raouf}, Mojtaba and {Peacock}, John A. and {Phillipps}, Steven and {Pimbblet}, Kevin A. and {Popescu}, Cristina and {Said}, Khaled and {Sansom}, Anne E. and {Seibert}, Mark and {Sutherland}, Will J. and {Thorne}, Jessica E. and {Tuffs}, Richard J. and {Turner}, Ryan and {van der Wel}, Arjen and {van Kampen}, Eelco and {Wilkins}, Steve M.},
        title = "{Galaxy And Mass Assembly (GAMA): Data Release 4 and the z < 0.1 total and z < 0.08 morphological galaxy stellar mass functions}",
      journal = {\mnras},
         year = 2022,
        month = jun,
       volume = {513},
       number = {1},
        pages = {439-467},
          doi = {10.1093/mnras/stac472},
archivePrefix = {arXiv},
       eprint = {2203.08539},
 primaryClass = {astro-ph.GA},
       adsurl = {https://ui.adsabs.harvard.edu/abs/2022MNRAS.513..439D}
}

@ARTICLE{robotham+20,
       author = {{Robotham}, A.~S.~G. and {Bellstedt}, S. and {Lagos}, C. del P. and {Thorne}, J.~E. and {Davies}, L.~J. and {Driver}, S.~P. and {Bravo}, M.},
        title = "{ProSpect: generating spectral energy distributions with complex star formation and metallicity histories}",
      journal = {\mnras},
         year = 2020,
        month = jun,
       volume = {495},
       number = {1},
        pages = {905-931},
          doi = {10.1093/mnras/staa1116},
archivePrefix = {arXiv},
       eprint = {2002.06980},
 primaryClass = {astro-ph.GA},
       adsurl = {https://ui.adsabs.harvard.edu/abs/2020MNRAS.495..905R}
}

@ARTICLE{planck+15,
       author = {{Planck Collaboration} and {Ade}, P.~A.~R. and {Aghanim}, N. and {Arnaud}, M. and {Ashdown}, M. and {Aumont}, J. and {Baccigalupi}, C. and {Banday}, A.~J. and {Barreiro}, R.~B. and {Bartlett}, J.~G. and {Bartolo}, N. and {Battaner}, E. and {Battye}, R. and {Benabed}, K. and {Beno{\^\i}t}, A. and {Benoit-L{\'e}vy}, A. and {Bernard}, J. -P. and {Bersanelli}, M. and {Bielewicz}, P. and {Bock}, J.~J. and {Bonaldi}, A. and {Bonavera}, L. and {Bond}, J.~R. and {Borrill}, J. and {Bouchet}, F.~R. and {Boulanger}, F. and {Bucher}, M. and {Burigana}, C. and {Butler}, R.~C. and {Calabrese}, E. and {Cardoso}, J. -F. and {Catalano}, A. and {Challinor}, A. and {Chamballu}, A. and {Chary}, R. -R. and {Chiang}, H.~C. and {Chluba}, J. and {Christensen}, P.~R. and {Church}, S. and {Clements}, D.~L. and {Colombi}, S. and {Colombo}, L.~P.~L. and {Combet}, C. and {Coulais}, A. and {Crill}, B.~P. and {Curto}, A. and {Cuttaia}, F. and {Danese}, L. and {Davies}, R.~D. and {Davis}, R.~J. and {de Bernardis}, P. and {de Rosa}, A. and {de Zotti}, G. and {Delabrouille}, J. and {D{\'e}sert}, F. -X. and {Di Valentino}, E. and {Dickinson}, C. and {Diego}, J.~M. and {Dolag}, K. and {Dole}, H. and {Donzelli}, S. and {Dor{\'e}}, O. and {Douspis}, M. and {Ducout}, A. and {Dunkley}, J. and {Dupac}, X. and {Efstathiou}, G. and {Elsner}, F. and {En{\ss}lin}, T.~A. and {Eriksen}, H.~K. and {Farhang}, M. and {Fergusson}, J. and {Finelli}, F. and {Forni}, O. and {Frailis}, M. and {Fraisse}, A.~A. and {Franceschi}, E. and {Frejsel}, A. and {Galeotta}, S. and {Galli}, S. and {Ganga}, K. and {Gauthier}, C. and {Gerbino}, M. and {Ghosh}, T. and {Giard}, M. and {Giraud-H{\'e}raud}, Y. and {Giusarma}, E. and {Gjerl{\o}w}, E. and {Gonz{\'a}lez-Nuevo}, J. and {G{\'o}rski}, K.~M. and {Gratton}, S. and {Gregorio}, A. and {Gruppuso}, A. and {Gudmundsson}, J.~E. and {Hamann}, J. and {Hansen}, F.~K. and {Hanson}, D. and {Harrison}, D.~L. and {Helou}, G. and {Henrot-Versill{\'e}}, S. and {Hern{\'a}ndez-Monteagudo}, C. and {Herranz}, D. and {Hildebrandt}, S.~R. and {Hivon}, E. and {Hobson}, M. and {Holmes}, W.~A. and {Hornstrup}, A. and {Hovest}, W. and {Huang}, Z. and {Huffenberger}, K.~M. and {Hurier}, G. and {Jaffe}, A.~H. and {Jaffe}, T.~R. and {Jones}, W.~C. and {Juvela}, M. and {Keih{\"a}nen}, E. and {Keskitalo}, R. and {Kisner}, T.~S. and {Kneissl}, R. and {Knoche}, J. and {Knox}, L. and {Kunz}, M. and {Kurki-Suonio}, H. and {Lagache}, G. and {L{\"a}hteenm{\"a}ki}, A. and {Lamarre}, J. -M. and {Lasenby}, A. and {Lattanzi}, M. and {Lawrence}, C.~R. and {Leahy}, J.~P. and {Leonardi}, R. and {Lesgourgues}, J. and {Levrier}, F. and {Lewis}, A. and {Liguori}, M. and {Lilje}, P.~B. and {Linden-V{\o}rnle}, M. and {L{\'o}pez-Caniego}, M. and {Lubin}, P.~M. and {Mac{\'\i}as-P{\'e}rez}, J.~F. and {Maggio}, G. and {Maino}, D. and {Mandolesi}, N. and {Mangilli}, A. and {Marchini}, A. and {Maris}, M. and {Martin}, P.~G. and {Martinelli}, M. and {Mart{\'\i}nez-Gonz{\'a}lez}, E. and {Masi}, S. and {Matarrese}, S. and {McGehee}, P. and {Meinhold}, P.~R. and {Melchiorri}, A. and {Melin}, J. -B. and {Mendes}, L. and {Mennella}, A. and {Migliaccio}, M. and {Millea}, M. and {Mitra}, S. and {Miville-Desch{\^e}nes}, M. -A. and {Moneti}, A. and {Montier}, L. and {Morgante}, G. and {Mortlock}, D. and {Moss}, A. and {Munshi}, D. and {Murphy}, J.~A. and {Naselsky}, P. and {Nati}, F. and {Natoli}, P. and {Netterfield}, C.~B. and {N{\o}rgaard-Nielsen}, H.~U. and {Noviello}, F. and {Novikov}, D. and {Novikov}, I. and {Oxborrow}, C.~A. and {Paci}, F. and {Pagano}, L. and {Pajot}, F. and {Paladini}, R. and {Paoletti}, D. and {Partridge}, B. and {Pasian}, F. and {Patanchon}, G. and {Pearson}, T.~J. and {Perdereau}, O. and {Perotto}, L. and {Perrotta}, F. and {Pettorino}, V. and {Piacentini}, F. and {Piat}, M. and {Pierpaoli}, E. and {Pietrobon}, D. and {Plaszczynski}, S. and {Pointecouteau}, E. and {Polenta}, G. and {Popa}, L. and {Pratt}, G.~W. and {Pr{\'e}zeau}, G.},
        title = "{Planck 2015 results. XIII. Cosmological parameters}",
      journal = {\aap},
         year = 2016,
        month = sep,
       volume = {594},
          eid = {A13},
        pages = {A13},
          doi = {10.1051/0004-6361/201525830},
archivePrefix = {arXiv},
       eprint = {1502.01589},
 primaryClass = {astro-ph.CO},
       adsurl = {https://ui.adsabs.harvard.edu/abs/2016A&A...594A..13P}
}

@ARTICLE{robotham+18,
       author = {{Robotham}, A.~S.~G. and {Davies}, L.~J.~M. and {Driver}, S.~P. and {Koushan}, S. and {Taranu}, D.~S. and {Casura}, S. and {Liske}, J.},
        title = "{ProFound: Source Extraction and Application to Modern Survey Data}",
      journal = {\mnras},
         year = 2018,
        month = may,
       volume = {476},
       number = {3},
        pages = {3137-3159},
          doi = {10.1093/mnras/sty440},
archivePrefix = {arXiv},
       eprint = {1802.00937},
 primaryClass = {astro-ph.IM},
       adsurl = {https://ui.adsabs.harvard.edu/abs/2018MNRAS.476.3137R}
}

@ARTICLE{davies+16,
       author = {{Davies}, L.~J.~M. and {Driver}, S.~P. and {Robotham}, A.~S.~G. and {Grootes}, M.~W. and {Popescu}, C.~C. and {Tuffs}, R.~J. and {Hopkins}, A. and {Alpaslan}, M. and {Andrews}, S.~K. and {Bland-Hawthorn}, J. and {Bremer}, M.~N. and {Brough}, S. and {Brown}, M.~J.~I. and {Cluver}, M.~E. and {Croom}, S. and {da Cunha}, E. and {Dunne}, L. and {Lara-L{\'o}pez}, M.~A. and {Liske}, J. and {Loveday}, J. and {Moffett}, A.~J. and {Owers}, M. and {Phillipps}, S. and {Sansom}, A.~E. and {Taylor}, E.~N. and {Michalowski}, M.~J. and {Ibar}, E. and {Smith}, M. and {Bourne}, N.},
        title = "{GAMA/H-ATLAS: a meta-analysis of SFR indicators - comprehensive measures of the SFR-M$_{*}$ relation and cosmic star formation history at z < 0.4}",
      journal = {\mnras},
         year = 2016,
        month = sep,
       volume = {461},
       number = {1},
        pages = {458-485},
          doi = {10.1093/mnras/stw1342},
archivePrefix = {arXiv},
       eprint = {1606.06299},
 primaryClass = {astro-ph.GA},
       adsurl = {https://ui.adsabs.harvard.edu/abs/2016MNRAS.461..458D}
}

@ARTICLE{hopkins+03,
       author = {{Hopkins}, A.~M. and {Miller}, C.~J. and {Nichol}, R.~C. and {Connolly}, A.~J. and {Bernardi}, M. and {G{\'o}mez}, P.~L. and {Goto}, T. and {Tremonti}, C.~A. and {Brinkmann}, J. and {Ivezi{\'c}}, {\v{Z}}. and {Lamb}, D.~Q.},
        title = "{Star Formation Rate Indicators in the Sloan Digital Sky Survey}",
      journal = {\apj},
         year = 2003,
        month = dec,
       volume = {599},
       number = {2},
        pages = {971-991},
          doi = {10.1086/379608},
archivePrefix = {arXiv},
       eprint = {astro-ph/0306621},
 primaryClass = {astro-ph},
       adsurl = {https://ui.adsabs.harvard.edu/abs/2003ApJ...599..971H}
}

@ARTICLE{robotham+11,
       author = {{Robotham}, A.~S.~G. and {Norberg}, P. and {Driver}, S.~P. and {Baldry}, I.~K. and {Bamford}, S.~P. and {Hopkins}, A.~M. and {Liske}, J. and {Loveday}, J. and {Merson}, A. and {Peacock}, J.~A. and {Brough}, S. and {Cameron}, E. and {Conselice}, C.~J. and {Croom}, S.~M. and {Frenk}, C.~S. and {Gunawardhana}, M. and {Hill}, D.~T. and {Jones}, D.~H. and {Kelvin}, L.~S. and {Kuijken}, K. and {Nichol}, R.~C. and {Parkinson}, H.~R. and {Pimbblet}, K.~A. and {Phillipps}, S. and {Popescu}, C.~C. and {Prescott}, M. and {Sharp}, R.~G. and {Sutherland}, W.~J. and {Taylor}, E.~N. and {Thomas}, D. and {Tuffs}, R.~J. and {van Kampen}, E. and {Wijesinghe}, D.},
        title = "{Galaxy and Mass Assembly (GAMA): the GAMA galaxy group catalogue (G$^{3}$Cv1)}",
      journal = {\mnras},
         year = 2011,
        month = oct,
       volume = {416},
       number = {4},
        pages = {2640-2668},
          doi = {10.1111/j.1365-2966.2011.19217.x},
archivePrefix = {arXiv},
       eprint = {1106.1994},
 primaryClass = {astro-ph.CO},
       adsurl = {https://ui.adsabs.harvard.edu/abs/2011MNRAS.416.2640R}
}

@ARTICLE{obreschkow+18,
       author = {{Obreschkow}, D. and {Murray}, S.~G. and {Robotham}, A.~S.~G. and {Westmeier}, T.},
        title = "{Eddington's demon: inferring galaxy mass functions and other distributions from uncertain data}",
      journal = {\mnras},
         year = 2018,
        month = mar,
       volume = {474},
       number = {4},
        pages = {5500-5522},
          doi = {10.1093/mnras/stx3155},
archivePrefix = {arXiv},
       eprint = {1712.00149},
 primaryClass = {astro-ph.GA},
       adsurl = {https://ui.adsabs.harvard.edu/abs/2018MNRAS.474.5500O}
}

@ARTICLE{baldry+08,
       author = {{Baldry}, I.~K. and {Glazebrook}, K. and {Driver}, S.~P.},
        title = "{On the galaxy stellar mass function, the mass-metallicity relation and the implied baryonic mass function}",
      journal = {\mnras},
         year = 2008,
        month = aug,
       volume = {388},
       number = {3},
        pages = {945-959},
          doi = {10.1111/j.1365-2966.2008.13348.x},
archivePrefix = {arXiv},
       eprint = {0804.2892},
 primaryClass = {astro-ph},
       adsurl = {https://ui.adsabs.harvard.edu/abs/2008MNRAS.388..945B}
}

@ARTICLE{baldry+12,
       author = {{Baldry}, I.~K. and {Driver}, S.~P. and {Loveday}, J. and {Taylor}, E.~N. and {Kelvin}, L.~S. and {Liske}, J. and {Norberg}, P. and {Robotham}, A.~S.~G. and {Brough}, S. and {Hopkins}, A.~M. and {Bamford}, S.~P. and {Peacock}, J.~A. and {Bland-Hawthorn}, J. and {Conselice}, C.~J. and {Croom}, S.~M. and {Jones}, D.~H. and {Parkinson}, H.~R. and {Popescu}, C.~C. and {Prescott}, M. and {Sharp}, R.~G. and {Tuffs}, R.~J.},
        title = "{Galaxy And Mass Assembly (GAMA): the galaxy stellar mass function at z < 0.06}",
      journal = {\mnras},
         year = 2012,
        month = mar,
       volume = {421},
       number = {1},
        pages = {621-634},
          doi = {10.1111/j.1365-2966.2012.20340.x},
archivePrefix = {arXiv},
       eprint = {1111.5707},
 primaryClass = {astro-ph.CO},
       adsurl = {https://ui.adsabs.harvard.edu/abs/2012MNRAS.421..621B}
}

@ARTICLE{pozzetti+10,
       author = {{Pozzetti}, L. and {Bolzonella}, M. and {Zucca}, E. and {Zamorani}, G. and {Lilly}, S. and {Renzini}, A. and {Moresco}, M. and {Mignoli}, M. and {Cassata}, P. and {Tasca}, L. and {Lamareille}, F. and {Maier}, C. and {Meneux}, B. and {Halliday}, C. and {Oesch}, P. and {Vergani}, D. and {Caputi}, K. and {Kova{\v{c}}}, K. and {Cimatti}, A. and {Cucciati}, O. and {Iovino}, A. and {Peng}, Y. and {Carollo}, M. and {Contini}, T. and {Kneib}, J. -P. and {Le F{\'e}vre}, O. and {Mainieri}, V. and {Scodeggio}, M. and {Bardelli}, S. and {Bongiorno}, A. and {Coppa}, G. and {de la Torre}, S. and {de Ravel}, L. and {Franzetti}, P. and {Garilli}, B. and {Kampczyk}, P. and {Knobel}, C. and {Le Borgne}, J. -F. and {Le Brun}, V. and {Pell{\`o}}, R. and {Perez Montero}, E. and {Ricciardelli}, E. and {Silverman}, J.~D. and {Tanaka}, M. and {Tresse}, L. and {Abbas}, U. and {Bottini}, D. and {Cappi}, A. and {Guzzo}, L. and {Koekemoer}, A.~M. and {Leauthaud}, A. and {Maccagni}, D. and {Marinoni}, C. and {McCracken}, H.~J. and {Memeo}, P. and {Porciani}, C. and {Scaramella}, R. and {Scarlata}, C. and {Scoville}, N.},
        title = "{zCOSMOS - 10k-bright spectroscopic sample. The bimodality in the galaxy stellar mass function: exploring its evolution with redshift}",
      journal = {\aap},
         year = 2010,
        month = nov,
       volume = {523},
          eid = {A13},
        pages = {A13},
          doi = {10.1051/0004-6361/200913020},
archivePrefix = {arXiv},
       eprint = {0907.5416},
 primaryClass = {astro-ph.CO},
       adsurl = {https://ui.adsabs.harvard.edu/abs/2010A&A...523A..13P}
}

@ARTICLE{vazquez+20,
       author = {{V{\'a}zquez-Mata}, J.~A. and {Loveday}, J. and {Riggs}, S.~D. and {Baldry}, I.~K. and {Davies}, L.~J.~M. and {Robotham}, A.~S.~G. and {Holwerda}, B.~W. and {Brown}, M.~J.~I. and {Cluver}, M.~E. and {Wang}, L. and {Alpaslan}, M. and {Bland-Hawthorn}, J. and {Brough}, S. and {Driver}, S.~P. and {Hopkins}, A.~M. and {Taylor}, E.~N. and {Wright}, A.~H.},
        title = "{Galaxy and mass assembly: luminosity and stellar mass functions in GAMA groups}",
      journal = {\mnras},
         year = 2020,
        month = nov,
       volume = {499},
       number = {1},
        pages = {631-652},
          doi = {10.1093/mnras/staa2889},
archivePrefix = {arXiv},
       eprint = {2009.08212},
 primaryClass = {astro-ph.GA},
       adsurl = {https://ui.adsabs.harvard.edu/abs/2020MNRAS.499..631V}
}

@ARTICLE{davies+19b,
       author = {{Davies}, L.~J.~M. and {Robotham}, A.~S.~G. and {Lagos}, C. del P. and {Driver}, S.~P. and {Stevens}, A.~R.~H. and {Bah{\'e}}, Y.~M. and {Alpaslan}, M. and {Bremer}, M.~N. and {Brown}, M.~J.~I. and {Brough}, S. and {Bland-Hawthorn}, J. and {Cortese}, L. and {Elahi}, P. and {Grootes}, M.~W. and {Holwerda}, B.~W. and {Ludlow}, A.~D. and {McGee}, S. and {Owers}, M. and {Phillipps}, S.},
        title = "{Galaxy and Mass Assembly (GAMA): environmental quenching of centrals and satellites in groups}",
      journal = {\mnras},
         year = 2019,
        month = mar,
       volume = {483},
       number = {4},
        pages = {5444-5458},
          doi = {10.1093/mnras/sty3393},
archivePrefix = {arXiv},
       eprint = {1901.01640},
 primaryClass = {astro-ph.GA},
       adsurl = {https://ui.adsabs.harvard.edu/abs/2019MNRAS.483.5444D}
}

@ARTICLE{kennicutt+98a,
       author = {{Kennicutt}, Jr., Robert C.},
        title = "{Star Formation in Galaxies Along the Hubble Sequence}",
      journal = {\araa},
         year = 1998,
        month = jan,
       volume = {36},
        pages = {189-232},
          doi = {10.1146/annurev.astro.36.1.189},
archivePrefix = {arXiv},
       eprint = {astro-ph/9807187},
 primaryClass = {astro-ph},
       adsurl = {https://ui.adsabs.harvard.edu/abs/1998ARA&A..36..189K}
}

@ARTICLE{kennicutt+98b,
       author = {{Kennicutt}, Jr., Robert C.},
        title = "{The Global Schmidt Law in Star-forming Galaxies}",
      journal = {\apj},
         year = 1998,
        month = may,
       volume = {498},
       number = {2},
        pages = {541-552},
          doi = {10.1086/305588},
archivePrefix = {arXiv},
       eprint = {astro-ph/9712213},
 primaryClass = {astro-ph},
       adsurl = {https://ui.adsabs.harvard.edu/abs/1998ApJ...498..541K}
}

@ARTICLE{hopkins+13,
       author = {{Hopkins}, A.~M. and {Driver}, S.~P. and {Brough}, S. and {Owers}, M.~S. and {Bauer}, A.~E. and {Gunawardhana}, M.~L.~P. and {Cluver}, M.~E. and {Colless}, M. and {Foster}, C. and {Lara-L{\'o}pez}, M.~A. and {Roseboom}, I. and {Sharp}, R. and {Steele}, O. and {Thomas}, D. and {Baldry}, I.~K. and {Brown}, M.~J.~I. and {Liske}, J. and {Norberg}, P. and {Robotham}, A.~S.~G. and {Bamford}, S. and {Bland-Hawthorn}, J. and {Drinkwater}, M.~J. and {Loveday}, J. and {Meyer}, M. and {Peacock}, J.~A. and {Tuffs}, R. and {Agius}, N. and {Alpaslan}, M. and {Andrae}, E. and {Cameron}, E. and {Cole}, S. and {Ching}, J.~H.~Y. and {Christodoulou}, L. and {Conselice}, C. and {Croom}, S. and {Cross}, N.~J.~G. and {De Propris}, R. and {Delhaize}, J. and {Dunne}, L. and {Eales}, S. and {Ellis}, S. and {Frenk}, C.~S. and {Graham}, Alister W. and {Grootes}, M.~W. and {H{\"a}u{\ss}ler}, B. and {Heymans}, C. and {Hill}, D. and {Hoyle}, B. and {Hudson}, M. and {Jarvis}, M. and {Johansson}, J. and {Jones}, D.~H. and {van Kampen}, E. and {Kelvin}, L. and {Kuijken}, K. and {L{\'o}pez-S{\'a}nchez}, {\'A}. and {Maddox}, S. and {Madore}, B. and {Maraston}, C. and {McNaught-Roberts}, T. and {Nichol}, R.~C. and {Oliver}, S. and {Parkinson}, H. and {Penny}, S. and {Phillipps}, S. and {Pimbblet}, K.~A. and {Ponman}, T. and {Popescu}, C.~C. and {Prescott}, M. and {Proctor}, R. and {Sadler}, E.~M. and {Sansom}, A.~E. and {Seibert}, M. and {Staveley-Smith}, L. and {Sutherland}, W. and {Taylor}, E. and {Van Waerbeke}, L. and {V{\'a}zquez-Mata}, J.~A. and {Warren}, S. and {Wijesinghe}, D.~B. and {Wild}, V. and {Wilkins}, S.},
        title = "{Galaxy And Mass Assembly (GAMA): spectroscopic analysis}",
      journal = {\mnras},
         year = 2013,
        month = apr,
       volume = {430},
       number = {3},
        pages = {2047-2066},
          doi = {10.1093/mnras/stt030},
archivePrefix = {arXiv},
       eprint = {1301.7127},
 primaryClass = {astro-ph.CO},
       adsurl = {https://ui.adsabs.harvard.edu/abs/2013MNRAS.430.2047H}
}

@ARTICLE{gunawardhana+11,
       author = {{Gunawardhana}, M.~L.~P. and {Hopkins}, A.~M. and {Sharp}, R.~G. and {Brough}, S. and {Taylor}, E. and {Bland-Hawthorn}, J. and {Maraston}, C. and {Tuffs}, R.~J. and {Popescu}, C.~C. and {Wijesinghe}, D. and {Jones}, D.~H. and {Croom}, S. and {Sadler}, E. and {Wilkins}, S. and {Driver}, S.~P. and {Liske}, J. and {Norberg}, P. and {Baldry}, I.~K. and {Bamford}, S.~P. and {Loveday}, J. and {Peacock}, J.~A. and {Robotham}, A.~S.~G. and {Zucker}, D.~B. and {Parker}, Q.~A. and {Conselice}, C.~J. and {Cameron}, E. and {Frenk}, C.~S. and {Hill}, D.~T. and {Kelvin}, L.~S. and {Kuijken}, K. and {Madore}, B.~F. and {Nichol}, B. and {Parkinson}, H.~R. and {Pimbblet}, K.~A. and {Prescott}, M. and {Sutherland}, W.~J. and {Thomas}, D. and {van Kampen}, E.},
        title = "{Galaxy and Mass Assembly (GAMA): the star formation rate dependence of the stellar initial mass function}",
      journal = {\mnras},
         year = 2011,
        month = aug,
       volume = {415},
       number = {2},
        pages = {1647-1662},
          doi = {10.1111/j.1365-2966.2011.18800.x},
archivePrefix = {arXiv},
       eprint = {1104.2379},
 primaryClass = {astro-ph.CO},
       adsurl = {https://ui.adsabs.harvard.edu/abs/2011MNRAS.415.1647G}
}

@ARTICLE{gunawardhana+13,
       author = {{Gunawardhana}, M.~L.~P. and {Hopkins}, A.~M. and {Bland-Hawthorn}, J. and {Brough}, S. and {Sharp}, R. and {Loveday}, J. and {Taylor}, E. and {Jones}, D.~H. and {Lara-L{\'o}pez}, M.~A. and {Bauer}, A.~E. and {Colless}, M. and {Owers}, M. and {Baldry}, I.~K. and {L{\'o}pez-S{\'a}nchez}, A.~R. and {Foster}, C. and {Bamford}, S. and {Brown}, M.~J.~I. and {Driver}, S.~P. and {Drinkwater}, M.~J. and {Liske}, J. and {Meyer}, M. and {Norberg}, P. and {Robotham}, A.~S.~G. and {Ching}, J.~H.~Y. and {Cluver}, M.~E. and {Croom}, S. and {Kelvin}, L. and {Prescott}, M. and {Steele}, O. and {Thomas}, D. and {Wang}, L.},
        title = "{Galaxy And Mass Assembly: evolution of the H{\ensuremath{\alpha}} luminosity function and star formation rate density up to z < 0.35}",
      journal = {\mnras},
         year = 2013,
        month = aug,
       volume = {433},
       number = {4},
        pages = {2764-2789},
          doi = {10.1093/mnras/stt890},
archivePrefix = {arXiv},
       eprint = {1305.5308},
 primaryClass = {astro-ph.CO},
       adsurl = {https://ui.adsabs.harvard.edu/abs/2013MNRAS.433.2764G}
}

@ARTICLE{driver+13,
       author = {{Driver}, S.~P. and {Robotham}, A.~S.~G. and {Bland-Hawthorn}, J. and {Brown}, M. and {Hopkins}, A. and {Liske}, J. and {Phillipps}, S. and {Wilkins}, S.},
        title = "{Two-phase galaxy evolution: the cosmic star formation histories of spheroids and discs}",
      journal = {\mnras},
         year = 2013,
        month = apr,
       volume = {430},
       number = {4},
        pages = {2622-2632},
          doi = {10.1093/mnras/sts717},
archivePrefix = {arXiv},
       eprint = {1301.0979},
 primaryClass = {astro-ph.CO},
       adsurl = {https://ui.adsabs.harvard.edu/abs/2013MNRAS.430.2622D}
}

@ARTICLE{davies+19a,
       author = {{Davies}, L.~J.~M. and {Lagos}, C. del P. and {Katsianis}, A. and {Robotham}, A.~S.~G. and {Cortese}, L. and {Driver}, S.~P. and {Bremer}, M.~N. and {Brown}, M.~J.~I. and {Brough}, S. and {Cluver}, M.~E. and {Grootes}, M.~W. and {Holwerda}, B.~W. and {Owers}, M. and {Phillipps}, S.},
        title = "{Galaxy And Mass Assembly (GAMA): The sSFR-M$_{*}$ relation part I - {\ensuremath{\sigma}}$_{sSFR}$-M$_{*}$ as a function of sample, SFR indicator, and environment}",
      journal = {\mnras},
         year = 2019,
        month = feb,
       volume = {483},
       number = {2},
        pages = {1881-1900},
          doi = {10.1093/mnras/sty2957},
archivePrefix = {arXiv},
       eprint = {1811.03712},
 primaryClass = {astro-ph.GA},
       adsurl = {https://ui.adsabs.harvard.edu/abs/2019MNRAS.483.1881D}
}

@ARTICLE{driver+16,
       author = {{Driver}, Simon P. and {Wright}, Angus H. and {Andrews}, Stephen K. and {Davies}, Luke J. and {Kafle}, Prajwal R. and {Lange}, Rebecca and {Moffett}, Amanda J. and {Mannering}, Elizabeth and {Robotham}, Aaron S.~G. and {Vinsen}, Kevin and {Alpaslan}, Mehmet and {Andrae}, Ellen and {Baldry}, Ivan K. and {Bauer}, Amanda E. and {Bamford}, Steven P. and {Bland-Hawthorn}, Joss and {Bourne}, Nathan and {Brough}, Sarah and {Brown}, Michael J.~I. and {Cluver}, Michelle E. and {Croom}, Scott and {Colless}, Matthew and {Conselice}, Christopher J. and {da Cunha}, Elisabete and {De Propris}, Roberto and {Drinkwater}, Michael and {Dunne}, Loretta and {Eales}, Steve and {Edge}, Alastair and {Frenk}, Carlos and {Graham}, Alister W. and {Grootes}, Meiert and {Holwerda}, Benne W. and {Hopkins}, Andrew M. and {Ibar}, Edo and {van Kampen}, Eelco and {Kelvin}, Lee S. and {Jarrett}, Tom and {Jones}, D. Heath and {Lara-Lopez}, Maritza A. and {Liske}, Jochen and {Lopez-Sanchez}, Angel R. and {Loveday}, Jon and {Maddox}, Steve J. and {Madore}, Barry and {Mahajan}, Smriti and {Meyer}, Martin and {Norberg}, Peder and {Penny}, Samantha J. and {Phillipps}, Steven and {Popescu}, Cristina and {Tuffs}, Richard J. and {Peacock}, John A. and {Pimbblet}, Kevin A. and {Prescott}, Matthew and {Rowlands}, Kate and {Sansom}, Anne E. and {Seibert}, Mark and {Smith}, Matthew W.~L. and {Sutherland}, Will J. and {Taylor}, Edward N. and {Valiante}, Elisabetta and {Vazquez-Mata}, J. Antonio and {Wang}, Lingyu and {Wilkins}, Stephen M. and {Williams}, Richard},
        title = "{Galaxy And Mass Assembly (GAMA): Panchromatic Data Release (far-UV-far-IR) and the low-z energy budget}",
      journal = {\mnras},
         year = 2016,
        month = feb,
       volume = {455},
       number = {4},
        pages = {3911-3942},
          doi = {10.1093/mnras/stv2505},
archivePrefix = {arXiv},
       eprint = {1508.02076},
 primaryClass = {astro-ph.GA},
       adsurl = {https://ui.adsabs.harvard.edu/abs/2016MNRAS.455.3911D}
}

@ARTICLE{sbaffoni+25,
       author = {{Sbaffoni}, A. and {Liske}, J. and {Driver}, S.~P. and {Robotham}, A.~S.~G. and {Taylor}, E.~N.},
        title = "{Galaxy And Mass Assembly (GAMA): Environment-dependent galaxy stellar mass functions in the low-redshift Universe}",
      journal = {\aap},
         year = 2025,
        month = apr,
       volume = {696},
          eid = {A89},
        pages = {A89},
          doi = {10.1051/0004-6361/202453570},
archivePrefix = {arXiv},
       eprint = {2503.21363},
 primaryClass = {astro-ph.GA},
       adsurl = {https://ui.adsabs.harvard.edu/abs/2025A&A...696A..89S},
        alias = {Paper I}
}

@ARTICLE{weigel+16,
       author = {{Weigel}, Anna K. and {Schawinski}, Kevin and {Bruderer}, Claudio},
        title = "{Stellar mass functions: methods, systematics and results for the local Universe}",
      journal = {\mnras},
         year = 2016,
        month = jun,
       volume = {459},
       number = {2},
        pages = {2150-2187},
          doi = {10.1093/mnras/stw756},
archivePrefix = {arXiv},
       eprint = {1604.00008},
 primaryClass = {astro-ph.GA},
       adsurl = {https://ui.adsabs.harvard.edu/abs/2016MNRAS.459.2150W}
}

@ARTICLE{peng+10,
       author = {{Peng}, Ying-jie and {Lilly}, Simon J. and {Kova{\v{c}}}, Katarina and {Bolzonella}, Micol and {Pozzetti}, Lucia and {Renzini}, Alvio and {Zamorani}, Gianni and {Ilbert}, Olivier and {Knobel}, Christian and {Iovino}, Angela and {Maier}, Christian and {Cucciati}, Olga and {Tasca}, Lidia and {Carollo}, C. Marcella and {Silverman}, John and {Kampczyk}, Pawel and {de Ravel}, Loic and {Sanders}, David and {Scoville}, Nicholas and {Contini}, Thierry and {Mainieri}, Vincenzo and {Scodeggio}, Marco and {Kneib}, Jean-Paul and {Le F{\`e}vre}, Olivier and {Bardelli}, Sandro and {Bongiorno}, Angela and {Caputi}, Karina and {Coppa}, Graziano and {de la Torre}, Sylvain and {Franzetti}, Paolo and {Garilli}, Bianca and {Lamareille}, Fabrice and {Le Borgne}, Jean-Francois and {Le Brun}, Vincent and {Mignoli}, Marco and {Perez Montero}, Enrique and {Pello}, Roser and {Ricciardelli}, Elena and {Tanaka}, Masayuki and {Tresse}, Laurence and {Vergani}, Daniela and {Welikala}, Niraj and {Zucca}, Elena and {Oesch}, Pascal and {Abbas}, Ummi and {Barnes}, Luke and {Bordoloi}, Rongmon and {Bottini}, Dario and {Cappi}, Alberto and {Cassata}, Paolo and {Cimatti}, Andrea and {Fumana}, Marco and {Hasinger}, Gunther and {Koekemoer}, Anton and {Leauthaud}, Alexei and {Maccagni}, Dario and {Marinoni}, Christian and {McCracken}, Henry and {Memeo}, Pierdomenico and {Meneux}, Baptiste and {Nair}, Preethi and {Porciani}, Cristiano and {Presotto}, Valentina and {Scaramella}, Roberto},
        title = "{Mass and Environment as Drivers of Galaxy Evolution in SDSS and zCOSMOS and the Origin of the Schechter Function}",
      journal = {\apj},
         year = 2010,
        month = sep,
       volume = {721},
       number = {1},
        pages = {193-221},
          doi = {10.1088/0004-637X/721/1/193},
archivePrefix = {arXiv},
       eprint = {1003.4747},
 primaryClass = {astro-ph.CO},
       adsurl = {https://ui.adsabs.harvard.edu/abs/2010ApJ...721..193P}
}

@ARTICLE{peng+12,
       author = {{Peng}, Ying-jie and {Lilly}, Simon J. and {Renzini}, Alvio and {Carollo}, Marcella},
        title = "{Mass and Environment as Drivers of Galaxy Evolution. II. The Quenching of Satellite Galaxies as the Origin of Environmental Effects}",
      journal = {\apj},
         year = 2012,
        month = sep,
       volume = {757},
       number = {1},
          eid = {4},
        pages = {4},
          doi = {10.1088/0004-637X/757/1/4},
archivePrefix = {arXiv},
       eprint = {1106.2546},
 primaryClass = {astro-ph.CO},
       adsurl = {https://ui.adsabs.harvard.edu/abs/2012ApJ...757....4P}
}

@ARTICLE{taylor+15,
       author = {{Taylor}, Edward N. and {Hopkins}, Andrew M. and {Baldry}, Ivan K. and {Bland-Hawthorn}, Joss and {Brown}, Michael J.~I. and {Colless}, Matthew and {Driver}, Simon and {Norberg}, Peder and {Robotham}, Aaron S.~G. and {Alpaslan}, Mehmet and {Brough}, Sarah and {Cluver}, Michelle E. and {Gunawardhana}, Madusha and {Kelvin}, Lee S. and {Liske}, Jochen and {Conselice}, Christopher J. and {Croom}, Scott and {Foster}, Caroline and {Jarrett}, Thomas H. and {Lara-Lopez}, Maritza and {Loveday}, Jon},
        title = "{Galaxy And Mass Assembly (GAMA): deconstructing bimodality - I. Red ones and blue ones}",
      journal = {\mnras},
         year = 2015,
        month = jan,
       volume = {446},
       number = {2},
        pages = {2144-2185},
          doi = {10.1093/mnras/stu1900},
archivePrefix = {arXiv},
       eprint = {1408.5984},
 primaryClass = {astro-ph.GA},
       adsurl = {https://ui.adsabs.harvard.edu/abs/2015MNRAS.446.2144T}
}

@ARTICLE{wright+17,
       author = {{Wright}, A.~H. and {Robotham}, A.~S.~G. and {Driver}, S.~P. and {Alpaslan}, M. and {Andrews}, S.~K. and {Baldry}, I.~K. and {Bland-Hawthorn}, J. and {Brough}, S. and {Brown}, M.~J.~I. and {Colless}, M. and {da Cunha}, E. and {Davies}, L.~J.~M. and {Graham}, Alister W. and {Holwerda}, B.~W. and {Hopkins}, A.~M. and {Kafle}, P.~R. and {Kelvin}, L.~S. and {Loveday}, J. and {Maddox}, S.~J. and {Meyer}, M.~J. and {Moffett}, A.~J. and {Norberg}, P. and {Phillipps}, S. and {Rowlands}, K. and {Taylor}, E.~N. and {Wang}, L. and {Wilkins}, S.~M.},
        title = "{Galaxy And Mass Assembly (GAMA): the galaxy stellar mass function to z = 0.1 from the r-band selected equatorial regions}",
      journal = {\mnras},
         year = 2017,
        month = sep,
       volume = {470},
       number = {1},
        pages = {283-302},
          doi = {10.1093/mnras/stx1149},
archivePrefix = {arXiv},
       eprint = {1705.04074},
 primaryClass = {astro-ph.GA},
       adsurl = {https://ui.adsabs.harvard.edu/abs/2017MNRAS.470..283W}
}

@ARTICLE{taylor+11,
       author = {{Taylor}, Edward N. and {Hopkins}, Andrew M. and {Baldry}, Ivan K. and {Brown}, Michael J.~I. and {Driver}, Simon P. and {Kelvin}, Lee S. and {Hill}, David T. and {Robotham}, Aaron S.~G. and {Bland-Hawthorn}, Joss and {Jones}, D.~H. and {Sharp}, R.~G. and {Thomas}, Daniel and {Liske}, Jochen and {Loveday}, Jon and {Norberg}, Peder and {Peacock}, J.~A. and {Bamford}, Steven P. and {Brough}, Sarah and {Colless}, Matthew and {Cameron}, Ewan and {Conselice}, Christopher J. and {Croom}, Scott M. and {Frenk}, C.~S. and {Gunawardhana}, Madusha and {Kuijken}, Konrad and {Nichol}, R.~C. and {Parkinson}, H.~R. and {Phillipps}, S. and {Pimbblet}, K.~A. and {Popescu}, C.~C. and {Prescott}, Matthew and {Sutherland}, W.~J. and {Tuffs}, R.~J. and {van Kampen}, Eelco and {Wijesinghe}, D.},
        title = "{Galaxy And Mass Assembly (GAMA): stellar mass estimates}",
      journal = {\mnras},
         year = 2011,
        month = dec,
       volume = {418},
       number = {3},
        pages = {1587-1620},
          doi = {10.1111/j.1365-2966.2011.19536.x},
archivePrefix = {arXiv},
       eprint = {1108.0635},
 primaryClass = {astro-ph.CO},
       adsurl = {https://ui.adsabs.harvard.edu/abs/2011MNRAS.418.1587T}
}

@ARTICLE{bell+03,
       author = {{Bell}, Eric F. and {McIntosh}, Daniel H. and {Katz}, Neal and {Weinberg}, Martin D.},
        title = "{The Optical and Near-Infrared Properties of Galaxies. I. Luminosity and Stellar Mass Functions}",
      journal = {\apjs},
         year = 2003,
        month = dec,
       volume = {149},
       number = {2},
        pages = {289-312},
          doi = {10.1086/378847},
archivePrefix = {arXiv},
       eprint = {astro-ph/0302543},
 primaryClass = {astro-ph},
       adsurl = {https://ui.adsabs.harvard.edu/abs/2003ApJS..149..289B}
}

@ARTICLE{baldry+04,
       author = {{Baldry}, Ivan K. and {Glazebrook}, Karl and {Brinkmann}, Jon and {Ivezi{\'c}}, {\v{Z}}eljko and {Lupton}, Robert H. and {Nichol}, Robert C. and {Szalay}, Alexander S.},
        title = "{Quantifying the Bimodal Color-Magnitude Distribution of Galaxies}",
      journal = {\apj},
         year = 2004,
        month = jan,
       volume = {600},
       number = {2},
        pages = {681-694},
          doi = {10.1086/380092},
archivePrefix = {arXiv},
       eprint = {astro-ph/0309710},
 primaryClass = {astro-ph},
       adsurl = {https://ui.adsabs.harvard.edu/abs/2004ApJ...600..681B}
}

@ARTICLE{yang+09,
       author = {{Yang}, Xiaohu and {Mo}, H.~J. and {van den Bosch}, Frank C.},
        title = "{Galaxy Groups in the SDSS DR4. III. The Luminosity and Stellar Mass Functions}",
      journal = {\apj},
         year = 2009,
        month = apr,
       volume = {695},
       number = {2},
        pages = {900-916},
          doi = {10.1088/0004-637X/695/2/900},
archivePrefix = {arXiv},
       eprint = {0808.0539},
 primaryClass = {astro-ph},
       adsurl = {https://ui.adsabs.harvard.edu/abs/2009ApJ...695..900Y}
}

@ARTICLE{baugh+06,
       author = {{Baugh}, C.~M.},
        title = "{A primer on hierarchical galaxy formation: the semi-analytical approach}",
      journal = {Reports on Progress in Physics},
         year = 2006,
        month = dec,
       volume = {69},
       number = {12},
        pages = {3101-3156},
          doi = {10.1088/0034-4885/69/12/R02},
archivePrefix = {arXiv},
       eprint = {astro-ph/0610031},
 primaryClass = {astro-ph},
       adsurl = {https://ui.adsabs.harvard.edu/abs/2006RPPh...69.3101B}
}

@ARTICLE{keres+05,
       author = {{Kere{\v{s}}}, Du{\v{s}}an and {Katz}, Neal and {Weinberg}, David H. and {Dav{\'e}}, Romeel},
        title = "{How do galaxies get their gas?}",
      journal = {\mnras},
         year = 2005,
        month = oct,
       volume = {363},
       number = {1},
        pages = {2-28},
          doi = {10.1111/j.1365-2966.2005.09451.x},
archivePrefix = {arXiv},
       eprint = {astro-ph/0407095},
 primaryClass = {astro-ph},
       adsurl = {https://ui.adsabs.harvard.edu/abs/2005MNRAS.363....2K}
}

@ARTICLE{brinchmann+04,
       author = {{Brinchmann}, J. and {Charlot}, S. and {White}, S.~D.~M. and {Tremonti}, C. and {Kauffmann}, G. and {Heckman}, T. and {Brinkmann}, J.},
        title = "{The physical properties of star-forming galaxies in the low-redshift Universe}",
      journal = {\mnras},
         year = 2004,
        month = jul,
       volume = {351},
       number = {4},
        pages = {1151-1179},
          doi = {10.1111/j.1365-2966.2004.07881.x},
archivePrefix = {arXiv},
       eprint = {astro-ph/0311060},
 primaryClass = {astro-ph},
       adsurl = {https://ui.adsabs.harvard.edu/abs/2004MNRAS.351.1151B}
}

@ARTICLE{daddi+07,
       author = {{Daddi}, E. and {Dickinson}, M. and {Morrison}, G. and {Chary}, R. and {Cimatti}, A. and {Elbaz}, D. and {Frayer}, D. and {Renzini}, A. and {Pope}, A. and {Alexander}, D.~M. and {Bauer}, F.~E. and {Giavalisco}, M. and {Huynh}, M. and {Kurk}, J. and {Mignoli}, M.},
        title = "{Multiwavelength Study of Massive Galaxies at z\raisebox{-0.5ex}\textasciitilde2. I. Star Formation and Galaxy Growth}",
      journal = {\apj},
         year = 2007,
        month = nov,
       volume = {670},
       number = {1},
        pages = {156-172},
          doi = {10.1086/521818},
archivePrefix = {arXiv},
       eprint = {0705.2831},
 primaryClass = {astro-ph},
       adsurl = {https://ui.adsabs.harvard.edu/abs/2007ApJ...670..156D}
}

@ARTICLE{elbaz+07,
       author = {{Elbaz}, D. and {Daddi}, E. and {Le Borgne}, D. and {Dickinson}, M. and {Alexander}, D.~M. and {Chary}, R. -R. and {Starck}, J. -L. and {Brandt}, W.~N. and {Kitzbichler}, M. and {MacDonald}, E. and {Nonino}, M. and {Popesso}, P. and {Stern}, D. and {Vanzella}, E.},
        title = "{The reversal of the star formation-density relation in the distant universe}",
      journal = {\aap},
         year = 2007,
        month = jun,
       volume = {468},
       number = {1},
        pages = {33-48},
          doi = {10.1051/0004-6361:20077525},
archivePrefix = {arXiv},
       eprint = {astro-ph/0703653},
 primaryClass = {astro-ph},
       adsurl = {https://ui.adsabs.harvard.edu/abs/2007A&A...468...33E}
}

@ARTICLE{noeske+07,
       author = {{Noeske}, K.~G. and {Weiner}, B.~J. and {Faber}, S.~M. and {Papovich}, C. and {Koo}, D.~C. and {Somerville}, R.~S. and {Bundy}, K. and {Conselice}, C.~J. and {Newman}, J.~A. and {Schiminovich}, D. and {Le Floc'h}, E. and {Coil}, A.~L. and {Rieke}, G.~H. and {Lotz}, J.~M. and {Primack}, J.~R. and {Barmby}, P. and {Cooper}, M.~C. and {Davis}, M. and {Ellis}, R.~S. and {Fazio}, G.~G. and {Guhathakurta}, P. and {Huang}, J. and {Kassin}, S.~A. and {Martin}, D.~C. and {Phillips}, A.~C. and {Rich}, R.~M. and {Small}, T.~A. and {Willmer}, C.~N.~A. and {Wilson}, G.},
        title = "{Star Formation in AEGIS Field Galaxies since z=1.1: The Dominance of Gradually Declining Star Formation, and the Main Sequence of Star-forming Galaxies}",
      journal = {\apjl},
         year = 2007,
        month = may,
       volume = {660},
       number = {1},
        pages = {L43-L46},
          doi = {10.1086/517926},
archivePrefix = {arXiv},
       eprint = {astro-ph/0701924},
 primaryClass = {astro-ph},
       adsurl = {https://ui.adsabs.harvard.edu/abs/2007ApJ...660L..43N}
}

@ARTICLE{dacunha+10,
       author = {{da Cunha}, E. and {Charmandaris}, V. and {D{\'\i}az-Santos}, T. and {Armus}, L. and {Marshall}, J.~A. and {Elbaz}, D.},
        title = "{Exploring the physical properties of local star-forming ULIRGs from the ultraviolet to the infrared}",
      journal = {\aap},
         year = 2010,
        month = nov,
       volume = {523},
          eid = {A78},
        pages = {A78},
          doi = {10.1051/0004-6361/201014498},
archivePrefix = {arXiv},
       eprint = {1008.2000},
 primaryClass = {astro-ph.CO},
       adsurl = {https://ui.adsabs.harvard.edu/abs/2010A&A...523A..78D}
}

@ARTICLE{kauffmann+03,
       author = {{Kauffmann}, Guinevere and {Heckman}, Timothy M. and {White}, Simon D.~M. and {Charlot}, St{\'e}phane and {Tremonti}, Christy and {Peng}, Eric W. and {Seibert}, Mark and {Brinkmann}, Jon and {Nichol}, Robert C. and {SubbaRao}, Mark and {York}, Don},
        title = "{The dependence of star formation history and internal structure on stellar mass for {}10$^{5}$ low-redshift galaxies}",
      journal = {\mnras},
         year = 2003,
        month = may,
       volume = {341},
       number = {1},
        pages = {54-69},
          doi = {10.1046/j.1365-8711.2003.06292.x},
archivePrefix = {arXiv},
       eprint = {astro-ph/0205070},
 primaryClass = {astro-ph},
       adsurl = {https://ui.adsabs.harvard.edu/abs/2003MNRAS.341...54K}
}

@ARTICLE{guglielmo+15,
       author = {{Guglielmo}, Valentina and {Poggianti}, Bianca M. and {Moretti}, Alessia and {Fritz}, Jacopo and {Calvi}, Rosa and {Vulcani}, Benedetta and {Fasano}, Giovanni and {Paccagnella}, Angela},
        title = "{The star formation history of galaxies: the role of galaxy mass, morphology and environment}",
      journal = {\mnras},
         year = 2015,
        month = jul,
       volume = {450},
       number = {3},
        pages = {2749-2763},
          doi = {10.1093/mnras/stv757},
archivePrefix = {arXiv},
       eprint = {1504.01594},
 primaryClass = {astro-ph.GA},
       adsurl = {https://ui.adsabs.harvard.edu/abs/2015MNRAS.450.2749G}
}

@ARTICLE{netzer+09,
       author = {{Netzer}, Hagai},
        title = "{Accretion and star formation rates in low-redshift type II active galactic nuclei}",
      journal = {\mnras},
         year = 2009,
        month = nov,
       volume = {399},
       number = {4},
        pages = {1907-1920},
          doi = {10.1111/j.1365-2966.2009.15434.x},
archivePrefix = {arXiv},
       eprint = {0907.3575},
 primaryClass = {astro-ph.GA},
       adsurl = {https://ui.adsabs.harvard.edu/abs/2009MNRAS.399.1907N}
}

@ARTICLE{thacker+14,
       author = {{Thacker}, Robert J. and {MacMackin}, C. and {Wurster}, James and {Hobbs}, Alexander},
        title = "{AGN feedback models: correlations with star formation and observational implications of time evolution}",
      journal = {\mnras},
         year = 2014,
        month = sep,
       volume = {443},
       number = {2},
        pages = {1125-1141},
          doi = {10.1093/mnras/stu1180},
archivePrefix = {arXiv},
       eprint = {1407.0685},
 primaryClass = {astro-ph.GA},
       adsurl = {https://ui.adsabs.harvard.edu/abs/2014MNRAS.443.1125T}
}

@ARTICLE{ellison+08,
       author = {{Ellison}, Sara L. and {Patton}, David R. and {Simard}, Luc and {McConnachie}, Alan W.},
        title = "{Galaxy Pairs in the Sloan Digital Sky Survey. I. Star Formation, Active Galactic Nucleus Fraction, and the Mass-Metallicity Relation}",
      journal = {\aj},
         year = 2008,
        month = may,
       volume = {135},
       number = {5},
        pages = {1877-1899},
          doi = {10.1088/0004-6256/135/5/1877},
archivePrefix = {arXiv},
       eprint = {0803.0161},
 primaryClass = {astro-ph},
       adsurl = {https://ui.adsabs.harvard.edu/abs/2008AJ....135.1877E}
}

@ARTICLE{mannucci+10,
       author = {{Mannucci}, F. and {Cresci}, G. and {Maiolino}, R. and {Marconi}, A. and {Gnerucci}, A.},
        title = "{A fundamental relation between mass, star formation rate and metallicity in local and high-redshift galaxies}",
      journal = {\mnras},
         year = 2010,
        month = nov,
       volume = {408},
       number = {4},
        pages = {2115-2127},
          doi = {10.1111/j.1365-2966.2010.17291.x},
archivePrefix = {arXiv},
       eprint = {1005.0006},
 primaryClass = {astro-ph.CO},
       adsurl = {https://ui.adsabs.harvard.edu/abs/2010MNRAS.408.2115M}
}

@ARTICLE{lara-lopez+13,
       author = {{Lara-L{\'o}pez}, M.~A. and {Hopkins}, A.~M. and {L{\'o}pez-S{\'a}nchez}, A.~R. and {Brough}, S. and {Gunawardhana}, M.~L.~P. and {Colless}, M. and {Robotham}, A.~S.~G. and {Bauer}, A.~E. and {Bland-Hawthorn}, J. and {Cluver}, M. and {Driver}, S. and {Foster}, C. and {Kelvin}, L.~S. and {Liske}, J. and {Loveday}, J. and {Owers}, M.~S. and {Ponman}, T.~J. and {Sharp}, R.~G. and {Steele}, O. and {Taylor}, E.~N. and {Thomas}, D.},
        title = "{Galaxy And Mass Assembly (GAMA): a deeper view of the mass, metallicity and SFR relationships}",
      journal = {\mnras},
         year = 2013,
        month = sep,
       volume = {434},
       number = {1},
        pages = {451-470},
          doi = {10.1093/mnras/stt1031},
archivePrefix = {arXiv},
       eprint = {1306.1583},
 primaryClass = {astro-ph.CO},
       adsurl = {https://ui.adsabs.harvard.edu/abs/2013MNRAS.434..451L}
}

@ARTICLE{mcintosh+08,
       author = {{McIntosh}, Daniel H. and {Guo}, Yicheng and {Hertzberg}, Jen and {Katz}, Neal and {Mo}, H.~J. and {van den Bosch}, Frank C. and {Yang}, Xiaohu},
        title = "{Ongoing assembly of massive galaxies by major merging in large groups and clusters from the SDSS}",
      journal = {\mnras},
         year = 2008,
        month = aug,
       volume = {388},
       number = {4},
        pages = {1537-1556},
          doi = {10.1111/j.1365-2966.2008.13531.x},
archivePrefix = {arXiv},
       eprint = {0710.2157},
 primaryClass = {astro-ph},
       adsurl = {https://ui.adsabs.harvard.edu/abs/2008MNRAS.388.1537M}
}

@ARTICLE{ellison+10,
       author = {{Ellison}, Sara L. and {Patton}, David R. and {Simard}, Luc and {McConnachie}, Alan W. and {Baldry}, Ivan K. and {Mendel}, J. Trevor},
        title = "{Galaxy pairs in the Sloan Digital Sky Survey - II. The effect of environment on interactions}",
      journal = {\mnras},
         year = 2010,
        month = sep,
       volume = {407},
       number = {3},
        pages = {1514-1528},
          doi = {10.1111/j.1365-2966.2010.17076.x},
archivePrefix = {arXiv},
       eprint = {1002.4418},
 primaryClass = {astro-ph.CO},
       adsurl = {https://ui.adsabs.harvard.edu/abs/2010MNRAS.407.1514E}
}

@ARTICLE{deravel+11,
       author = {{de Ravel}, L. and {Kampczyk}, P. and {Le F{\`e}vre}, O. and {Lilly}, S.~J. and {Tasca}, L. and {Tresse}, L. and {Lopez-Sanjuan}, C. and {Bolzonella}, M. and {Kovac}, K. and {Abbas}, U. and {Bardelli}, S. and {Bongiorno}, A. and {Caputi}, K. and {Contini}, T. and {Coppa}, G. and {Cucciati}, O. and {de la Torre}, S. and {Dunlop}, J.~S. and {Franzetti}, P. and {Garilli}, B. and {Iovino}, A. and {Kneib}, J. -P. and {Koekemoer}, A.~M. and {Knobel}, C. and {Lamareille}, F. and {Le Borgne}, J. -F. and {Le Brun}, V. and {Leauthaud}, A. and {Maier}, C. and {Mainieri}, V. and {Mignoli}, M. and {Pello}, R. and {Peng}, Y. and {Perez Montero}, E. and {Ricciardelli}, E. and {Scodeggio}, M. and {Silverman}, J.~D. and {Tanaka}, M. and {Vergani}, D. and {Zamorani}, G. and {Zucca}, E. and {Bottini}, D. and {Cappi}, A. and {Carollo}, C.~M. and {Cassata}, P. and {Cimatti}, A. and {Fumana}, M. and {Guzzo}, L. and {Maccagni}, D. and {Marinoni}, C. and {McCracken}, H.~J. and {Memeo}, P. and {Meneux}, B. and {Oesch}, P. and {Porciani}, C. and {Pozzetti}, L. and {Renzini}, A. and {Scaramella}, R. and {Scarlata}, C.},
        title = "{The zCOSMOS redshift survey : Influence of luminosity, mass and environment on the galaxy merger rate}",
      journal = {arXiv e-prints},
         year = 2011,
        month = apr,
          eid = {arXiv:1104.5470},
        pages = {arXiv:1104.5470},
          doi = {10.48550/arXiv.1104.5470},
archivePrefix = {arXiv},
       eprint = {1104.5470},
 primaryClass = {astro-ph.CO},
       adsurl = {https://ui.adsabs.harvard.edu/abs/2011arXiv1104.5470D}
}

@ARTICLE{shen+03,
       author = {{Shen}, Shiyin and {Mo}, H.~J. and {White}, Simon D.~M. and {Blanton}, Michael R. and {Kauffmann}, Guinevere and {Voges}, Wolfgang and {Brinkmann}, J. and {Csabai}, Istvan},
        title = "{The size distribution of galaxies in the Sloan Digital Sky Survey}",
      journal = {\mnras},
         year = 2003,
        month = aug,
       volume = {343},
       number = {3},
        pages = {978-994},
          doi = {10.1046/j.1365-8711.2003.06740.x},
archivePrefix = {arXiv},
       eprint = {astro-ph/0301527},
 primaryClass = {astro-ph},
       adsurl = {https://ui.adsabs.harvard.edu/abs/2003MNRAS.343..978S}
}

@ARTICLE{blanton+03,
       author = {{Blanton}, Michael R. and {Hogg}, David W. and {Bahcall}, Neta A. and {Baldry}, Ivan K. and {Brinkmann}, J. and {Csabai}, Istv{\'a}n and {Eisenstein}, Daniel and {Fukugita}, Masataka and {Gunn}, James E. and {Ivezi{\'c}}, {\v{Z}}eljko and {Lamb}, D.~Q. and {Lupton}, Robert H. and {Loveday}, Jon and {Munn}, Jeffrey A. and {Nichol}, R.~C. and {Okamura}, Sadanori and {Schlegel}, David J. and {Shimasaku}, Kazuhiro and {Strauss}, Michael A. and {Vogeley}, Michael S. and {Weinberg}, David H.},
        title = "{The Broadband Optical Properties of Galaxies with Redshifts 0.02<z<0.22}",
      journal = {\apj},
         year = 2003,
        month = sep,
       volume = {594},
       number = {1},
        pages = {186-207},
          doi = {10.1086/375528},
archivePrefix = {arXiv},
       eprint = {astro-ph/0209479},
 primaryClass = {astro-ph},
       adsurl = {https://ui.adsabs.harvard.edu/abs/2003ApJ...594..186B}
}

@ARTICLE{blanton+05,
       author = {{Blanton}, Michael R. and {Eisenstein}, Daniel and {Hogg}, David W. and {Schlegel}, David J. and {Brinkmann}, J.},
        title = "{Relationship between Environment and the Broadband Optical Properties of Galaxies in the Sloan Digital Sky Survey}",
      journal = {\apj},
         year = 2005,
        month = aug,
       volume = {629},
       number = {1},
        pages = {143-157},
          doi = {10.1086/422897},
archivePrefix = {arXiv},
       eprint = {astro-ph/0310453},
 primaryClass = {astro-ph},
       adsurl = {https://ui.adsabs.harvard.edu/abs/2005ApJ...629..143B}
}

@ARTICLE{bell+04a,
       author = {{Bell}, Eric F. and {McIntosh}, Daniel H. and {Barden}, Marco and {Wolf}, Christian and {Caldwell}, John A.~R. and {Rix}, Hans-Walter and {Beckwith}, Steven V.~W. and {Borch}, Andrea and {H{\"a}ussler}, Boris and {Jahnke}, Knud and {Jogee}, Shardha and {Meisenheimer}, Klaus and {Peng}, Chien and {Sanchez}, Sebastian F. and {Somerville}, Rachel S. and {Wisotzki}, Lutz},
        title = "{GEMS Imaging of Red-Sequence Galaxies at z\raisebox{-0.5ex}\textasciitilde0.7: Dusty or Old?}",
      journal = {\apjl},
         year = 2004,
        month = jan,
       volume = {600},
       number = {1},
        pages = {L11-L14},
          doi = {10.1086/381388},
archivePrefix = {arXiv},
       eprint = {astro-ph/0308272},
 primaryClass = {astro-ph},
       adsurl = {https://ui.adsabs.harvard.edu/abs/2004ApJ...600L..11B}
}

@ARTICLE{bell+04b,
       author = {{Bell}, Eric F. and {Wolf}, Christian and {Meisenheimer}, Klaus and {Rix}, Hans-Walter and {Borch}, Andrea and {Dye}, Simon and {Kleinheinrich}, Martina and {Wisotzki}, Lutz and {McIntosh}, Daniel H.},
        title = "{Nearly 5000 Distant Early-Type Galaxies in COMBO-17: A Red Sequence and Its Evolution since z\raisebox{-0.5ex}\textasciitilde1}",
      journal = {\apj},
         year = 2004,
        month = jun,
       volume = {608},
       number = {2},
        pages = {752-767},
          doi = {10.1086/420778},
archivePrefix = {arXiv},
       eprint = {astro-ph/0303394},
 primaryClass = {astro-ph},
       adsurl = {https://ui.adsabs.harvard.edu/abs/2004ApJ...608..752B}
}

@ARTICLE{ellis+05,
       author = {{Ellis}, S.~C. and {Driver}, S.~P. and {Allen}, P.~D. and {Liske}, J. and {Bland-Hawthorn}, J. and {De Propris}, R.},
        title = "{The Millennium Galaxy Catalogue: on the natural subdivision of galaxies}",
      journal = {\mnras},
         year = 2005,
        month = nov,
       volume = {363},
       number = {4},
        pages = {1257-1271},
          doi = {10.1111/j.1365-2966.2005.09521.x},
archivePrefix = {arXiv},
       eprint = {astro-ph/0508365},
 primaryClass = {astro-ph},
       adsurl = {https://ui.adsabs.harvard.edu/abs/2005MNRAS.363.1257E}
}

@ARTICLE{driver+06,
       author = {{Driver}, S.~P. and {Allen}, P.~D. and {Graham}, Alister. W. and {Cameron}, E. and {Liske}, J. and {Ellis}, S.~C. and {Cross}, N.~J.~G. and {De Propris}, R. and {Phillipps}, S. and {Couch}, W.~J.},
        title = "{The Millennium Galaxy Catalogue: morphological classification and bimodality in the colour-concentration plane}",
      journal = {\mnras},
         year = 2006,
        month = may,
       volume = {368},
       number = {1},
        pages = {414-434},
          doi = {10.1111/j.1365-2966.2006.10126.x},
archivePrefix = {arXiv},
       eprint = {astro-ph/0602240},
 primaryClass = {astro-ph},
       adsurl = {https://ui.adsabs.harvard.edu/abs/2006MNRAS.368..414D}
}

@ARTICLE{papovich+12,
       author = {{Papovich}, C. and {Bassett}, R. and {Lotz}, J.~M. and {van der Wel}, A. and {Tran}, K. -V. and {Finkelstein}, S.~L. and {Bell}, E.~F. and {Conselice}, C.~J. and {Dekel}, A. and {Dunlop}, J.~S. and {Guo}, Yicheng and {Faber}, S.~M. and {Farrah}, D. and {Ferguson}, H.~C. and {Finkelstein}, K.~D. and {H{\"a}ussler}, B. and {Kocevski}, D.~D. and {Koekemoer}, A.~M. and {Koo}, D.~C. and {McGrath}, E.~J. and {McLure}, R.~J. and {McIntosh}, D.~H. and {Momcheva}, I. and {Newman}, J.~A. and {Rudnick}, G. and {Weiner}, B. and {Willmer}, C.~N.~A. and {Wuyts}, S.},
        title = "{CANDELS Observations of the Structural Properties of Cluster Galaxies at z = 1.62}",
      journal = {\apj},
         year = 2012,
        month = may,
       volume = {750},
       number = {2},
          eid = {93},
        pages = {93},
          doi = {10.1088/0004-637X/750/2/93},
archivePrefix = {arXiv},
       eprint = {1110.3794},
 primaryClass = {astro-ph.CO},
       adsurl = {https://ui.adsabs.harvard.edu/abs/2012ApJ...750...93P}
}

@ARTICLE{dressler+80,
       author = {{Dressler}, A.},
        title = "{Galaxy morphology in rich clusters: implications for the formation and evolution of galaxies.}",
      journal = {\apj},
         year = 1980,
        month = mar,
       volume = {236},
        pages = {351-365},
          doi = {10.1086/157753},
       adsurl = {https://ui.adsabs.harvard.edu/abs/1980ApJ...236..351D}
}

@ARTICLE{baldry+06,
       author = {{Baldry}, I.~K. and {Balogh}, M.~L. and {Bower}, R.~G. and {Glazebrook}, K. and {Nichol}, R.~C. and {Bamford}, S.~P. and {Budavari}, T.},
        title = "{Galaxy bimodality versus stellar mass and environment}",
      journal = {\mnras},
         year = 2006,
        month = dec,
       volume = {373},
       number = {2},
        pages = {469-483},
          doi = {10.1111/j.1365-2966.2006.11081.x},
archivePrefix = {arXiv},
       eprint = {astro-ph/0607648},
 primaryClass = {astro-ph},
       adsurl = {https://ui.adsabs.harvard.edu/abs/2006MNRAS.373..469B}
}

@ARTICLE{vanderwel+08,
       author = {{van der Wel}, Arjen},
        title = "{The Dependence of Galaxy Morphology and Structure on Environment and Stellar Mass}",
      journal = {\apjl},
         year = 2008,
        month = mar,
       volume = {675},
       number = {1},
        pages = {L13},
          doi = {10.1086/529432},
archivePrefix = {arXiv},
       eprint = {0801.1995},
 primaryClass = {astro-ph},
       adsurl = {https://ui.adsabs.harvard.edu/abs/2008ApJ...675L..13V}
}

@ARTICLE{kauffmann+04,
       author = {{Kauffmann}, Guinevere and {White}, Simon D.~M. and {Heckman}, Timothy M. and {M{\'e}nard}, Brice and {Brinchmann}, Jarle and {Charlot}, St{\'e}phane and {Tremonti}, Christy and {Brinkmann}, Jon},
        title = "{The environmental dependence of the relations between stellar mass, structure, star formation and nuclear activity in galaxies}",
      journal = {\mnras},
         year = 2004,
        month = sep,
       volume = {353},
       number = {3},
        pages = {713-731},
          doi = {10.1111/j.1365-2966.2004.08117.x},
archivePrefix = {arXiv},
       eprint = {astro-ph/0402030},
 primaryClass = {astro-ph},
       adsurl = {https://ui.adsabs.harvard.edu/abs/2004MNRAS.353..713K}
}

@ARTICLE{balogh+04a,
       author = {{Balogh}, Michael and {Eke}, Vince and {Miller}, Chris and {Lewis}, Ian and {Bower}, Richard and {Couch}, Warrick and {Nichol}, Robert and {Bland-Hawthorn}, Joss and {Baldry}, Ivan K. and {Baugh}, Carlton and {Bridges}, Terry and {Cannon}, Russell and {Cole}, Shaun and {Colless}, Matthew and {Collins}, Chris and {Cross}, Nicholas and {Dalton}, Gavin and {de Propris}, Roberto and {Driver}, Simon P. and {Efstathiou}, George and {Ellis}, Richard S. and {Frenk}, Carlos S. and {Glazebrook}, Karl and {Gomez}, Percy and {Gray}, Alex and {Hawkins}, Edward and {Jackson}, Carole and {Lahav}, Ofer and {Lumsden}, Stuart and {Maddox}, Steve and {Madgwick}, Darren and {Norberg}, Peder and {Peacock}, John A. and {Percival}, Will and {Peterson}, Bruce A. and {Sutherland}, Will and {Taylor}, Keith},
        title = "{Galaxy ecology: groups and low-density environments in the SDSS and 2dFGRS}",
      journal = {\mnras},
         year = 2004,
        month = mar,
       volume = {348},
       number = {4},
        pages = {1355-1372},
          doi = {10.1111/j.1365-2966.2004.07453.x},
archivePrefix = {arXiv},
       eprint = {astro-ph/0311379},
 primaryClass = {astro-ph},
       adsurl = {https://ui.adsabs.harvard.edu/abs/2004MNRAS.348.1355B}
}

@ARTICLE{moustakas+13,
       author = {{Moustakas}, John and {Coil}, Alison L. and {Aird}, James and {Blanton}, Michael R. and {Cool}, Richard J. and {Eisenstein}, Daniel J. and {Mendez}, Alexander J. and {Wong}, Kenneth C. and {Zhu}, Guangtun and {Arnouts}, St{\'e}phane},
        title = "{PRIMUS: Constraints on Star Formation Quenching and Galaxy Merging, and the Evolution of the Stellar Mass Function from z = 0-1}",
      journal = {\apj},
         year = 2013,
        month = apr,
       volume = {767},
       number = {1},
          eid = {50},
        pages = {50},
          doi = {10.1088/0004-637X/767/1/50},
archivePrefix = {arXiv},
       eprint = {1301.1688},
 primaryClass = {astro-ph.CO},
       adsurl = {https://ui.adsabs.harvard.edu/abs/2013ApJ...767...50M}
}

@ARTICLE{arnouts+07,
       author = {{Arnouts}, S. and {Walcher}, C.~J. and {Le F{\`e}vre}, O. and {Zamorani}, G. and {Ilbert}, O. and {Le Brun}, V. and {Pozzetti}, L. and {Bardelli}, S. and {Tresse}, L. and {Zucca}, E. and {Charlot}, S. and {Lamareille}, F. and {McCracken}, H.~J. and {Bolzonella}, M. and {Iovino}, A. and {Lonsdale}, C. and {Polletta}, M. and {Surace}, J. and {Bottini}, D. and {Garilli}, B. and {Maccagni}, D. and {Picat}, J.~P. and {Scaramella}, R. and {Scodeggio}, M. and {Vettolani}, G. and {Zanichelli}, A. and {Adami}, C. and {Cappi}, A. and {Ciliegi}, P. and {Contini}, T. and {de la Torre}, S. and {Foucaud}, S. and {Franzetti}, P. and {Gavignaud}, I. and {Guzzo}, L. and {Marano}, B. and {Marinoni}, C. and {Mazure}, A. and {Meneux}, B. and {Merighi}, R. and {Paltani}, S. and {Pell{\`o}}, R. and {Pollo}, A. and {Radovich}, M. and {Temporin}, S. and {Vergani}, D.},
        title = "{The SWIRE-VVDS-CFHTLS surveys: stellar mass assembly over the last 10 Gyr. Evidence for a major build up of the red sequence between z = 2 and z = 1}",
      journal = {\aap},
         year = 2007,
        month = dec,
       volume = {476},
       number = {1},
        pages = {137-150},
          doi = {10.1051/0004-6361:20077632},
archivePrefix = {arXiv},
       eprint = {0705.2438},
 primaryClass = {astro-ph},
       adsurl = {https://ui.adsabs.harvard.edu/abs/2007A&A...476..137A}
}

@ARTICLE{foltz+18,
       author = {{Foltz}, R. and {Wilson}, G. and {Muzzin}, A. and {Cooper}, M.~C. and {Nantais}, J. and {van der Burg}, R.~F.~J. and {Cerulo}, P. and {Chan}, J. and {Fillingham}, S.~P. and {Surace}, J. and {Webb}, T. and {Noble}, A. and {Lacy}, M. and {McDonald}, M. and {Rudnick}, G. and {Lidman}, C. and {Demarco}, R. and {Hlavacek-Larrondo}, J. and {Yee}, H.~K.~C. and {Perlmutter}, S. and {Hayden}, B.},
        title = "{The Evolution of Environmental Quenching Timescales to z {\ensuremath{\sim}} 1.6: Evidence for Dynamically Driven Quenching of the Cluster Galaxy Population}",
      journal = {\apj},
         year = 2018,
        month = oct,
       volume = {866},
       number = {2},
          eid = {136},
        pages = {136},
          doi = {10.3847/1538-4357/aad80d},
archivePrefix = {arXiv},
       eprint = {1803.03305},
 primaryClass = {astro-ph.GA},
       adsurl = {https://ui.adsabs.harvard.edu/abs/2018ApJ...866..136F}
}

@ARTICLE{balogh+04b,
       author = {{Balogh}, Michael L. and {Baldry}, Ivan K. and {Nichol}, Robert and {Miller}, Chris and {Bower}, Richard and {Glazebrook}, Karl},
        title = "{The Bimodal Galaxy Color Distribution: Dependence on Luminosity and Environment}",
      journal = {\apjl},
         year = 2004,
        month = nov,
       volume = {615},
       number = {2},
        pages = {L101-L104},
          doi = {10.1086/426079},
archivePrefix = {arXiv},
       eprint = {astro-ph/0406266},
 primaryClass = {astro-ph},
       adsurl = {https://ui.adsabs.harvard.edu/abs/2004ApJ...615L.101B}
}

@ARTICLE{vanderburg+18,
       author = {{van der Burg}, Remco F.~J. and {McGee}, Sean and {Aussel}, Herv{\'e} and {Dahle}, H{\r{a}}kon and {Arnaud}, Monique and {Pratt}, Gabriel W. and {Muzzin}, Adam},
        title = "{The stellar mass function of galaxies in Planck-selected clusters at 0.5 < z < 0.7: new constraints on the timescale and location of satellite quenching}",
      journal = {\aap},
         year = 2018,
        month = oct,
       volume = {618},
          eid = {A140},
        pages = {A140},
          doi = {10.1051/0004-6361/201833572},
archivePrefix = {arXiv},
       eprint = {1807.00820},
 primaryClass = {astro-ph.GA},
       adsurl = {https://ui.adsabs.harvard.edu/abs/2018A&A...618A.140V}
}

@ARTICLE{reeves+21,
       author = {{Reeves}, Andrew M.~M. and {Balogh}, Michael L. and {van der Burg}, Remco F.~J. and {Finoguenov}, Alexis and {Kukstas}, Egidijus and {McCarthy}, Ian G. and {Webb}, Kristi and {Muzzin}, Adam and {McGee}, Sean and {Rudnick}, Gregory and {Biviano}, Andrea and {Cerulo}, Pierluigi and {Chan}, Jeffrey C.~C. and {Cooper}, M.~C. and {Demarco}, Ricardo and {Jablonka}, Pascale and {De Lucia}, Gabriella and {Vulcani}, Benedetta and {Wilson}, Gillian and {Yee}, Howard K.~C. and {Zaritsky}, Dennis},
        title = "{The GOGREEN survey: dependence of galaxy properties on halo mass at z > 1 and implications for environmental quenching}",
      journal = {\mnras},
         year = 2021,
        month = sep,
       volume = {506},
       number = {3},
        pages = {3364-3384},
          doi = {10.1093/mnras/stab1955},
archivePrefix = {arXiv},
       eprint = {2107.03425},
 primaryClass = {astro-ph.GA},
       adsurl = {https://ui.adsabs.harvard.edu/abs/2021MNRAS.506.3364R}
}

@ARTICLE{woo+13,
       author = {{Woo}, Joanna and {Dekel}, Avishai and {Faber}, S.~M. and {Noeske}, Kai and {Koo}, David C. and {Gerke}, Brian F. and {Cooper}, Michael C. and {Salim}, Samir and {Dutton}, Aaron A. and {Newman}, Jeffrey and {Weiner}, Benjamin J. and {Bundy}, Kevin and {Willmer}, Christopher N.~A. and {Davis}, Marc and {Yan}, Renbin},
        title = "{Dependence of galaxy quenching on halo mass and distance from its centre}",
      journal = {\mnras},
         year = 2013,
        month = feb,
       volume = {428},
       number = {4},
        pages = {3306-3326},
          doi = {10.1093/mnras/sts274},
archivePrefix = {arXiv},
       eprint = {1203.1625},
 primaryClass = {astro-ph.CO},
       adsurl = {https://ui.adsabs.harvard.edu/abs/2013MNRAS.428.3306W}
}

@ARTICLE{schaefer+17,
       author = {{Schaefer}, A.~L. and {Croom}, S.~M. and {Allen}, J.~T. and {Brough}, S. and {Medling}, A.~M. and {Ho}, I. -T. and {Scott}, N. and {Richards}, S.~N. and {Pracy}, M.~B. and {Gunawardhana}, M.~L.~P. and {Norberg}, P. and {Alpaslan}, M. and {Bauer}, A.~E. and {Bekki}, K. and {Bland-Hawthorn}, J. and {Bloom}, J.~V. and {Bryant}, J.~J. and {Couch}, W.~J. and {Driver}, S.~P. and {Fogarty}, L.~M.~R. and {Foster}, C. and {Goldstein}, G. and {Green}, A.~W. and {Hopkins}, A.~M. and {Konstantopoulos}, I.~S. and {Lawrence}, J.~S. and {L{\'o}pez-S{\'a}nchez}, A.~R. and {Lorente}, N.~P.~F. and {Owers}, M.~S. and {Sharp}, R. and {Sweet}, S.~M. and {Taylor}, E.~N. and {van de Sande}, J. and {Walcher}, C.~J. and {Wong}, O.~I.},
        title = "{The SAMI Galaxy Survey: spatially resolving the environmental quenching of star formation in GAMA galaxies}",
      journal = {\mnras},
         year = 2017,
        month = jan,
       volume = {464},
       number = {1},
        pages = {121-142},
          doi = {10.1093/mnras/stw2289},
archivePrefix = {arXiv},
       eprint = {1609.02635},
 primaryClass = {astro-ph.GA},
       adsurl = {https://ui.adsabs.harvard.edu/abs/2017MNRAS.464..121S}
}

@ARTICLE{schaefer+19,
       author = {{Schaefer}, A.~L. and {Croom}, S.~M. and {Scott}, N. and {Brough}, S. and {Allen}, J.~T. and {Bekki}, K. and {Bland-Hawthorn}, J. and {Bloom}, J.~V. and {Bryant}, J.~J. and {Cortese}, L. and {Davies}, L.~J.~M. and {Federrath}, C. and {Fogarty}, L.~M.~R. and {Green}, A.~W. and {Groves}, B. and {Hopkins}, A.~M. and {Konstantopoulos}, I.~S. and {L{\'o}pez-S{\'a}nchez}, A.~R. and {Lawrence}, J.~S. and {McElroy}, R.~E. and {Medling}, A.~M. and {Owers}, M.~S. and {Pracy}, M.~B. and {Richards}, S.~N. and {Robotham}, A.~S.~G. and {van de Sande}, J. and {Tonini}, C. and {Yi}, S.~K.},
        title = "{The SAMI Galaxy Survey: observing the environmental quenching of star formation in GAMA groups}",
      journal = {\mnras},
         year = 2019,
        month = mar,
       volume = {483},
       number = {3},
        pages = {2851-2870},
          doi = {10.1093/mnras/sty3258},
archivePrefix = {arXiv},
       eprint = {1811.11676},
 primaryClass = {astro-ph.GA},
       adsurl = {https://ui.adsabs.harvard.edu/abs/2019MNRAS.483.2851S}
}

@ARTICLE{barsanti+18,
       author = {{Barsanti}, S. and {Owers}, M.~S. and {Brough}, S. and {Davies}, L.~J.~M. and {Driver}, S.~P. and {Gunawardhana}, M.~L.~P. and {Holwerda}, B.~W. and {Liske}, J. and {Loveday}, J. and {Pimbblet}, K.~A. and {Robotham}, A.~S.~G. and {Taylor}, E.~N.},
        title = "{Galaxy and Mass Assembly (GAMA): Impact of the Group Environment on Galaxy Star Formation}",
      journal = {\apj},
         year = 2018,
        month = apr,
       volume = {857},
       number = {1},
          eid = {71},
        pages = {71},
          doi = {10.3847/1538-4357/aab61a},
archivePrefix = {arXiv},
       eprint = {1803.05076},
 primaryClass = {astro-ph.GA},
       adsurl = {https://ui.adsabs.harvard.edu/abs/2018ApJ...857...71B}
}

@ARTICLE{vandesande+21,
       author = {{van de Sande}, Jesse and {Croom}, Scott M. and {Bland-Hawthorn}, Joss and {Cortese}, Luca and {Scott}, Nicholas and {Lagos}, Claudia D.~P. and {D'Eugenio}, Francesco and {Bryant}, Julia J. and {Brough}, Sarah and {Catinella}, Barbara and {Foster}, Caroline and {Groves}, Brent and {Harborne}, Katherine E. and {L{\'o}pez-S{\'a}nchez}, {\'A}ngel R. and {McDermid}, Richard and {Medling}, Anne and {Owers}, Matt S. and {Richards}, Samuel N. and {Sweet}, Sarah M. and {Vaughan}, Sam P.},
        title = "{The SAMI galaxy survey: Mass and environment as independent drivers of galaxy dynamics}",
      journal = {\mnras},
         year = 2021,
        month = dec,
       volume = {508},
       number = {2},
        pages = {2307-2328},
          doi = {10.1093/mnras/stab2647},
archivePrefix = {arXiv},
       eprint = {2109.06189},
 primaryClass = {astro-ph.GA},
       adsurl = {https://ui.adsabs.harvard.edu/abs/2021MNRAS.508.2307V}
}

@ARTICLE{kovac+14,
       author = {{Kova{\v{c}}}, K. and {Lilly}, S.~J. and {Knobel}, C. and {Bschorr}, T.~J. and {Peng}, Y. and {Carollo}, C.~M. and {Contini}, T. and {Kneib}, J. -P. and {Le F{\'e}vre}, O. and {Mainieri}, V. and {Renzini}, A. and {Scodeggio}, M. and {Zamorani}, G. and {Bardelli}, S. and {Bolzonella}, M. and {Bongiorno}, A. and {Caputi}, K. and {Cucciati}, O. and {de la Torre}, S. and {de Ravel}, L. and {Franzetti}, P. and {Garilli}, B. and {Iovino}, A. and {Kampczyk}, P. and {Lamareille}, F. and {Le Borgne}, J. -F. and {Le Brun}, V. and {Maier}, C. and {Mignoli}, M. and {Oesch}, P. and {Pello}, R. and {Montero}, E. Perez and {Presotto}, V. and {Silverman}, J. and {Tanaka}, M. and {Tasca}, L. and {Tresse}, L. and {Vergani}, D. and {Zucca}, E. and {Aussel}, H. and {Koekemoer}, A.~M. and {Le Floc'h}, E. and {Moresco}, M. and {Pozzetti}, L.},
        title = "{zCOSMOS 20k: satellite galaxies are the main drivers of environmental effects in the galaxy population at least to z {\ensuremath{\sim}} 0.7}",
      journal = {\mnras},
         year = 2014,
        month = feb,
       volume = {438},
       number = {1},
        pages = {717-738},
          doi = {10.1093/mnras/stt2241},
archivePrefix = {arXiv},
       eprint = {1307.4402},
 primaryClass = {astro-ph.CO},
       adsurl = {https://ui.adsabs.harvard.edu/abs/2014MNRAS.438..717K}
}

@ARTICLE{kawinwanichakij+17,
       author = {{Kawinwanichakij}, Lalitwadee and {Papovich}, Casey and {Quadri}, Ryan F. and {Glazebrook}, Karl and {Kacprzak}, Glenn G. and {Allen}, Rebecca J. and {Bell}, Eric F. and {Croton}, Darren J. and {Dekel}, Avishai and {Ferguson}, Henry C. and {Forrest}, Ben and {Grogin}, Norman A. and {Guo}, Yicheng and {Kocevski}, Dale D. and {Koekemoer}, Anton M. and {Labb{\'e}}, Ivo and {Lucas}, Ray A. and {Nanayakkara}, Themiya and {Spitler}, Lee R. and {Straatman}, Caroline M.~S. and {Tran}, Kim-Vy H. and {Tomczak}, Adam and {van Dokkum}, Pieter},
        title = "{Effect of Local Environment and Stellar Mass on Galaxy Quenching and Morphology at 0.5 < z < 2.0}",
      journal = {\apj},
         year = 2017,
        month = oct,
       volume = {847},
       number = {2},
          eid = {134},
        pages = {134},
          doi = {10.3847/1538-4357/aa8b75},
archivePrefix = {arXiv},
       eprint = {1706.03780},
 primaryClass = {astro-ph.GA},
       adsurl = {https://ui.adsabs.harvard.edu/abs/2017ApJ...847..134K}
}

@ARTICLE{oppenheimer+10,
       author = {{Oppenheimer}, Benjamin D. and {Dav{\'e}}, Romeel and {Kere{\v{s}}}, Du{\v{s}}an and {Fardal}, Mark and {Katz}, Neal and {Kollmeier}, Juna A. and {Weinberg}, David H.},
        title = "{Feedback and recycled wind accretion: assembling the z = 0 galaxy mass function}",
      journal = {\mnras},
         year = 2010,
        month = aug,
       volume = {406},
       number = {4},
        pages = {2325-2338},
          doi = {10.1111/j.1365-2966.2010.16872.x},
archivePrefix = {arXiv},
       eprint = {0912.0519},
 primaryClass = {astro-ph.CO},
       adsurl = {https://ui.adsabs.harvard.edu/abs/2010MNRAS.406.2325O}
}

@ARTICLE{bower+06,
       author = {{Bower}, R.~G. and {Benson}, A.~J. and {Malbon}, R. and {Helly}, J.~C. and {Frenk}, C.~S. and {Baugh}, C.~M. and {Cole}, S. and {Lacey}, C.~G.},
        title = "{Breaking the hierarchy of galaxy formation}",
      journal = {\mnras},
         year = 2006,
        month = aug,
       volume = {370},
       number = {2},
        pages = {645-655},
          doi = {10.1111/j.1365-2966.2006.10519.x},
archivePrefix = {arXiv},
       eprint = {astro-ph/0511338},
 primaryClass = {astro-ph},
       adsurl = {https://ui.adsabs.harvard.edu/abs/2006MNRAS.370..645B}
}

@ARTICLE{croton+06,
       author = {{Croton}, Darren J. and {Springel}, Volker and {White}, Simon D.~M. and {De Lucia}, G. and {Frenk}, C.~S. and {Gao}, L. and {Jenkins}, A. and {Kauffmann}, G. and {Navarro}, J.~F. and {Yoshida}, N.},
        title = "{The many lives of active galactic nuclei: cooling flows, black holes and the luminosities and colours of galaxies}",
      journal = {\mnras},
         year = 2006,
        month = jan,
       volume = {365},
       number = {1},
        pages = {11-28},
          doi = {10.1111/j.1365-2966.2005.09675.x},
archivePrefix = {arXiv},
       eprint = {astro-ph/0508046},
 primaryClass = {astro-ph},
       adsurl = {https://ui.adsabs.harvard.edu/abs/2006MNRAS.365...11C}
}

@ARTICLE{tremonti+07,
       author = {{Tremonti}, Christy A. and {Moustakas}, John and {Diamond-Stanic}, Aleksandar M.},
        title = "{The Discovery of 1000 km s$^{-1}$ Outflows in Massive Poststarburst Galaxies at z=0.6}",
      journal = {\apjl},
         year = 2007,
        month = jul,
       volume = {663},
       number = {2},
        pages = {L77-L80},
          doi = {10.1086/520083},
archivePrefix = {arXiv},
       eprint = {0706.0527},
 primaryClass = {astro-ph},
       adsurl = {https://ui.adsabs.harvard.edu/abs/2007ApJ...663L..77T}
}

@ARTICLE{dekel+06,
       author = {{Dekel}, Avishai and {Birnboim}, Yuval},
        title = "{Galaxy bimodality due to cold flows and shock heating}",
      journal = {\mnras},
         year = 2006,
        month = may,
       volume = {368},
       number = {1},
        pages = {2-20},
          doi = {10.1111/j.1365-2966.2006.10145.x},
archivePrefix = {arXiv},
       eprint = {astro-ph/0412300},
 primaryClass = {astro-ph},
       adsurl = {https://ui.adsabs.harvard.edu/abs/2006MNRAS.368....2D}
}

@ARTICLE{cattaneo+08,
       author = {{Cattaneo}, A. and {Dekel}, A. and {Faber}, S.~M. and {Guiderdoni}, B.},
        title = "{Downsizing by shutdown in red galaxies}",
      journal = {\mnras},
         year = 2008,
        month = sep,
       volume = {389},
       number = {2},
        pages = {567-584},
          doi = {10.1111/j.1365-2966.2008.13562.x},
archivePrefix = {arXiv},
       eprint = {0801.1673},
 primaryClass = {astro-ph},
       adsurl = {https://ui.adsabs.harvard.edu/abs/2008MNRAS.389..567C}
}

@ARTICLE{vandenbosch+08,
       author = {{van den Bosch}, Frank C. and {Aquino}, Daniel and {Yang}, Xiaohu and {Mo}, H.~J. and {Pasquali}, Anna and {McIntosh}, Daniel H. and {Weinmann}, Simone M. and {Kang}, Xi},
        title = "{The importance of satellite quenching for the build-up of the red sequence of present-day galaxies}",
      journal = {\mnras},
         year = 2008,
        month = jun,
       volume = {387},
       number = {1},
        pages = {79-91},
          doi = {10.1111/j.1365-2966.2008.13230.x},
archivePrefix = {arXiv},
       eprint = {0710.3164},
 primaryClass = {astro-ph},
       adsurl = {https://ui.adsabs.harvard.edu/abs/2008MNRAS.387...79V}
}

@ARTICLE{menci+06,
       author = {{Menci}, N. and {Fontana}, A. and {Giallongo}, E. and {Grazian}, A. and {Salimbeni}, S.},
        title = "{The Abundance of Distant and Extremely Red Galaxies: The Role of AGN Feedback in Hierarchical Models}",
      journal = {\apj},
         year = 2006,
        month = aug,
       volume = {647},
       number = {2},
        pages = {753-762},
          doi = {10.1086/505528},
archivePrefix = {arXiv},
       eprint = {astro-ph/0605123},
 primaryClass = {astro-ph},
       adsurl = {https://ui.adsabs.harvard.edu/abs/2006ApJ...647..753M}
}

@ARTICLE{bower+08,
       author = {{Bower}, R.~G. and {McCarthy}, I.~G. and {Benson}, A.~J.},
        title = "{The flip side of galaxy formation: a combined model of galaxy formation and cluster heating}",
      journal = {\mnras},
         year = 2008,
        month = nov,
       volume = {390},
       number = {4},
        pages = {1399-1410},
          doi = {10.1111/j.1365-2966.2008.13869.x},
archivePrefix = {arXiv},
       eprint = {0808.2994},
 primaryClass = {astro-ph},
       adsurl = {https://ui.adsabs.harvard.edu/abs/2008MNRAS.390.1399B}
}

@ARTICLE{somerville+08,
       author = {{Somerville}, Rachel S. and {Hopkins}, Philip F. and {Cox}, Thomas J. and {Robertson}, Brant E. and {Hernquist}, Lars},
        title = "{A semi-analytic model for the co-evolution of galaxies, black holes and active galactic nuclei}",
      journal = {\mnras},
         year = 2008,
        month = dec,
       volume = {391},
       number = {2},
        pages = {481-506},
          doi = {10.1111/j.1365-2966.2008.13805.x},
archivePrefix = {arXiv},
       eprint = {0808.1227},
 primaryClass = {astro-ph},
       adsurl = {https://ui.adsabs.harvard.edu/abs/2008MNRAS.391..481S}
}

@ARTICLE{benson+03,
       author = {{Benson}, A.~J. and {Bower}, R.~G. and {Frenk}, C.~S. and {Lacey}, C.~G. and {Baugh}, C.~M. and {Cole}, S.},
        title = "{What Shapes the Luminosity Function of Galaxies?}",
      journal = {\apj},
         year = 2003,
        month = dec,
       volume = {599},
       number = {1},
        pages = {38-49},
          doi = {10.1086/379160},
archivePrefix = {arXiv},
       eprint = {astro-ph/0302450},
 primaryClass = {astro-ph},
       adsurl = {https://ui.adsabs.harvard.edu/abs/2003ApJ...599...38B}
}

@ARTICLE{gunn+72,
       author = {{Gunn}, James E. and {Gott}, III, J. Richard},
        title = "{On the Infall of Matter Into Clusters of Galaxies and Some Effects on Their Evolution}",
      journal = {\apj},
         year = 1972,
        month = aug,
       volume = {176},
        pages = {1},
          doi = {10.1086/151605},
       adsurl = {https://ui.adsabs.harvard.edu/abs/1972ApJ...176....1G}
}

@ARTICLE{larson+80,
       author = {{Larson}, R.~B. and {Tinsley}, B.~M. and {Caldwell}, C.~N.},
        title = "{The evolution of disk galaxies and the origin of S0 galaxies}",
      journal = {\apj},
         year = 1980,
        month = may,
       volume = {237},
        pages = {692-707},
          doi = {10.1086/157917},
       adsurl = {https://ui.adsabs.harvard.edu/abs/1980ApJ...237..692L}
}

@ARTICLE{moore+96,
       author = {{Moore}, Ben and {Katz}, Neal and {Lake}, George and {Dressler}, Alan and {Oemler}, Augustus},
        title = "{Galaxy harassment and the evolution of clusters of galaxies}",
      journal = {\nat},
         year = 1996,
        month = feb,
       volume = {379},
       number = {6566},
        pages = {613-616},
          doi = {10.1038/379613a0},
archivePrefix = {arXiv},
       eprint = {astro-ph/9510034},
 primaryClass = {astro-ph},
       adsurl = {https://ui.adsabs.harvard.edu/abs/1996Natur.379..613M}
}

@ARTICLE{bialas+15,
       author = {{Bialas}, D. and {Lisker}, T. and {Olczak}, C. and {Spurzem}, R. and {Kotulla}, R.},
        title = "{On the occurrence of galaxy harassment}",
      journal = {\aap},
         year = 2015,
        month = apr,
       volume = {576},
          eid = {A103},
        pages = {A103},
          doi = {10.1051/0004-6361/201425235},
archivePrefix = {arXiv},
       eprint = {1503.01965},
 primaryClass = {astro-ph.GA},
       adsurl = {https://ui.adsabs.harvard.edu/abs/2015A&A...576A.103B}
}

@ARTICLE{peng+15,
       author = {{Peng}, Y. and {Maiolino}, R. and {Cochrane}, R.},
        title = "{Strangulation as the primary mechanism for shutting down star formation in galaxies}",
      journal = {\nat},
         year = 2015,
        month = may,
       volume = {521},
       number = {7551},
        pages = {192-195},
          doi = {10.1038/nature14439},
archivePrefix = {arXiv},
       eprint = {1505.03143},
 primaryClass = {astro-ph.GA},
       adsurl = {https://ui.adsabs.harvard.edu/abs/2015Natur.521..192P}
}

@ARTICLE{brough+13,
       author = {{Brough}, S. and {Croom}, S. and {Sharp}, R. and {Hopkins}, A.~M. and {Taylor}, E.~N. and {Baldry}, I.~K. and {Gunawardhana}, M.~L.~P. and {Liske}, J. and {Norberg}, P. and {Robotham}, A.~S.~G. and {Bauer}, A.~E. and {Bland-Hawthorn}, J. and {Colless}, M. and {Foster}, C. and {Kelvin}, L.~S. and {Lara-Lopez}, M.~A. and {L{\'o}pez-S{\'a}nchez}, {\'A}. R. and {Loveday}, J. and {Owers}, M. and {Pimbblet}, K.~A. and {Prescott}, M.},
        title = "{Galaxy And Mass Assembly: resolving the role of environment in galaxy evolution}",
      journal = {\mnras},
         year = 2013,
        month = nov,
       volume = {435},
       number = {4},
        pages = {2903-2917},
          doi = {10.1093/mnras/stt1489},
archivePrefix = {arXiv},
       eprint = {1308.2985},
 primaryClass = {astro-ph.CO},
       adsurl = {https://ui.adsabs.harvard.edu/abs/2013MNRAS.435.2903B}
}

@ARTICLE{trussler+20,
       author = {{Trussler}, James and {Maiolino}, Roberto and {Maraston}, Claudia and {Peng}, Yingjie and {Thomas}, Daniel and {Goddard}, Daniel and {Lian}, Jianhui},
        title = "{Both starvation and outflows drive galaxy quenching}",
      journal = {\mnras},
         year = 2020,
        month = feb,
       volume = {491},
       number = {4},
        pages = {5406-5434},
          doi = {10.1093/mnras/stz3286},
archivePrefix = {arXiv},
       eprint = {1811.09283},
 primaryClass = {astro-ph.GA},
       adsurl = {https://ui.adsabs.harvard.edu/abs/2020MNRAS.491.5406T}
}

@ARTICLE{sotilloramos+21,
       author = {{Sotillo-Ramos}, D. and {Lara-L{\'o}pez}, M.~A. and {P{\'e}rez-Garc{\'\i}a}, A.~M. and {P{\'e}rez-Mart{\'\i}nez}, R. and {Hopkins}, A.~M. and {Holwerda}, B.~W. and {Liske}, J. and {L{\'o}pez-S{\'a}nchez}, A.~R. and {Owers}, M.~S. and {Pimbblet}, K.~A.},
        title = "{Galaxy and mass assembly (GAMA): The environmental impact on SFR and metallicity in galaxy groups}",
      journal = {\mnras},
         year = 2021,
        month = dec,
       volume = {508},
       number = {2},
        pages = {1817-1830},
          doi = {10.1093/mnras/stab2641},
archivePrefix = {arXiv},
       eprint = {2109.12078},
 primaryClass = {astro-ph.GA},
       adsurl = {https://ui.adsabs.harvard.edu/abs/2021MNRAS.508.1817S}
}

@ARTICLE{weinmann+09,
       author = {{Weinmann}, Simone M. and {Kauffmann}, Guinevere and {van den Bosch}, Frank C. and {Pasquali}, Anna and {McIntosh}, Daniel H. and {Mo}, Houjun and {Yang}, Xiaohu and {Guo}, Yicheng},
        title = "{Environmental effects on satellite galaxies: the link between concentration, size and colour profile}",
      journal = {\mnras},
         year = 2009,
        month = apr,
       volume = {394},
       number = {3},
        pages = {1213-1228},
          doi = {10.1111/j.1365-2966.2009.14412.x},
archivePrefix = {arXiv},
       eprint = {0809.2283},
 primaryClass = {astro-ph},
       adsurl = {https://ui.adsabs.harvard.edu/abs/2009MNRAS.394.1213W}
}

@ARTICLE{wetzel+12,
       author = {{Wetzel}, Andrew R. and {Tinker}, Jeremy L. and {Conroy}, Charlie},
        title = "{Galaxy evolution in groups and clusters: star formation rates, red sequence fractions and the persistent bimodality}",
      journal = {\mnras},
         year = 2012,
        month = jul,
       volume = {424},
       number = {1},
        pages = {232-243},
          doi = {10.1111/j.1365-2966.2012.21188.x},
archivePrefix = {arXiv},
       eprint = {1107.5311},
 primaryClass = {astro-ph.CO},
       adsurl = {https://ui.adsabs.harvard.edu/abs/2012MNRAS.424..232W}
}

@ARTICLE{knobel+13,
       author = {{Knobel}, C. and {Lilly}, S.~J. and {Kova{\v{c}}}, K. and {Peng}, Y. and {Bschorr}, T.~J. and {Carollo}, C.~M. and {Contini}, T. and {Kneib}, J. -P. and {Le Fevre}, O. and {Mainieri}, V. and {Renzini}, A. and {Scodeggio}, M. and {Zamorani}, G. and {Bardelli}, S. and {Bolzonella}, M. and {Bongiorno}, A. and {Caputi}, K. and {Cucciati}, O. and {de la Torre}, S. and {de Ravel}, L. and {Franzetti}, P. and {Garilli}, B. and {Iovino}, A. and {Kampczyk}, P. and {Lamareille}, F. and {Le Borgne}, J. -F. and {Le Brun}, V. and {Maier}, C. and {Mignoli}, M. and {Pello}, R. and {Perez Montero}, E. and {Presotto}, V. and {Silverman}, J. and {Tanaka}, M. and {Tasca}, L. and {Tresse}, L. and {Vergani}, D. and {Zucca}, E. and {Barnes}, L. and {Bordoloi}, R. and {Cappi}, A. and {Cimatti}, A. and {Coppa}, G. and {Koekemoer}, A.~M. and {L{\'o}pez-Sanjuan}, C. and {McCracken}, H.~J. and {Moresco}, M. and {Nair}, P. and {Pozzetti}, L. and {Welikala}, N.},
        title = "{The Colors of Central and Satellite Galaxies in zCOSMOS Out to z \raisebox{-0.5ex}\textasciitilde= 0.8 and Implications for Quenching}",
      journal = {\apj},
         year = 2013,
        month = may,
       volume = {769},
       number = {1},
          eid = {24},
        pages = {24},
          doi = {10.1088/0004-637X/769/1/24},
archivePrefix = {arXiv},
       eprint = {1211.5607},
 primaryClass = {astro-ph.CO},
       adsurl = {https://ui.adsabs.harvard.edu/abs/2013ApJ...769...24K}
}

@ARTICLE{robotham+14,
       author = {{Robotham}, A.~S.~G. and {Driver}, S.~P. and {Davies}, L.~J.~M. and {Hopkins}, A.~M. and {Baldry}, I.~K. and {Agius}, N.~K. and {Bauer}, A.~E. and {Bland-Hawthorn}, J. and {Brough}, S. and {Brown}, M.~J.~I. and {Cluver}, M. and {De Propris}, R. and {Drinkwater}, M.~J. and {Holwerda}, B.~W. and {Kelvin}, L.~S. and {Lara-Lopez}, M.~A. and {Liske}, J. and {L{\'o}pez-S{\'a}nchez}, {\'A}. R. and {Loveday}, J. and {Mahajan}, S. and {McNaught-Roberts}, T. and {Moffett}, A. and {Norberg}, P. and {Obreschkow}, D. and {Owers}, M.~S. and {Penny}, S.~J. and {Pimbblet}, K. and {Prescott}, M. and {Taylor}, E.~N. and {van Kampen}, E. and {Wilkins}, S.~M.},
        title = "{Galaxy And Mass Assembly (GAMA): galaxy close pairs, mergers and the future fate of stellar mass}",
      journal = {\mnras},
         year = 2014,
        month = nov,
       volume = {444},
       number = {4},
        pages = {3986-4008},
          doi = {10.1093/mnras/stu1604},
archivePrefix = {arXiv},
       eprint = {1408.1476},
 primaryClass = {astro-ph.GA},
       adsurl = {https://ui.adsabs.harvard.edu/abs/2014MNRAS.444.3986R}
}

@ARTICLE{grootes+17,
       author = {{Grootes}, M.~W. and {Tuffs}, R.~J. and {Popescu}, C.~C. and {Norberg}, P. and {Robotham}, A.~S.~G. and {Liske}, J. and {Andrae}, E. and {Baldry}, I.~K. and {Gunawardhana}, M. and {Kelvin}, L.~S. and {Madore}, B.~F. and {Seibert}, M. and {Taylor}, E.~N. and {Alpaslan}, M. and {Brown}, M.~J.~I. and {Cluver}, M.~E. and {Driver}, S.~P. and {Bland-Hawthorn}, J. and {Holwerda}, B.~W. and {Hopkins}, A.~M. and {Lopez-Sanchez}, A.~R. and {Loveday}, J. and {Rushton}, M.},
        title = "{Galaxy And Mass Assembly (GAMA): Gas Fueling of Spiral Galaxies in the Local Universe. I. The Effect of the Group Environment on Star Formation in Spiral Galaxies}",
      journal = {\aj},
         year = 2017,
        month = mar,
       volume = {153},
       number = {3},
          eid = {111},
        pages = {111},
          doi = {10.3847/1538-3881/153/3/111},
archivePrefix = {arXiv},
       eprint = {1612.07322},
 primaryClass = {astro-ph.GA},
       adsurl = {https://ui.adsabs.harvard.edu/abs/2017AJ....153..111G}
}

@ARTICLE{wetzel+13,
       author = {{Wetzel}, Andrew R. and {Tinker}, Jeremy L. and {Conroy}, Charlie and {van den Bosch}, Frank C.},
        title = "{Galaxy evolution in groups and clusters: satellite star formation histories and quenching time-scales in a hierarchical Universe}",
      journal = {\mnras},
         year = 2013,
        month = jun,
       volume = {432},
       number = {1},
        pages = {336-358},
          doi = {10.1093/mnras/stt469},
archivePrefix = {arXiv},
       eprint = {1206.3571},
 primaryClass = {astro-ph.CO},
       adsurl = {https://ui.adsabs.harvard.edu/abs/2013MNRAS.432..336W}
}

@ARTICLE{treyer+18,
       author = {{Treyer}, M. and {Kraljic}, K. and {Arnouts}, S. and {de la Torre}, S. and {Pichon}, C. and {Dubois}, Y. and {Vibert}, D. and {Milliard}, B. and {Laigle}, C. and {Seibert}, M. and {Brown}, M.~J.~I. and {Grootes}, M.~W. and {Wright}, A.~H. and {Liske}, J. and {Lara-Lopez}, M.~A. and {Bland-Hawthorn}, J.},
        title = "{Group quenching and galactic conformity at low redshift}",
      journal = {\mnras},
         year = 2018,
        month = jun,
       volume = {477},
       number = {2},
        pages = {2684-2704},
          doi = {10.1093/mnras/sty769},
archivePrefix = {arXiv},
       eprint = {1712.05318},
 primaryClass = {astro-ph.GA},
       adsurl = {https://ui.adsabs.harvard.edu/abs/2018MNRAS.477.2684T}
}

@ARTICLE{moore+99,
       author = {{Moore}, Ben and {Lake}, George and {Quinn}, Thomas and {Stadel}, Joachim},
        title = "{On the survival and destruction of spiral galaxies in clusters}",
      journal = {\mnras},
         year = 1999,
        month = apr,
       volume = {304},
       number = {3},
        pages = {465-474},
          doi = {10.1046/j.1365-8711.1999.02345.x},
archivePrefix = {arXiv},
       eprint = {astro-ph/9811127},
 primaryClass = {astro-ph},
       adsurl = {https://ui.adsabs.harvard.edu/abs/1999MNRAS.304..465M}
}

@ARTICLE{nichols+11,
       author = {{Nichols}, Matthew and {Bland-Hawthorn}, Joss},
        title = "{Gas Depletion in Local Group Dwarfs on \raisebox{-0.5ex}\textasciitilde250 kpc Scales: Ram Pressure Stripping Assisted by Internal Heating at Early Times}",
      journal = {\apj},
         year = 2011,
        month = may,
       volume = {732},
       number = {1},
          eid = {17},
        pages = {17},
          doi = {10.1088/0004-637X/732/1/17},
archivePrefix = {arXiv},
       eprint = {1102.4849},
 primaryClass = {astro-ph.GA},
       adsurl = {https://ui.adsabs.harvard.edu/abs/2011ApJ...732...17N}
}

@ARTICLE{poggianti+17,
       author = {{Poggianti}, Bianca M. and {Moretti}, Alessia and {Gullieuszik}, Marco and {Fritz}, Jacopo and {Jaff{\'e}}, Yara and {Bettoni}, Daniela and {Fasano}, Giovanni and {Bellhouse}, Callum and {Hau}, George and {Vulcani}, Benedetta and {Biviano}, Andrea and {Omizzolo}, Alessandro and {Paccagnella}, Angela and {D'Onofrio}, Mauro and {Cava}, Antonio and {Sheen}, Y. -K. and {Couch}, Warrick and {Owers}, Matt},
        title = "{GASP. I. Gas Stripping Phenomena in Galaxies with MUSE}",
      journal = {\apj},
         year = 2017,
        month = jul,
       volume = {844},
       number = {1},
          eid = {48},
        pages = {48},
          doi = {10.3847/1538-4357/aa78ed},
archivePrefix = {arXiv},
       eprint = {1704.05086},
 primaryClass = {astro-ph.GA},
       adsurl = {https://ui.adsabs.harvard.edu/abs/2017ApJ...844...48P}
}

@ARTICLE{brown+17,
       author = {{Brown}, Toby and {Catinella}, Barbara and {Cortese}, Luca and {Lagos}, Claudia del P. and {Dav{\'e}}, Romeel and {Kilborn}, Virginia and {Haynes}, Martha P. and {Giovanelli}, Riccardo and {Rafieferantsoa}, Mika},
        title = "{Cold gas stripping in satellite galaxies: from pairs to clusters}",
      journal = {\mnras},
         year = 2017,
        month = apr,
       volume = {466},
       number = {2},
        pages = {1275-1289},
          doi = {10.1093/mnras/stw2991},
archivePrefix = {arXiv},
       eprint = {1611.00896},
 primaryClass = {astro-ph.GA},
       adsurl = {https://ui.adsabs.harvard.edu/abs/2017MNRAS.466.1275B}
}

@ARTICLE{faber+07,
       author = {{Faber}, S.~M. and {Willmer}, C.~N.~A. and {Wolf}, C. and {Koo}, D.~C. and {Weiner}, B.~J. and {Newman}, J.~A. and {Im}, M. and {Coil}, A.~L. and {Conroy}, C. and {Cooper}, M.~C. and {Davis}, M. and {Finkbeiner}, D.~P. and {Gerke}, B.~F. and {Gebhardt}, K. and {Groth}, E.~J. and {Guhathakurta}, P. and {Harker}, J. and {Kaiser}, N. and {Kassin}, S. and {Kleinheinrich}, M. and {Konidaris}, N.~P. and {Kron}, R.~G. and {Lin}, L. and {Luppino}, G. and {Madgwick}, D.~S. and {Meisenheimer}, K. and {Noeske}, K.~G. and {Phillips}, A.~C. and {Sarajedini}, V.~L. and {Schiavon}, R.~P. and {Simard}, L. and {Szalay}, A.~S. and {Vogt}, N.~P. and {Yan}, R.},
        title = "{Galaxy Luminosity Functions to z\raisebox{-0.5ex}\textasciitilde1 from DEEP2 and COMBO-17: Implications for Red Galaxy Formation}",
      journal = {\apj},
         year = 2007,
        month = aug,
       volume = {665},
       number = {1},
        pages = {265-294},
          doi = {10.1086/519294},
archivePrefix = {arXiv},
       eprint = {astro-ph/0506044},
 primaryClass = {astro-ph},
       adsurl = {https://ui.adsabs.harvard.edu/abs/2007ApJ...665..265F}
}

@ARTICLE{martin+07,
       author = {{Martin}, D. Christopher and {Wyder}, Ted K. and {Schiminovich}, David and {Barlow}, Tom A. and {Forster}, Karl and {Friedman}, Peter G. and {Morrissey}, Patrick and {Neff}, Susan G. and {Seibert}, Mark and {Small}, Todd and {Welsh}, Barry Y. and {Bianchi}, Luciana and {Donas}, Jos{\'e} and {Heckman}, Timothy M. and {Lee}, Young-Wook and {Madore}, Barry F. and {Milliard}, Bruno and {Rich}, R. Michael and {Szalay}, Alex S. and {Yi}, Sukyoung K.},
        title = "{The UV-Optical Galaxy Color-Magnitude Diagram. III. Constraints on Evolution from the Blue to the Red Sequence}",
      journal = {\apjs},
         year = 2007,
        month = dec,
       volume = {173},
       number = {2},
        pages = {342-356},
          doi = {10.1086/516639},
archivePrefix = {arXiv},
       eprint = {astro-ph/0703281},
 primaryClass = {astro-ph},
       adsurl = {https://ui.adsabs.harvard.edu/abs/2007ApJS..173..342M}
}

@ARTICLE{schawinski+14,
       author = {{Schawinski}, Kevin and {Urry}, C. Megan and {Simmons}, Brooke D. and {Fortson}, Lucy and {Kaviraj}, Sugata and {Keel}, William C. and {Lintott}, Chris J. and {Masters}, Karen L. and {Nichol}, Robert C. and {Sarzi}, Marc and {Skibba}, Ramin and {Treister}, Ezequiel and {Willett}, Kyle W. and {Wong}, O. Ivy and {Yi}, Sukyoung K.},
        title = "{The green valley is a red herring: Galaxy Zoo reveals two evolutionary pathways towards quenching of star formation in early- and late-type galaxies}",
      journal = {\mnras},
         year = 2014,
        month = may,
       volume = {440},
       number = {1},
        pages = {889-907},
          doi = {10.1093/mnras/stu327},
archivePrefix = {arXiv},
       eprint = {1402.4814},
 primaryClass = {astro-ph.GA},
       adsurl = {https://ui.adsabs.harvard.edu/abs/2014MNRAS.440..889S}
}

@ARTICLE{marinoni+02,
       author = {{Marinoni}, Christian and {Hudson}, Michael J.},
        title = "{The Mass-to-Light Function of Virialized Systems and the Relationship between Their Optical and X-Ray Properties}",
      journal = {\apj},
         year = 2002,
        month = apr,
       volume = {569},
       number = {1},
        pages = {101-111},
          doi = {10.1086/339319},
archivePrefix = {arXiv},
       eprint = {astro-ph/0109134},
 primaryClass = {astro-ph},
       adsurl = {https://ui.adsabs.harvard.edu/abs/2002ApJ...569..101M}
}

@ARTICLE{shankar+06,
       author = {{Shankar}, F. and {Lapi}, A. and {Salucci}, P. and {De Zotti}, G. and {Danese}, L.},
        title = "{New Relationships between Galaxy Properties and Host Halo Mass, and the Role of Feedbacks in Galaxy Formation}",
      journal = {\apj},
         year = 2006,
        month = may,
       volume = {643},
       number = {1},
        pages = {14-25},
          doi = {10.1086/502794},
archivePrefix = {arXiv},
       eprint = {astro-ph/0601577},
 primaryClass = {astro-ph},
       adsurl = {https://ui.adsabs.harvard.edu/abs/2006ApJ...643...14S}
}

@ARTICLE{conroy+09,
       author = {{Conroy}, Charlie and {Wechsler}, Risa H.},
        title = "{Connecting Galaxies, Halos, and Star Formation Rates Across Cosmic Time}",
      journal = {\apj},
         year = 2009,
        month = may,
       volume = {696},
       number = {1},
        pages = {620-635},
          doi = {10.1088/0004-637X/696/1/620},
archivePrefix = {arXiv},
       eprint = {0805.3346},
 primaryClass = {astro-ph},
       adsurl = {https://ui.adsabs.harvard.edu/abs/2009ApJ...696..620C}
}

@ARTICLE{guo+10,
       author = {{Guo}, Qi and {White}, Simon and {Li}, Cheng and {Boylan-Kolchin}, Michael},
        title = "{How do galaxies populate dark matter haloes?}",
      journal = {\mnras},
         year = 2010,
        month = may,
       volume = {404},
       number = {3},
        pages = {1111-1120},
          doi = {10.1111/j.1365-2966.2010.16341.x},
archivePrefix = {arXiv},
       eprint = {0909.4305},
 primaryClass = {astro-ph.CO},
       adsurl = {https://ui.adsabs.harvard.edu/abs/2010MNRAS.404.1111G}
}

@ARTICLE{moster+10,
       author = {{Moster}, Benjamin P. and {Somerville}, Rachel S. and {Maulbetsch}, Christian and {van den Bosch}, Frank C. and {Macci{\`o}}, Andrea V. and {Naab}, Thorsten and {Oser}, Ludwig},
        title = "{Constraints on the Relationship between Stellar Mass and Halo Mass at Low and High Redshift}",
      journal = {\apj},
         year = 2010,
        month = feb,
       volume = {710},
       number = {2},
        pages = {903-923},
          doi = {10.1088/0004-637X/710/2/903},
archivePrefix = {arXiv},
       eprint = {0903.4682},
 primaryClass = {astro-ph.CO},
       adsurl = {https://ui.adsabs.harvard.edu/abs/2010ApJ...710..903M}
}

@ARTICLE{press+74,
       author = {{Press}, William H. and {Schechter}, Paul},
        title = "{Formation of Galaxies and Clusters of Galaxies by Self-Similar Gravitational Condensation}",
      journal = {\apj},
         year = 1974,
        month = feb,
       volume = {187},
        pages = {425-438},
          doi = {10.1086/152650},
       adsurl = {https://ui.adsabs.harvard.edu/abs/1974ApJ...187..425P}
}

@ARTICLE{plionis+06,
       author = {{Plionis}, M. and {Basilakos}, S. and {Ragone-Figueroa}, C.},
        title = "{Morphological and Dynamical Properties of Low-Redshift Two Degree Field Galaxy Redshift Survey Groups}",
      journal = {\apj},
         year = 2006,
        month = oct,
       volume = {650},
       number = {2},
        pages = {770-776},
          doi = {10.1086/507445},
archivePrefix = {arXiv},
       eprint = {astro-ph/0608457},
 primaryClass = {astro-ph},
       adsurl = {https://ui.adsabs.harvard.edu/abs/2006ApJ...650..770P}
}

@ARTICLE{robotham+08,
       author = {{Robotham}, Aaron and {Phillipps}, Steven and {De Propris}, Roberto},
        title = "{The Shapes of Galaxy Groups: Footballs or Frisbees?}",
      journal = {\apj},
         year = 2008,
        month = jan,
       volume = {672},
       number = {2},
        pages = {834-848},
          doi = {10.1086/523885},
       adsurl = {https://ui.adsabs.harvard.edu/abs/2008ApJ...672..834R}
}

@ARTICLE{cooray+02,
       author = {{Cooray}, Asantha and {Sheth}, Ravi},
        title = "{Halo models of large scale structure}",
      journal = {\physrep},
         year = 2002,
        month = dec,
       volume = {372},
       number = {1},
        pages = {1-129},
          doi = {10.1016/S0370-1573(02)00276-4},
archivePrefix = {arXiv},
       eprint = {astro-ph/0206508},
 primaryClass = {astro-ph},
       adsurl = {https://ui.adsabs.harvard.edu/abs/2002PhR...372....1C}
}

@ARTICLE{yang+03,
       author = {{Yang}, Xiaohu and {Mo}, H.~J. and {van den Bosch}, Frank C.},
        title = "{Constraining galaxy formation and cosmology with the conditional luminosity function of galaxies}",
      journal = {\mnras},
         year = 2003,
        month = mar,
       volume = {339},
       number = {4},
        pages = {1057-1080},
          doi = {10.1046/j.1365-8711.2003.06254.x},
archivePrefix = {arXiv},
       eprint = {astro-ph/0207019},
 primaryClass = {astro-ph},
       adsurl = {https://ui.adsabs.harvard.edu/abs/2003MNRAS.339.1057Y}
}

@ARTICLE{cooray+06,
       author = {{Cooray}, Asantha},
        title = "{Halo model at its best: constraints on conditional luminosity functions from measured galaxy statistics}",
      journal = {\mnras},
         year = 2006,
        month = jan,
       volume = {365},
       number = {3},
        pages = {842-866},
          doi = {10.1111/j.1365-2966.2005.09747.x},
archivePrefix = {arXiv},
       eprint = {astro-ph/0509033},
 primaryClass = {astro-ph},
       adsurl = {https://ui.adsabs.harvard.edu/abs/2006MNRAS.365..842C}
}

@ARTICLE{robotham+06,
       author = {{Robotham}, Aaron and {Wallace}, Christopher and {Phillipps}, Steven and {De Propris}, Roberto},
        title = "{Galaxy Luminosities in 2dF Percolation-Inferred Galaxy (2PIGG) Groups}",
      journal = {\apj},
         year = 2006,
        month = dec,
       volume = {652},
       number = {2},
        pages = {1077-1084},
          doi = {10.1086/508130},
archivePrefix = {arXiv},
       eprint = {astro-ph/0608027},
 primaryClass = {astro-ph},
       adsurl = {https://ui.adsabs.harvard.edu/abs/2006ApJ...652.1077R}
}

@ARTICLE{robotham+10a,
       author = {{Robotham}, A. and {Driver}, S.~P. and {Norberg}, P. and {Baldry}, I.~K. and {Bamford}, S.~P. and {Hopkins}, A.~M. and {Liske}, J. and {Loveday}, J. and {Peacock}, J.~A. and {Cameron}, E. and {Croom}, S.~M. and {Doyle}, I.~F. and {Frenk}, C.~S. and {Hill}, D.~T. and {Jones}, D.~H. and {van Kampen}, E. and {Kelvin}, L.~S. and {Kuijken}, K. and {Nichol}, R.~C. and {Parkinson}, H.~R. and {Popescu}, C.~C. and {Prescott}, M. and {Sharp}, R.~G. and {Sutherland}, W.~J. and {Thomas}, D. and {Tuffs}, R.~J.},
        title = "{Galaxy and Mass Assembly (GAMA): Optimal Tiling of Dense Surveys with a Multi-Object Spectrograph}",
      journal = {\pasa},
         year = 2010,
        month = mar,
       volume = {27},
       number = {1},
        pages = {76-90},
          doi = {10.1071/AS09053},
archivePrefix = {arXiv},
       eprint = {0910.5121},
 primaryClass = {astro-ph.CO},
       adsurl = {https://ui.adsabs.harvard.edu/abs/2010PASA...27...76R}
}

@ARTICLE{robotham+10b,
       author = {{Robotham}, Aaron and {Phillipps}, Steven and {de Propris}, Roberto},
        title = "{The variation of the galaxy luminosity function with group properties}",
      journal = {\mnras},
         year = 2010,
        month = apr,
       volume = {403},
       number = {4},
        pages = {1812-1828},
          doi = {10.1111/j.1365-2966.2010.16252.x},
archivePrefix = {arXiv},
       eprint = {1003.1981},
 primaryClass = {astro-ph.CO},
       adsurl = {https://ui.adsabs.harvard.edu/abs/2010MNRAS.403.1812R}
}

@ARTICLE{schwarz+78,
       author = {{Schwarz}, Gideon},
        title = "{Estimating the Dimension of a Model}",
      journal = {Annals of Statistics},
         year = 1978,
        month = jul,
       volume = {6},
       number = {2},
        pages = {461-464},
       adsurl = {https://ui.adsabs.harvard.edu/abs/1978AnSta...6..461S}
}

@ARTICLE{akaike+74,
       author = {{Akaike}, H.},
        title = "{A New Look at the Statistical Model Identification}",
      journal = {IEEE Transactions on Automatic Control},
         year = 1974,
        month = jan,
       volume = {19},
        pages = {716-723},
       adsurl = {https://ui.adsabs.harvard.edu/abs/1974ITAC...19..716A}
}

@BOOK{jeffreys+39,
       author = {{Jeffreys}, Harold},
        title = "{Theory of Probability}",
         year = 1939,
       adsurl = {https://ui.adsabs.harvard.edu/abs/1939thpr.book.....J}
}

@ARTICLE{colless+01,
       author = {{Colless}, Matthew and {Dalton}, Gavin and {Maddox}, Steve and {Sutherland}, Will and {Norberg}, Peder and {Cole}, Shaun and {Bland-Hawthorn}, Joss and {Bridges}, Terry and {Cannon}, Russell and {Collins}, Chris and {Couch}, Warrick and {Cross}, Nicholas and {Deeley}, Kathryn and {De Propris}, Roberto and {Driver}, Simon P. and {Efstathiou}, George and {Ellis}, Richard S. and {Frenk}, Carlos S. and {Glazebrook}, Karl and {Jackson}, Carole and {Lahav}, Ofer and {Lewis}, Ian and {Lumsden}, Stuart and {Madgwick}, Darren and {Peacock}, John A. and {Peterson}, Bruce A. and {Price}, Ian and {Seaborne}, Mark and {Taylor}, Keith},
        title = "{The 2dF Galaxy Redshift Survey: spectra and redshifts}",
      journal = {\mnras},
         year = 2001,
        month = dec,
       volume = {328},
       number = {4},
        pages = {1039-1063},
          doi = {10.1046/j.1365-8711.2001.04902.x},
archivePrefix = {arXiv},
       eprint = {astro-ph/0106498},
 primaryClass = {astro-ph},
       adsurl = {https://ui.adsabs.harvard.edu/abs/2001MNRAS.328.1039C}
}

@ARTICLE{abazajian+09,
       author = {{Abazajian}, Kevork N. and {Adelman-McCarthy}, Jennifer K. and {Ag{\"u}eros}, Marcel A. and {Allam}, Sahar S. and {Allende Prieto}, Carlos and {An}, Deokkeun and {Anderson}, Kurt S.~J. and {Anderson}, Scott F. and {Annis}, James and {Bahcall}, Neta A. and {Bailer-Jones}, C.~A.~L. and {Barentine}, J.~C. and {Bassett}, Bruce A. and {Becker}, Andrew C. and {Beers}, Timothy C. and {Bell}, Eric F. and {Belokurov}, Vasily and {Berlind}, Andreas A. and {Berman}, Eileen F. and {Bernardi}, Mariangela and {Bickerton}, Steven J. and {Bizyaev}, Dmitry and {Blakeslee}, John P. and {Blanton}, Michael R. and {Bochanski}, John J. and {Boroski}, William N. and {Brewington}, Howard J. and {Brinchmann}, Jarle and {Brinkmann}, J. and {Brunner}, Robert J. and {Budav{\'a}ri}, Tam{\'a}s and {Carey}, Larry N. and {Carliles}, Samuel and {Carr}, Michael A. and {Castander}, Francisco J. and {Cinabro}, David and {Connolly}, A.~J. and {Csabai}, Istv{\'a}n and {Cunha}, Carlos E. and {Czarapata}, Paul C. and {Davenport}, James R.~A. and {de Haas}, Ernst and {Dilday}, Ben and {Doi}, Mamoru and {Eisenstein}, Daniel J. and {Evans}, Michael L. and {Evans}, N.~W. and {Fan}, Xiaohui and {Friedman}, Scott D. and {Frieman}, Joshua A. and {Fukugita}, Masataka and {G{\"a}nsicke}, Boris T. and {Gates}, Evalyn and {Gillespie}, Bruce and {Gilmore}, G. and {Gonzalez}, Belinda and {Gonzalez}, Carlos F. and {Grebel}, Eva K. and {Gunn}, James E. and {Gy{\"o}ry}, Zsuzsanna and {Hall}, Patrick B. and {Harding}, Paul and {Harris}, Frederick H. and {Harvanek}, Michael and {Hawley}, Suzanne L. and {Hayes}, Jeffrey J.~E. and {Heckman}, Timothy M. and {Hendry}, John S. and {Hennessy}, Gregory S. and {Hindsley}, Robert B. and {Hoblitt}, J. and {Hogan}, Craig J. and {Hogg}, David W. and {Holtzman}, Jon A. and {Hyde}, Joseph B. and {Ichikawa}, Shin-ichi and {Ichikawa}, Takashi and {Im}, Myungshin and {Ivezi{\'c}}, {\v{Z}}eljko and {Jester}, Sebastian and {Jiang}, Linhua and {Johnson}, Jennifer A. and {Jorgensen}, Anders M. and {Juri{\'c}}, Mario and {Kent}, Stephen M. and {Kessler}, R. and {Kleinman}, S.~J. and {Knapp}, G.~R. and {Konishi}, Kohki and {Kron}, Richard G. and {Krzesinski}, Jurek and {Kuropatkin}, Nikolay and {Lampeitl}, Hubert and {Lebedeva}, Svetlana and {Lee}, Myung Gyoon and {Lee}, Young Sun and {French Leger}, R. and {L{\'e}pine}, S{\'e}bastien and {Li}, Nolan and {Lima}, Marcos and {Lin}, Huan and {Long}, Daniel C. and {Loomis}, Craig P. and {Loveday}, Jon and {Lupton}, Robert H. and {Magnier}, Eugene and {Malanushenko}, Olena and {Malanushenko}, Viktor and {Mandelbaum}, Rachel and {Margon}, Bruce and {Marriner}, John P. and {Mart{\'\i}nez-Delgado}, David and {Matsubara}, Takahiko and {McGehee}, Peregrine M. and {McKay}, Timothy A. and {Meiksin}, Avery and {Morrison}, Heather L. and {Mullally}, Fergal and {Munn}, Jeffrey A. and {Murphy}, Tara and {Nash}, Thomas and {Nebot}, Ada and {Neilsen}, Eric H., Jr. and {Newberg}, Heidi Jo and {Newman}, Peter R. and {Nichol}, Robert C. and {Nicinski}, Tom and {Nieto-Santisteban}, Maria and {Nitta}, Atsuko and {Okamura}, Sadanori and {Oravetz}, Daniel J. and {Ostriker}, Jeremiah P. and {Owen}, Russell and {Padmanabhan}, Nikhil and {Pan}, Kaike and {Park}, Changbom and {Pauls}, George and {Peoples}, John, Jr. and {Percival}, Will J. and {Pier}, Jeffrey R. and {Pope}, Adrian C. and {Pourbaix}, Dimitri and {Price}, Paul A. and {Purger}, Norbert and {Quinn}, Thomas and {Raddick}, M. Jordan and {Re Fiorentin}, Paola and {Richards}, Gordon T. and {Richmond}, Michael W. and {Riess}, Adam G. and {Rix}, Hans-Walter and {Rockosi}, Constance M. and {Sako}, Masao and {Schlegel}, David J. and {Schneider}, Donald P. and {Scholz}, Ralf-Dieter and {Schreiber}, Matthias R. and {Schwope}, Axel D. and {Seljak}, Uro{\v{s}} and {Sesar}, Branimir and {Sheldon}, Erin and {Shimasaku}, Kazu and {Sibley}, Valena C. and {Simmons}, A.~E. and {Sivarani}, Thirupathi and {Allyn Smith}, J. and {Smith}, Martin C. and {Smol{\v{c}}i{\'c}}, Vernesa and {Snedden}, Stephanie A. and {Stebbins}, Albert and {Steinmetz}, Matthias and {Stoughton}, Chris and {Strauss}, Michael A. and {SubbaRao}, Mark and {Suto}, Yasushi and {Szalay}, Alexander S. and {Szapudi}, Istv{\'a}n and {Szkody}, Paula and {Tanaka}, Masayuki and {Tegmark}, Max and {Teodoro}, Luis F.~A. and {Thakar}, Aniruddha R. and {Tremonti}, Christy A. and {Tucker}, Douglas L. and {Uomoto}, Alan and {Vanden Berk}, Daniel E. and {Vandenberg}, Jan and {Vidrih}, S. and {Vogeley}, Michael S. and {Voges}, Wolfgang and {Vogt}, Nicole P. and {Wadadekar}, Yogesh and {Watters}, Shannon and {Weinberg}, David H. and {West}, Andrew A. and {White}, Simon D.~M. and {Wilhite}, Brian C. and {Wonders}, Alainna C. and {Yanny}, Brian and {Yocum}, D.~R. and {York}, Donald G. and {Zehavi}, Idit and {Zibetti}, Stefano and {Zucker}, Daniel B.},
        title = "{The Seventh Data Release of the Sloan Digital Sky Survey}",
      journal = {\apjs},
         year = 2009,
        month = jun,
       volume = {182},
       number = {2},
        pages = {543-558},
          doi = {10.1088/0067-0049/182/2/543},
archivePrefix = {arXiv},
       eprint = {0812.0649},
 primaryClass = {astro-ph},
       adsurl = {https://ui.adsabs.harvard.edu/abs/2009ApJS..182..543A}
}

@ARTICLE{brammer+11,
       author = {{Brammer}, Gabriel B. and {Whitaker}, K.~E. and {van Dokkum}, P.~G. and {Marchesini}, D. and {Franx}, M. and {Kriek}, M. and {Labb{\'e}}, I. and {Lee}, K. -S. and {Muzzin}, A. and {Quadri}, R.~F. and {Rudnick}, G. and {Williams}, R.},
        title = "{The Number Density and Mass Density of Star-forming and Quiescent Galaxies at 0.4 <= z <= 2.2}",
      journal = {\apj},
         year = 2011,
        month = sep,
       volume = {739},
       number = {1},
          eid = {24},
        pages = {24},
          doi = {10.1088/0004-637X/739/1/24},
archivePrefix = {arXiv},
       eprint = {1104.2595},
 primaryClass = {astro-ph.CO},
       adsurl = {https://ui.adsabs.harvard.edu/abs/2011ApJ...739...24B}
}

@ARTICLE{ilbert+10,
       author = {{Ilbert}, O. and {Salvato}, M. and {Le Floc'h}, E. and {Aussel}, H. and {Capak}, P. and {McCracken}, H.~J. and {Mobasher}, B. and {Kartaltepe}, J. and {Scoville}, N. and {Sanders}, D.~B. and {Arnouts}, S. and {Bundy}, K. and {Cassata}, P. and {Kneib}, J. -P. and {Koekemoer}, A. and {Le F{\`e}vre}, O. and {Lilly}, S. and {Surace}, J. and {Taniguchi}, Y. and {Tasca}, L. and {Thompson}, D. and {Tresse}, L. and {Zamojski}, M. and {Zamorani}, G. and {Zucca}, E.},
        title = "{Galaxy Stellar Mass Assembly Between 0.2 < z < 2 from the S-COSMOS Survey}",
      journal = {\apj},
         year = 2010,
        month = feb,
       volume = {709},
       number = {2},
        pages = {644-663},
          doi = {10.1088/0004-637X/709/2/644},
archivePrefix = {arXiv},
       eprint = {0903.0102},
 primaryClass = {astro-ph.CO},
       adsurl = {https://ui.adsabs.harvard.edu/abs/2010ApJ...709..644I}
}

@ARTICLE{drory+09,
       author = {{Drory}, N. and {Bundy}, K. and {Leauthaud}, A. and {Scoville}, N. and {Capak}, P. and {Ilbert}, O. and {Kartaltepe}, J.~S. and {Kneib}, J.~P. and {McCracken}, H.~J. and {Salvato}, M. and {Sanders}, D.~B. and {Thompson}, D. and {Willott}, C.~J.},
        title = "{The Bimodal Galaxy Stellar Mass Function in the COSMOS Survey to z \raisebox{-0.5ex}\textasciitilde 1: A Steep Faint End and a New Galaxy Dichotomy}",
      journal = {\apj},
         year = 2009,
        month = dec,
       volume = {707},
       number = {2},
        pages = {1595-1609},
          doi = {10.1088/0004-637X/707/2/1595},
archivePrefix = {arXiv},
       eprint = {0910.5720},
 primaryClass = {astro-ph.CO},
       adsurl = {https://ui.adsabs.harvard.edu/abs/2009ApJ...707.1595D}
}

@ARTICLE{zheng+24,
       author = {{Zheng}, Haonan and {Bose}, Sownak and {Frenk}, Carlos S. and {Gao}, Liang and {Jenkins}, Adrian and {Liao}, Shihong and {Liu}, Yizhou and {Wang}, Jie},
        title = "{The abundance of dark matter haloes down to Earth mass}",
      journal = {\mnras},
         year = 2024,
        month = mar,
       volume = {528},
       number = {4},
        pages = {7300-7309},
          doi = {10.1093/mnras/stae289},
archivePrefix = {arXiv},
       eprint = {2310.16093},
 primaryClass = {astro-ph.GA},
       adsurl = {https://ui.adsabs.harvard.edu/abs/2024MNRAS.528.7300Z}
}

@ARTICLE{davies+25a,
       author = {{Davies}, L.~J.~M. and {Thorne}, J.~E. and {Bellstedt}, S. and {Cook}, R.~H.~W. and {Bravo}, M. and {Robotham}, A.~S.~G. and {Lagos}, C. del P. and {Phillipps}, S. and {Siudek}, M. and {Holwerda}, B.~W. and {Bremer}, M.~N. and {D'Silva}, J. and {Driver}, S.~P.},
        title = "{Deep Extragalactic VIsible Legacy Survey (DEVILS): the sSFR{\textendash}M$_{{\ensuremath{\star}}}$ plane {\textendash} II. Starbursts, SFHs, and AGN feedback}",
      journal = {\mnras},
         year = 2025,
        month = jul,
       volume = {541},
       number = {1},
        pages = {573-600},
          doi = {10.1093/mnras/staf872},
archivePrefix = {arXiv},
       eprint = {2505.21948},
 primaryClass = {astro-ph.GA},
       adsurl = {https://ui.adsabs.harvard.edu/abs/2025MNRAS.541..573D}
}

@ARTICLE{davies+25b,
       author = {{Davies}, L.~J.~M. and {Fuentealba-Fuentes}, M.~F. and {Wright}, R.~J. and {Bravo}, M. and {Wagh}, S. and {Siudek}, M.},
        title = "{Deep Extragalactic VIsible Legacy Survey (DEVILS): satellite quenching at intermediate redshift}",
      journal = {\mnras},
         year = 2025,
        month = aug,
       volume = {541},
       number = {4},
        pages = {3220-3235},
          doi = {10.1093/mnras/staf1205},
archivePrefix = {arXiv},
       eprint = {2507.20822},
 primaryClass = {astro-ph.GA},
       adsurl = {https://ui.adsabs.harvard.edu/abs/2025MNRAS.541.3220D}
}

@ARTICLE{davies+15,
       author = {{Davies}, L.~J.~M. and {Robotham}, A.~S.~G. and {Driver}, S.~P. and {Alpaslan}, M. and {Baldry}, I.~K. and {Bland-Hawthorn}, J. and {Brough}, S. and {Brown}, M.~J.~I. and {Cluver}, M.~E. and {Drinkwater}, M.~J. and {Foster}, C. and {Grootes}, M.~W. and {Konstantopoulos}, I.~S. and {Lara-L{\'o}pez}, M.~A. and {L{\'o}pez-S{\'a}nchez}, {\'A}. R. and {Loveday}, J. and {Meyer}, M.~J. and {Moffett}, A.~J. and {Norberg}, P. and {Owers}, M.~S. and {Popescu}, C.~C. and {De Propris}, R. and {Sharp}, R. and {Tuffs}, R.~J. and {Wang}, L. and {Wilkins}, S.~M. and {Dunne}, L. and {Bourne}, N. and {Smith}, M.~W.~L.},
        title = "{Galaxy And Mass Assembly (GAMA): the effect of close interactions on star formation in galaxies}",
      journal = {\mnras},
         year = 2015,
        month = sep,
       volume = {452},
       number = {1},
        pages = {616-636},
          doi = {10.1093/mnras/stv1241},
archivePrefix = {arXiv},
       eprint = {1507.04447},
 primaryClass = {astro-ph.GA},
       adsurl = {https://ui.adsabs.harvard.edu/abs/2015MNRAS.452..616D}
}

\appendix
\section{Parameter degeneracies in the best-fit double Schechter GSMF fits}
\label{app:cornerplots}
\FloatBarrier
In this Appendix, we present the joint likelihood distributions of the double Schechter function parameters $M^{\star}$, $\alpha_{1}$ and $\alpha_{2}$ for all GSMFs discussed in Sects.~\ref{6.1}--\ref{6.3}. These plots illustrate possible parameter degeneracies and provide a visual check of the robustness of our fits. 

\begin{figure}[!h]
\centering
\includegraphics[width=0.4\textwidth]{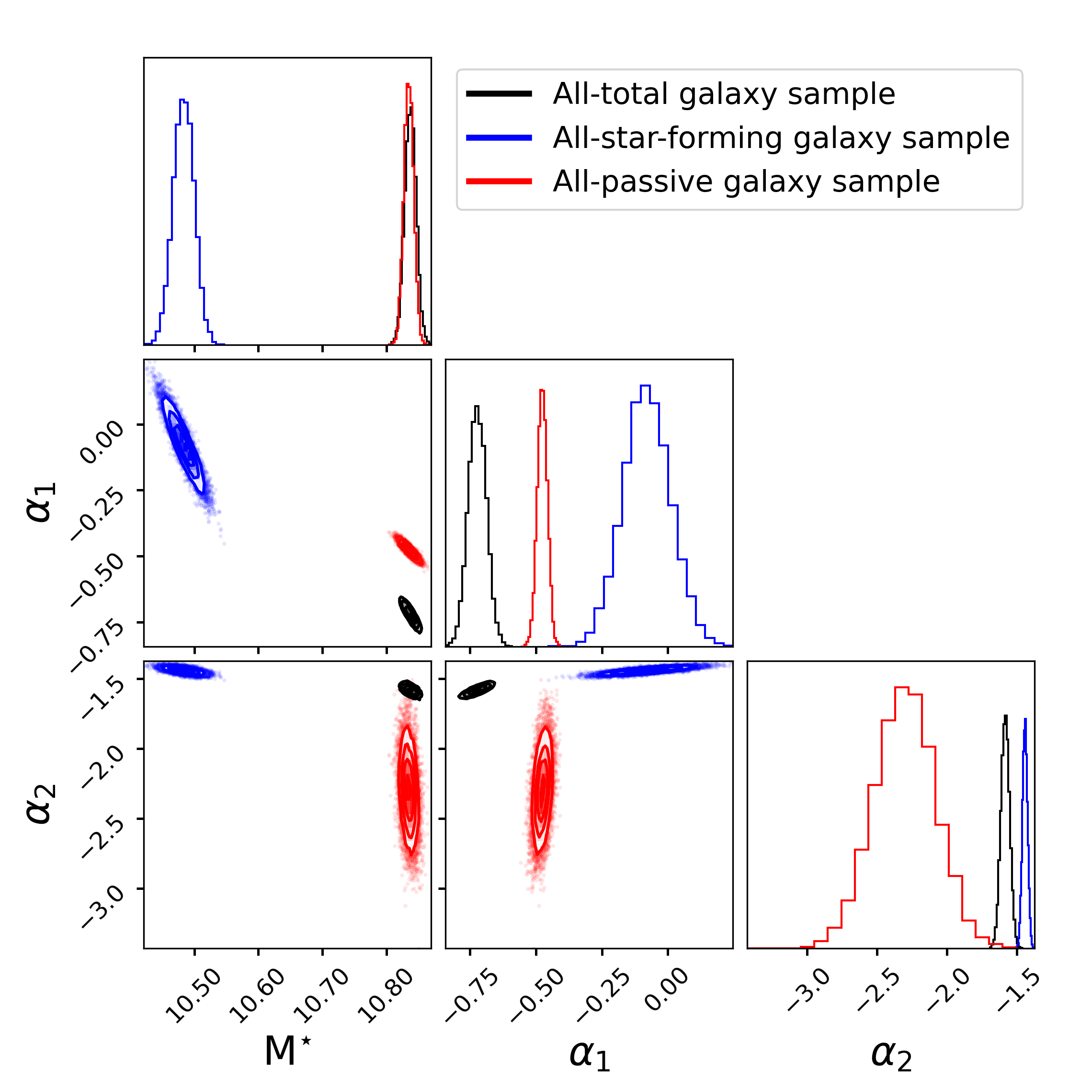}
\caption{$1$-$\sigma$, $2$-$\sigma$ and $3$-$\sigma$ likelihood contours of the GSMFs for our all-total, all-star-forming, and all-passive galaxy sample, as indicated in the legend. For each distribution, $10^{4}$ random samples were generated.}
\label{fig:appendix1}
\end{figure}

\begin{figure}[!h]
\centering
\includegraphics[width=0.4\textwidth]{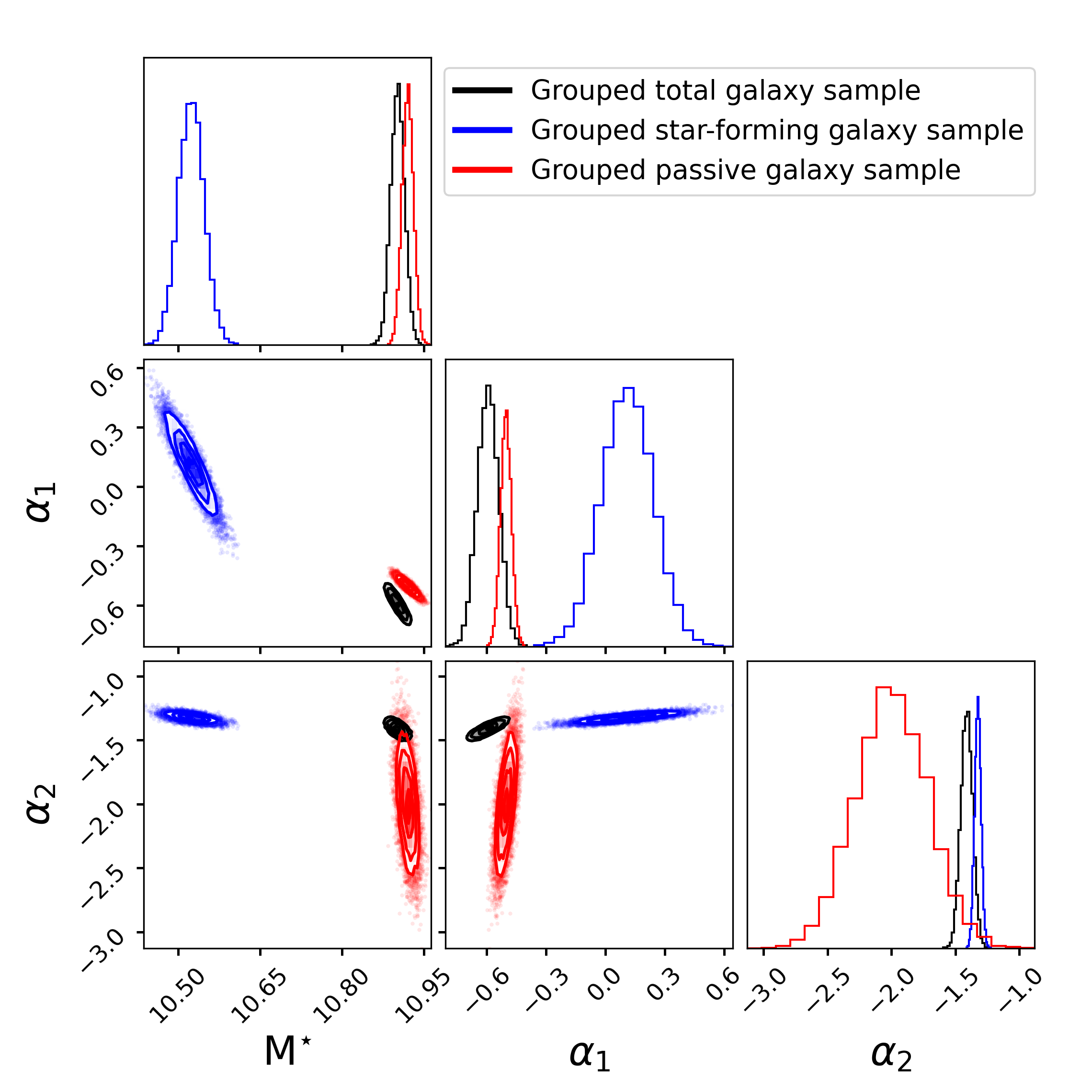}
\caption{Same as Fig.~\ref{fig:appendix1}, but for the subsamples indicated.}
\label{fig:appendix2}
\end{figure}

\begin{figure}
\centering
\vspace{-0.4cm}
\includegraphics[width=0.4\textwidth]{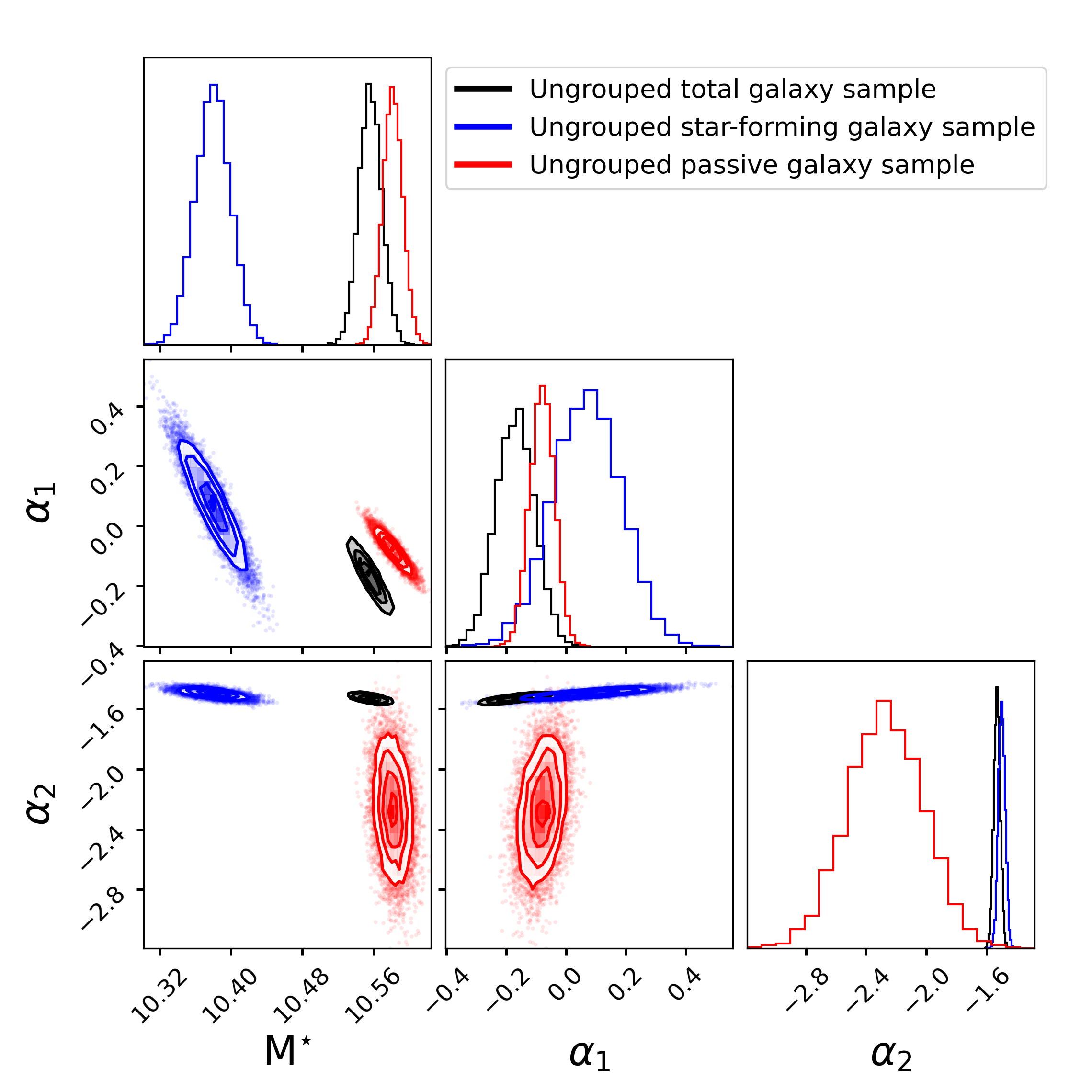}
\caption{Same as Fig.~\ref{fig:appendix1}, but for the subsamples indicated.}
\label{fig:appendix3}
\end{figure}

\begin{figure}
\centering
\vspace{-0.3cm}
\includegraphics[width=0.4\textwidth]{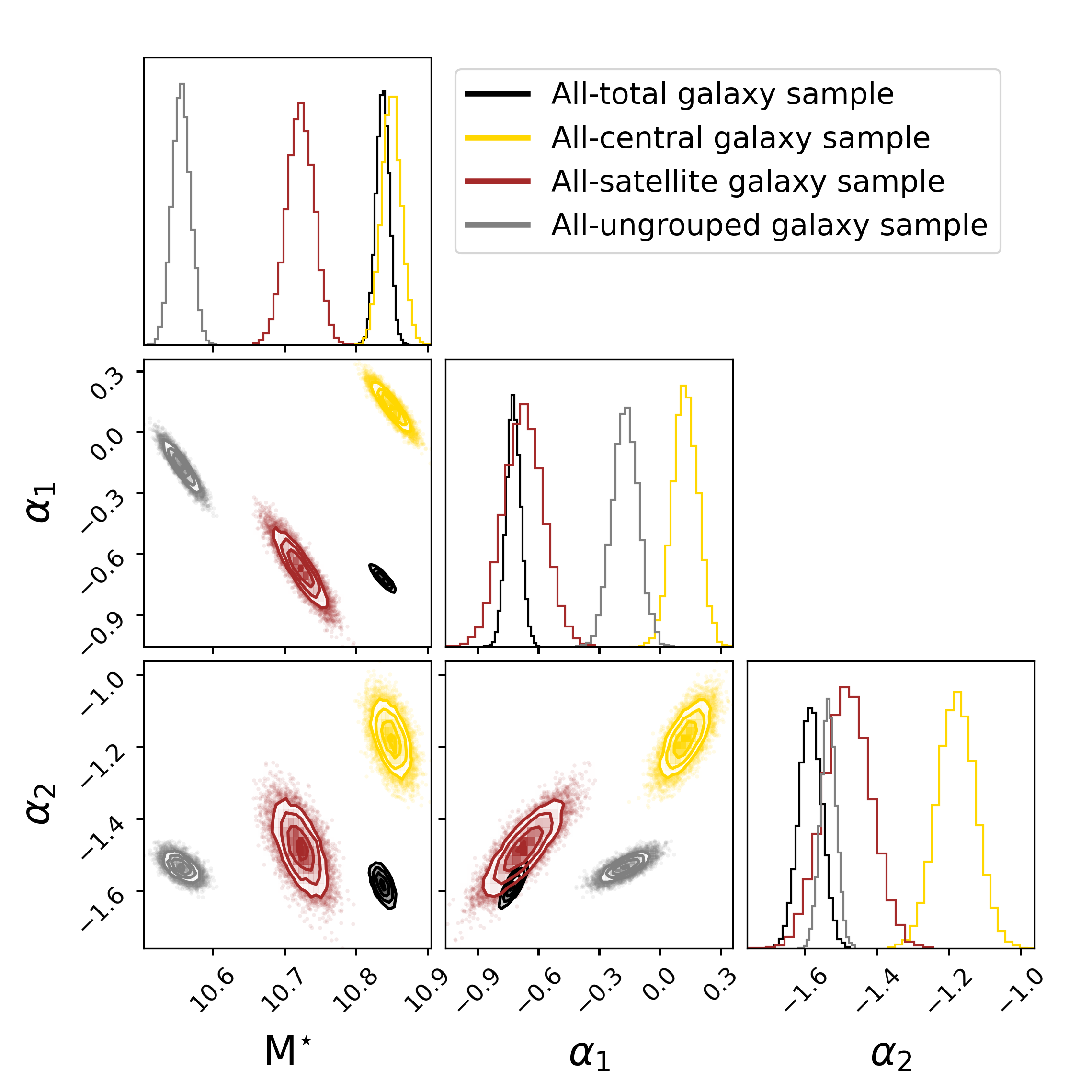}
\caption{Same as Fig.~\ref{fig:appendix1}, but for the subsamples indicated.}
\label{fig:appendix4}
\end{figure}

\begin{figure}
\centering
\vspace{-0.3cm}
\includegraphics[width=0.4\textwidth]{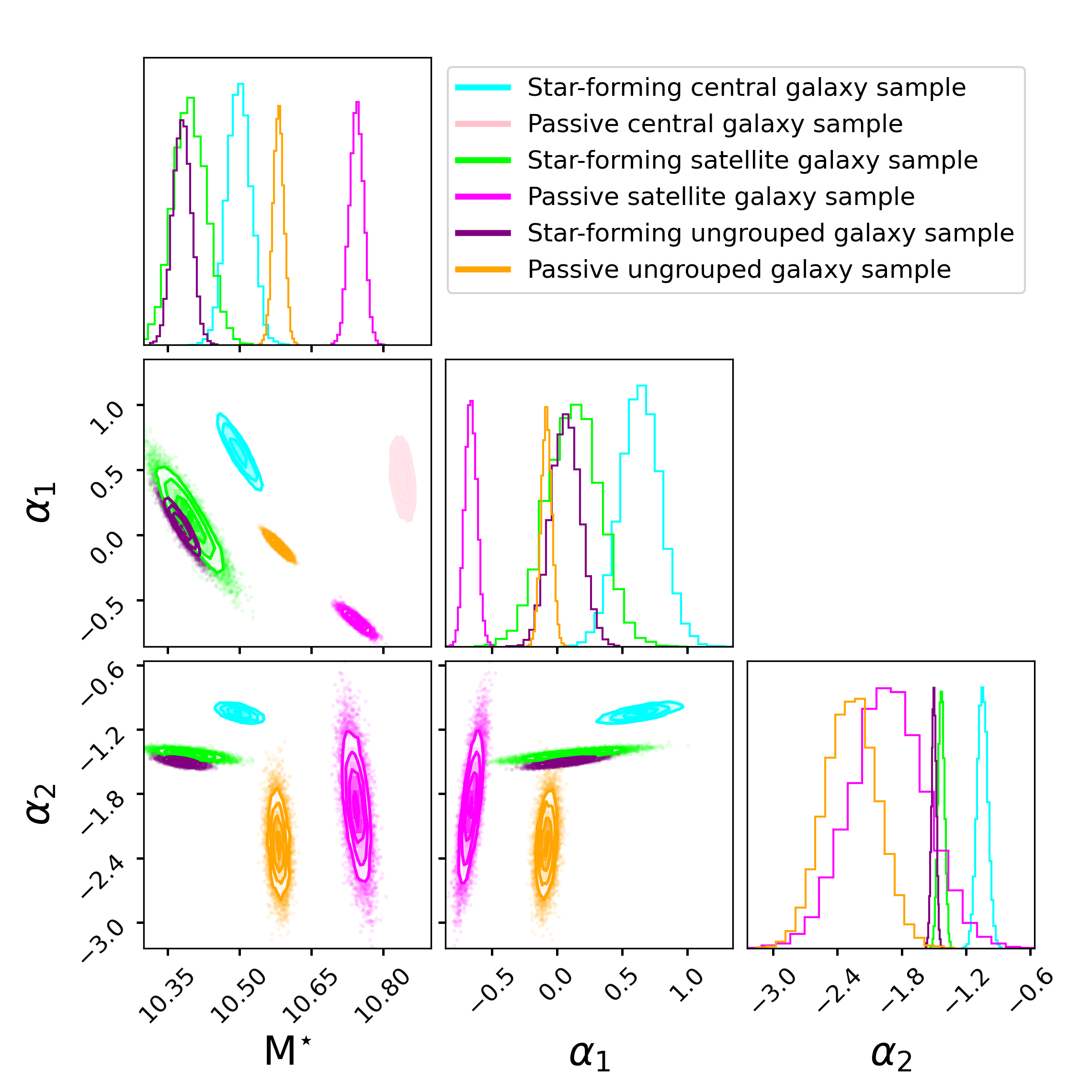}
\caption{Same as Fig.~\ref{fig:appendix1}, but for the subsamples indicated.}
\label{fig:appendix5}
\end{figure}

\begin{figure}
\centering
\vspace{-0.4cm}
\includegraphics[width=0.4\textwidth]{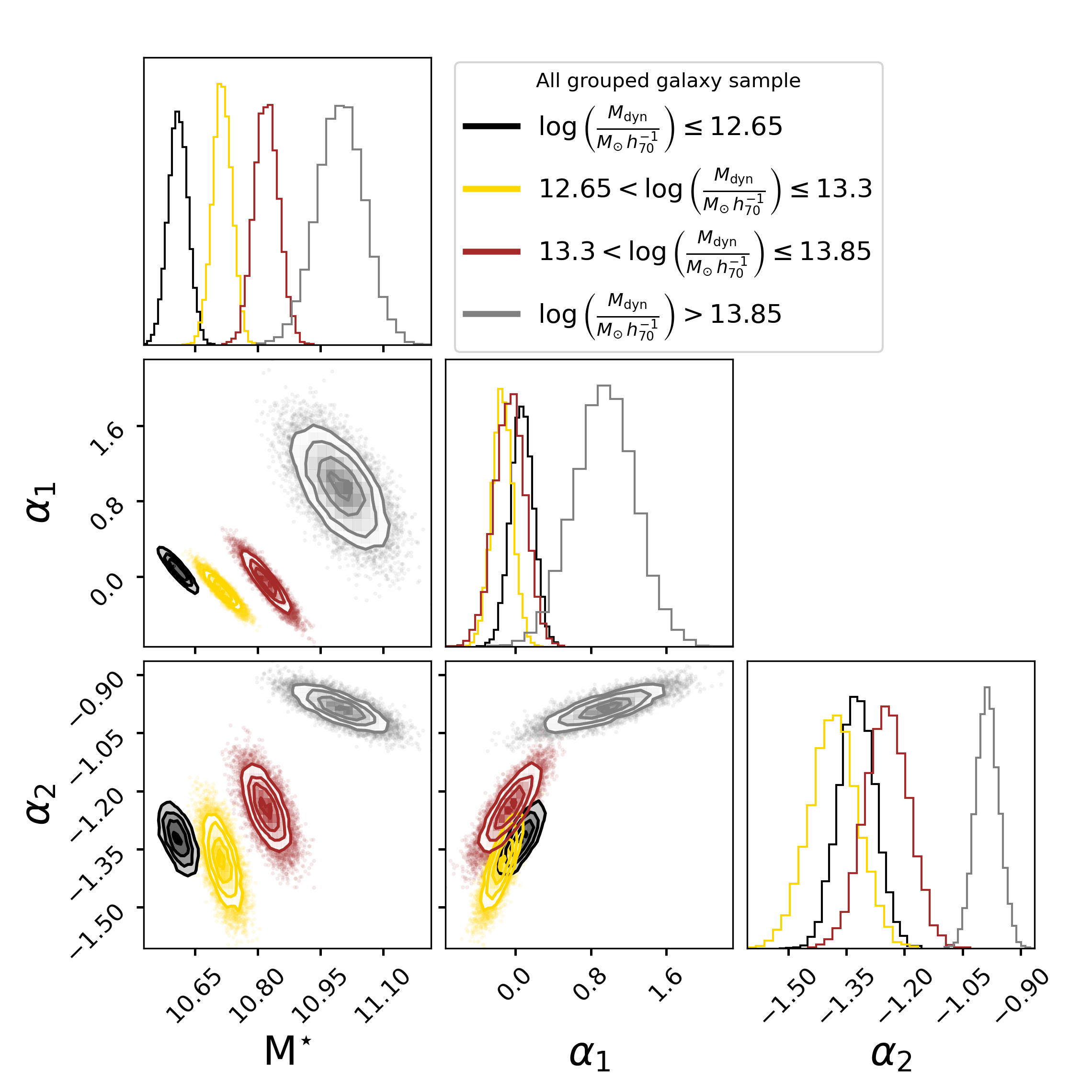}
\caption{Same as Fig.~\ref{fig:appendix1}, but for the subsamples indicated.}
\label{fig:appendix6}
\end{figure}

\begin{figure}
\centering
\vspace{-0.3cm}
\includegraphics[width=0.4\textwidth]{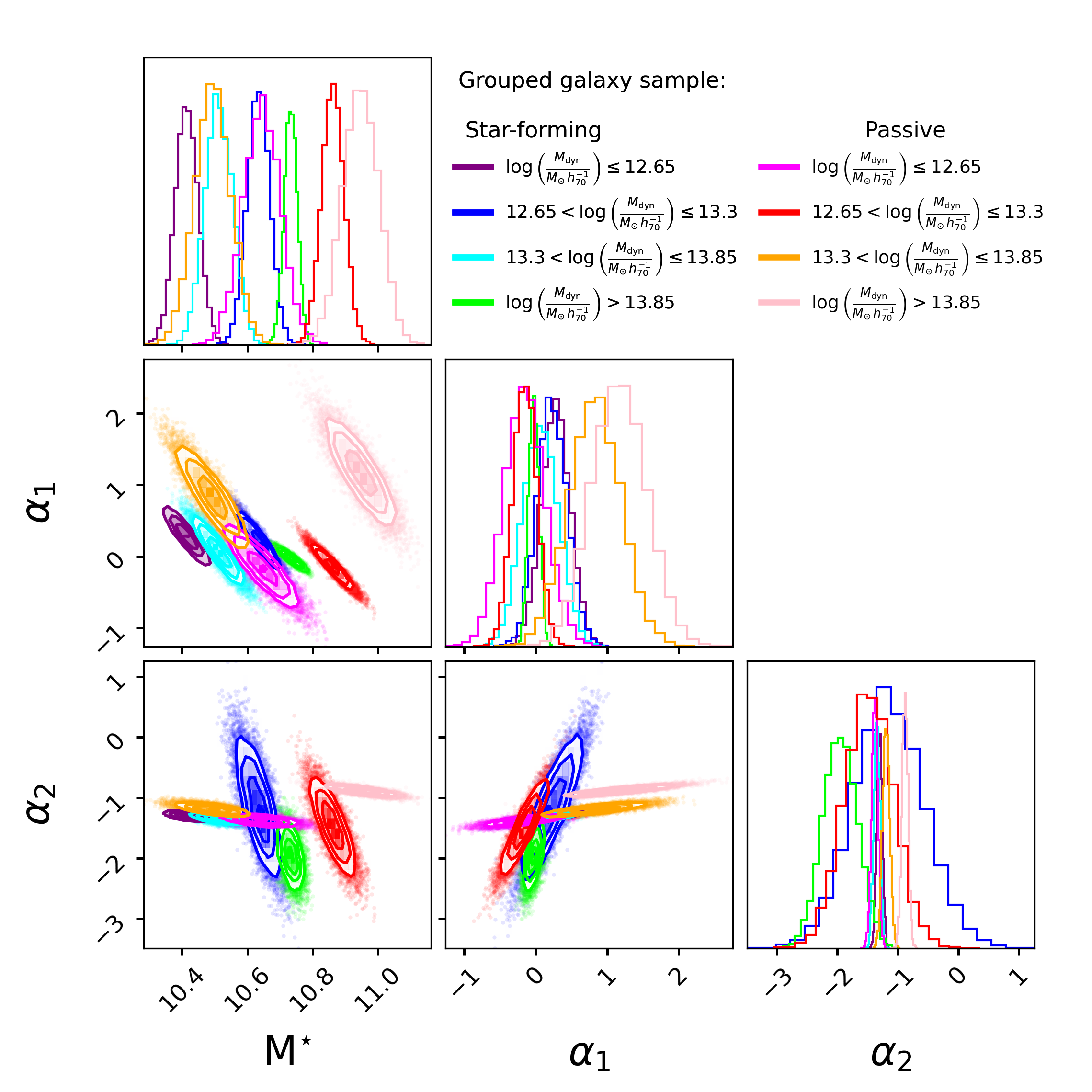}
\caption{Same as Fig.~\ref{fig:appendix1}, but for the subsamples indicated.}
\label{fig:appendix7}
\end{figure}

\begin{figure}
\centering
\vspace{-0.3cm}
\includegraphics[width=0.4\textwidth]{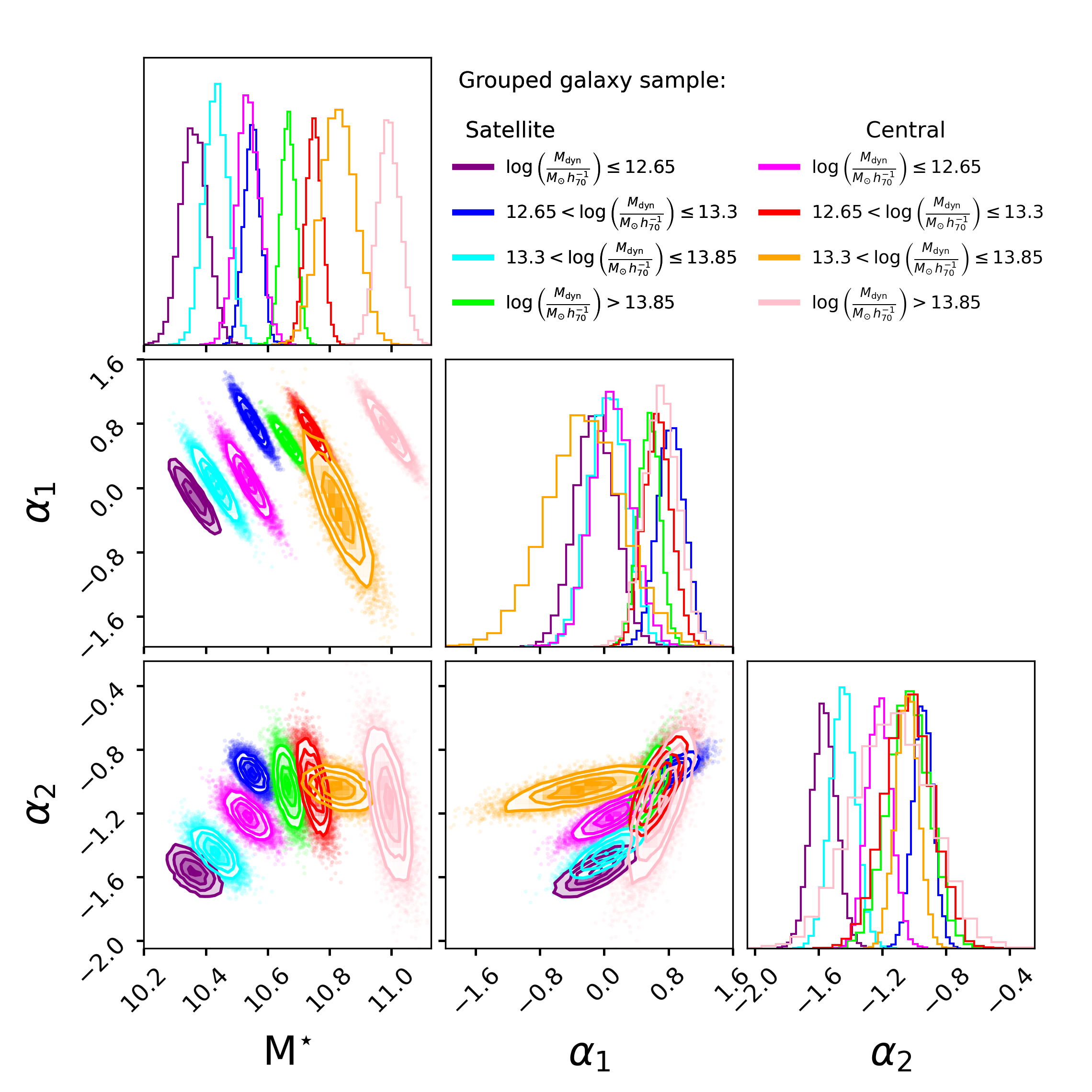}
\caption{Same as Fig.~\ref{fig:appendix1}, but for the subsamples indicated..}
\label{fig:appendix8}
\end{figure}

For each subsample, the contours correspond to the $1$-$\sigma$, $2$-$\sigma$ and $3$-$\sigma$ confidence levels. Figures~\ref{fig:appendix1}--\ref{fig:appendix8} are grouped according to the corresponding sections of the main text: Figs.~\ref{fig:appendix1}--\ref{fig:appendix3} for Sect.~\ref{6.1}, Figs.~\ref{fig:appendix4} and \ref{fig:appendix5} for Sect.~\ref{6.2}, and Figs.~\ref{fig:appendix6}--\ref{fig:appendix8} for Sect.~\ref{6.3}.

\section{Robustness of the GSMF fits to the star-forming/passive classification}
\label{app:robustness}
In this Appendix, we investigate to what extent our results depend on the specific division between star-forming and passive galaxies, that is, how robust our best-fit double Schechter parameters are against small changes in the dividing line used to separate the two populations in Sect.~\ref{4.1}. To address this, we perform the following experiment. We first identify the 1$\sigma$ confidence region based on the posterior distribution of the slope and intercept defining our fiducial dividing line in the $\log M-\log$ SFR plane. From this region, we randomly extract 100 different (slope, intercept) pairs. Each of these defines a slightly different separation between star-forming and passive galaxies. For each resulting star-forming subsample, we compute the corresponding GSMF and perform a double Schechter fit. This procedure allows us to estimate the sensitivity of the fitted parameters to small variations in the classification between the two populations. We find that the resulting uncertainty on $\log M^{\star}$ is $\sim 0.02$, i.e.\ of the same order of magnitude as the random error obtained using our fiducial dividing line (cf. Table~\ref{tab:ds1}), while the errors on $\alpha_{1}$ and $\alpha_{2}$ are subdominant in comparison ($0.03$ and $0.01$, respectively). This test, performed on the star-forming population, suggests that our results are not significantly affected by small perturbations in the classification.

\end{document}